\documentclass[]{aa}
\usepackage{graphicx}
\usepackage{latexsym}
\usepackage{color}
\usepackage{longtable}    
\usepackage{lscape}
\usepackage{natbib}
\bibpunct{(}{)}{;}{a}{}{,} 
\newcommand{\teff}{$T_{\rm eff}$}
\newcommand{\logg}{$\log g$}
\newcommand{\vsini}{$v\sin i$}

\newcommand{\kms}{km\,s$^{-1}$}

\newcommand{\Msun}{{$M_{\odot}$}}
\newcommand{\Lsun}{{$L_{\odot}$}}

\newcommand{\Mstar}{{$M_{\star}$}}
\newcommand{\Lstar}{{$L_{\star}$}}

\newcommand{\Macc}{{$\dot{M}_{\rm acc}$}}

\begin{document}

\title{X-Shooter spectroscopy of young stellar objects in Lupus.}
\subtitle{Atmospheric parameters, membership and activity diagnostics\thanks{Based on observations collected at the Very Large Telescope 
of the European Southern Observatory at Paranal, under programs 084.C-0269(A), 085.C-0238(A), 086.C-0173(A), 087.C-0244(A), 089.C-0143(A),
095.C-0134(A), 097.C-0349(A), and archive data of programmes 085.C-0764(A) and 093.C-0506(A).
}~\fnmsep\thanks{Tables~\ref{Tab:param}, \ref{Tab:fluxes}, and \ref{Tab:NIR} are also available at the CDS via anonymous ftp to {\tt cdsarc.u-strasbg.fr (130.79.128.5)} or via 
{\tt http://cdsarc.u-strasbg.fr/viz-bin/qcat?J/A+A/?/?}. }}

\author{A. Frasca\inst{1}
\and	K. Biazzo\inst{1}
\and 	J. M. Alcal\'a\inst{2}
\and	C. F. Manara\inst{3}
\and	B. Stelzer\inst{4}
\and	E. Covino\inst{2}
\and    S. Antoniucci\inst{5}
}

\offprints{A. Frasca\\ \email{antonio.frasca@oact.inaf.it}}

\institute{INAF - Osservatorio Astrofisico di Catania, via S. Sofia, 78, 95123 Catania, Italy
\and
INAF -- Osservatorio Astronomico di Capodimonte, via Moiariello, 16, 80131 Napoli, Italy
\and
Scientific Support Office, Directorate of Science, European Space Research and Technology Centre
(ESA/ESTEC), Keplerlaan 1, 2201 AZ Noordwijk, The Netherlands
\and
INAF -- Osservatorio Astronomico di Palermo,  Piazza del Parlamento 1,
  I-90134 Palermo, Italy
\and
INAF -- Osservatorio Astronomico di Roma, via Frascati 33, I-00078 Monte Porzio Catone, Italy
}

\date{Received  22 Nov 2016 / Accepted 3 March 2017}

\abstract 
{} 
{A homogeneous determination of basic stellar parameters of young stellar object (YSO) candidates is needed to confirm their
pre-main sequence evolutionary stage, membership to star forming regions (SFRs), and to get 
reliable values of the quantities related to chromospheric activity and accretion. 
}
{We used the code ROTFIT and synthetic BT-Settl spectra for the determination of the atmospheric parameters (\teff\ and \logg),  
the veiling ($r$), the radial (RV) and projected rotational velocity (\vsini), from X-Shooter spectra of 
102 YSO candidates (95 of infrared Class\,II and seven Class\,III) in the Lupus SFR. 
The spectral subtraction of inactive templates, rotationally broadened to match the $v\sin i$ of the targets, 
enabled us to measure the line fluxes for several diagnostics of both chromospheric activity and accretion, such as H$\alpha$, H$\beta$, 
\ion{Ca}{ii} and \ion{Na}{i} lines.
} 
{We have shown that 13 candidates can be rejected as Lupus members based on their discrepant RV with respect to Lupus and/or the very low \logg\ values. At 
least 11 of them are background giants, two of which turned out to be lithium-rich giants.
Regarding the members, we found that all Class\,III sources  have H$\alpha$ fluxes compatible with a pure chromospheric activity, 
while objects with disks lie mostly above the boundary between chromospheres and accretion. 
YSOs with transitional disks displays both high and low H$\alpha$ fluxes.
We found that the line fluxes per unit surface are tightly correlated with the accretion luminosity ($L_{\rm acc}$) derived from the Balmer continuum 
excess. This rules out that the relationships between $L_{\rm acc}$ and line luminosities found in previous works are simply due to calibration effects.
We also found that the \ion{Ca}{ii}-IRT flux ratio, $F_{\rm CaII 8542}/F_{\rm CaII 8498}$, is always small, indicating an optically thick emission
source. The latter can be identified with the accretion shock near the stellar photosphere. 
The Balmer decrement reaches instead, for several accretors, high values typical of optically thin emission, suggesting that the Balmer emission originates 
in different parts of the accretion funnels with a smaller optical depth. 
}
{}

   \titlerunning{X-Shooter spectroscopy of young stellar objects in Lupus}
      \authorrunning{A. Frasca et al.}

\maketitle

\section{Introduction}
\label{Sec:Intro}

A full characterization of members of young open clusters (OCs) and star forming regions (SFRs)
is necessary to understand the mechanisms of star and planet formation, as well as the 
different phenomena occurring during the early phases of stellar evolution. 
It is therefore mandatory to derive, as homogeneously as possible, the basic physical parameters 
of young stellar objects (YSOs), such as effective temperature (\teff), surface gravity (\logg), mass and 
projected rotational velocity (\vsini). These quantities are closely related to the 
evolutionary stage of the YSO and are also directly involved in the generation of stellar
magnetic fields by the dynamo processes. 	
In the context of the magnetospheric accretion model, magnetic fields play a key role in the infall of 
matter from the disk onto the central star \citep[][and references therein]{Hartmann1994,Hartmann2016}.	
They are also responsible for the strong activity observed from the photospheres to the chromospheres 
and coronae of YSOs and late-type stars in general \citep[e.g.,][and references therein]{Feigelson1999, Berdyugina2009}.
 Circumstellar disks, stellar rotation, magnetic activity and accretion are all  
strongly involved in the process of planet formation \citep[e.g.,][and references therein]{HawleyBalbus1991,Brandenburg1995,Larson2003}.  

Deriving the basic stellar parameters is particularly hard for objects belonging to very young OCs and SFRs where 
several phenomena, such as fast rotation, cloud extinction and mass accretion, affect their spectra, making  
the analysis of their photospheres difficult. 
In particular, the veiling of the spectra, arising from accretion shocks  and, possibly, from disk gas 
inside the dust sublimation radius, dilutes the photospheric lines and it must be properly taken into account 
to derive reliable stellar parameters \citep[see, e.g.,][and references therein]{Frasca2015}.	
Moreover, emission lines, which are numerous and very strong and broad in the spectra of highly accreting
objects, need to be masked, leaving only limited spectral ranges suitable for the analysis of photospheric lines.

 On the other hand, it is well known that the luminosity of several emission lines of the Balmer series, 
the \ion{He}{i} and \ion{Ca}{ii} lines are well correlated with the accretion luminosity 
\citep[e.g.,][and references therein]{Herczeg2008,Alcala2014,Alcala2017},
as well as with hydrogen recombination lines in the near-infrared \citep{Muzerolle1998,Calvet2004,Alcala2017},
highlighting the importance of these emission features as accretion diagnostics.

Spectra with both a very wide wavelength coverage and a high or intermediate resolution are particularly helpful,  as they
offer the possibility to find spectral intervals free from or little affected by these issues, where fully resolved 
absorption lines can be modeled to consistently derive atmospheric parameters, radial velocity, \vsini, and veiling.  
For this reason,  the VLT/X-Shooter spectrograph \citep{Vernet11}, with its wide wavelength coverage, is an ideal instrument to reach these goals.

Thanks to their relatively close distance to the Sun (150--200 pc, \citealt{Comeron2008}) and the large numbers of Class\,II and 
Class\,III infrared sources they contain, the dark clouds in Lupus are among the best sites for which a spectroscopic determination 
of the atmospheric parameters for YSOs can be carried out down to the hydrogen burning mass limit. In addition, extinction to the 
individual YSOs is relatively low in comparison with many other SFRs.

The selection of candidates and the characterization of some of the confirmed Lupus members, both physical and kinematical, has 
been performed in recent years \citep[see, e.g.,][]{Evans2009,Comeron2009,Comeron2013,Galli2013,Herczeg2014,Bustamante2015}. 
However, most of these works are based on non-simultaneous optical and IR photometry and low-resolution spectroscopy. Moreover, the atmospheric 
parameters are derived mostly for the Class\,III sources.	 
A homogeneous determination of the atmospheric parameters, \vsini, and veiling  for a large	
sample of Class\,II sources in Lupus is still lacking. 
In this work we aimed at filling-in this gap. Our study is based on X-Shooter spectra of more than one hundred YSO candidates
in the Lupus I, II, III, and IV clouds. 

The high quality and the intermediate resolution of these spectra allowed us also to investigate the behavior of 
hydrogen, calcium and sodium emission lines that are diagnostics of both chromospheric and accretion activity.

The paper is organized as follows. In Sect.~\ref{Sec:Data} we present the data sample. The analysis of the spectra for the
determination of stellar parameters and the radial velocities is described in Sect.~\ref{Sec:Analysis}. The results are
discussed in Sect.~\ref{Sec:Results}, where the properties of the targets, the membership to the Lupus SFR, and the line fluxes 
per unit surface are discussed. 
The main conclusions are summarized in Sect.~\ref{Sec:Conclusions}.

\section{Data sample}
\label{Sec:Data}

In total, we analyzed X-Shooter spectra of 102 objects, 43 of which were obtained in 2010, 2011, and 2012, during the INAF guaranteed 
time observations \citep[GTO,][]{Alcala2011a}. 
The remaining sources were observed in the framework of the ESO-P95 proposal from March to August 2015 (48 objects) and the ESO-P97 
proposal from May to June 2016 (5 objects).
In addition,  X-Shooter spectra of six YSOs were retrieved from the ESO archive. 
Our sample comprises sources belonging to the infrared (IR) Class II. However, seven Class~III objects were also 
included. Six of them have been already investigated by \citet{Stelzer2013} and \citet{Manara2013}. 
The sample accounts for roughly half of the entire known population of YSOs in Lupus, but $\sim$\,95\,\% of disky YSOs. 
For the criteria of target selection and  details about the data reduction	
the reader is referred to \citet{Alcala2014, Alcala2017}.

We remark that 13 out of the 103 objects studied here were suspected to be non-members based on the appearance of their spectrum 
\citep{Alcala2017}. The analysis we performed in this work allowed us to definitely reject them as members. 

The target list is given in Table~\ref{Tab:param} together with the heliocentric Julian Date of  observation 
and some of the parameters derived in this work.
The spectra of Sz\,68, Sz\,74, Sz\,83, Sz\,102, Sz\,105, and the archive data (i.e., IM~Lup, Sz\,77, EX~Lup, GQ~Lup, Sz\,76, 
RXJ1556.1-3655) were acquired using the 0$\farcs$5/0$\farcs$4/0$\farcs$4 slits in the UVB/VIS/NIR arms, respectively. 
All the other targets were instead observed with the 1$\farcs$0/0$\farcs$9/0$\farcs$9 slits in the UVB/VIS/NIR arms, respectively.

For our analysis, particularly for the \vsini\  measurement, it is important to know the actual spectral resolution. 	
To this purpose we used Th-Ar calibration spectra taken during our observing runs.
We found that the resolving power of VIS spectra taken with the 0$\farcs$9 slit was $R\simeq 8400$ in the 
spectral region used for the \vsini\  determination ($\lambda\sim$\,9700\AA), while in the highest resolution mode 
(0$\farcs$4 slit) it is about 17\,000. 
More details can be found in  Appendix\,\ref{sec:resolution}. 

\section{Atmospheric parameters, veiling, radial and projected rotational velocities}
\label{Sec:Analysis}

The determination of the atmospheric parameters ($T_{\rm eff}$ and $\log g$), the veiling $r$, the projected rotational velocity 
($v \sin i$) and the radial velocity (RV) was accomplished through the code ROTFIT \citep[e.g.,][]{Frasca2006,Frasca2015}. 

We adopted as templates a grid of synthetic 
BT-Settl spectra \citep{Allard2012} with a solar iron abundance and effective temperature in the range 2000--6000\,K (in 
steps of 100\,K) and $\log g$ from 5.5 to 0.5 dex (in steps of 0.5 dex).  
 This grid covers the wide range in spectral type (SpT), from early-K to late-M, spanned by our targets and extends also
to lower gravities typical of  giants.

The code ROTFIT was already applied to X-Shooter spectra of Class~III sources in \citet[][]{Stelzer2013}. However, 
the code was improved since then with the inclusion of further spectral 
segments to be analyzed both in the VIS and UVB spectra. Furthermore, in this version we implemented an accurate 
search for the minimum of the chi-square ($\chi^2$) in the space of parameters and the evaluation of the errors based on the 1$\sigma$ 
confidence levels. The veiling, $r$, was also considered as a free parameter.
Details on the application of the code to the X-Shooter spectra are given in Appendix~\ref{sec:ROTFIT}.

\subsection{Radial velocity}\label{subsec:analysis_rv}

The RV was needed to align the synthetic template spectra with the observed ones. 
The peak of the cross-correlation function (CCF) was fitted with a Gaussian for a more accurate determination 
of its center (see the insets in Fig.~\ref{fig:ROTFIT_vsini}). For each i$^{th}$ spectral segment, the error of the $RV_{\rm i}$ value, 
$\sigma_{RV_{\rm i}}$, was estimated by the fitting procedure \textsc{curvefit} \citep{Bevington}, taking into account the CCF noise, 
which was evaluated as the standard deviation of the CCF values outside the peak.  Then, the heliocentric correction of the RV measurements 
was performed with the IDL procedure {\sc helcorr}.
The final RV for each star, reported in Table~\ref{Tab:param}, is the weighted mean of the values obtained from all analyzed 
spectral segments, using as weights $w_{\rm i}=1/\sigma_{RV_{\rm i}}^2$.
The standard error of the weighted mean was adopted  as the RV uncertainty.
These errors range from 0.7 to 12 \kms, with a median value of 2.3\,\kms.
We note that, as expected, the RV errors tend to be larger for the stars with higher $v\sin i$ and/or with low S/N spectra, or strongly 
affected by veiling (see Table~\ref{Tab:param}). 

\subsection{Accuracy of \teff\ and \logg\ determinations and comparisons with the literature}
\label{subsec:accuracy}

The \teff\ errors reported in Table~\ref{Tab:param} range from about 30 to 250\,K, depending on \teff\  (and on the S/N),
with a median value of 70\,K. The only exception is  SSTc2dJ161045.4-385455 ($\sigma_{T_{\rm eff}}$\,=\,514\,K), a likely non-member 
with a low-signal spectrum.

\begin{figure}[th]
\hspace{-.3cm}
\includegraphics[width=9cm]{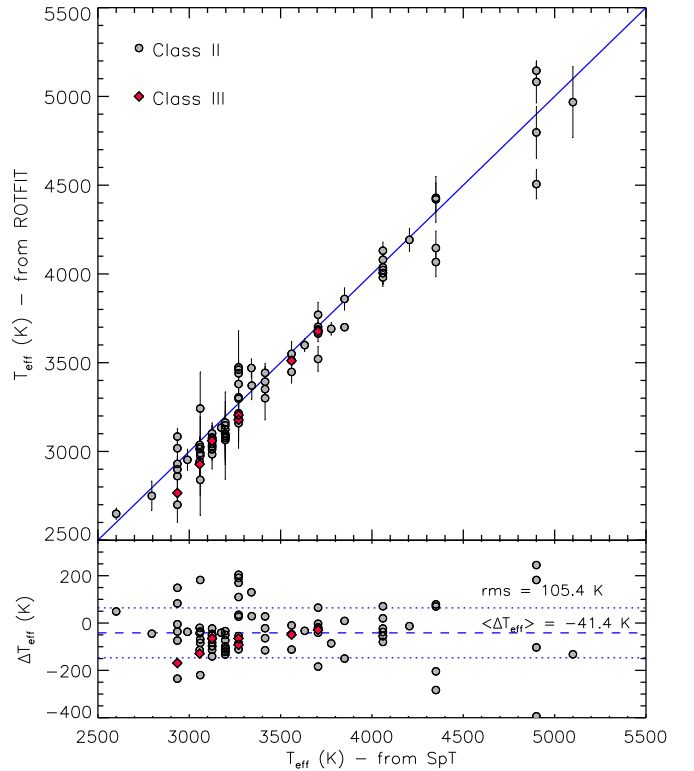}	
\caption{{\it Top panel)} Comparison between the effective temperature determined with ROTFIT and that derived from the MK spectral 
classification performed for the Lupus members by \citet{Manara2013} and \citet{Alcala2014,Alcala2017}.
The continuous line in the top panel represents the 1:1 relationship.
The \teff\ differences ({\it Bottom panel}) show a mean value of about $-40$\,K (dashed line) and a standard 
deviation of about 100~K (dotted lines).}
\label{Fig:Teff_comp}
\end{figure}

To estimate the external accuracy of \teff, we have compared our values with those derived
from the spectral classification performed by \citet{Manara2013} for the Class~III objects and by \citet{Alcala2014,Alcala2017}
for the Class~II sources, who adopted the temperature scales of \citet{Luhman2003} and \citet{KenyonHartmannn1995} for the 
M- and K-type stars, respectively. 
The comparison is shown in Fig.~\ref{Fig:Teff_comp}. The general agreement of the two \teff\ determinations is evident,
with a scatter of the residuals $\simeq$100\,K rms. 
Nevertheless, this comparison displays a systematic offset between the two \teff\  determinations, which is more evident
in the M-type domain, with the \teff\  values derived from SpT being about 40\,K higher, on average, than those from ROTFIT. 
This offset is likely due to the different \teff\  scales between the BT-Settl models and the empirical SpT--\teff\  calibration
defined for M-type stars by \citet[][]{Luhman2003}.   
We note that the \teff\  values of \citet{Luhman2003} for early M-type stars are systematically higher by $\sim$\,70\,K than those 
reported by \citet[][]{PecautMamajek2013} for young stars, which are based on the fitting of the optical/IR spectral energy distributions.  

\begin{figure}[th]
\hspace{-.3cm}
\includegraphics[width=9cm]{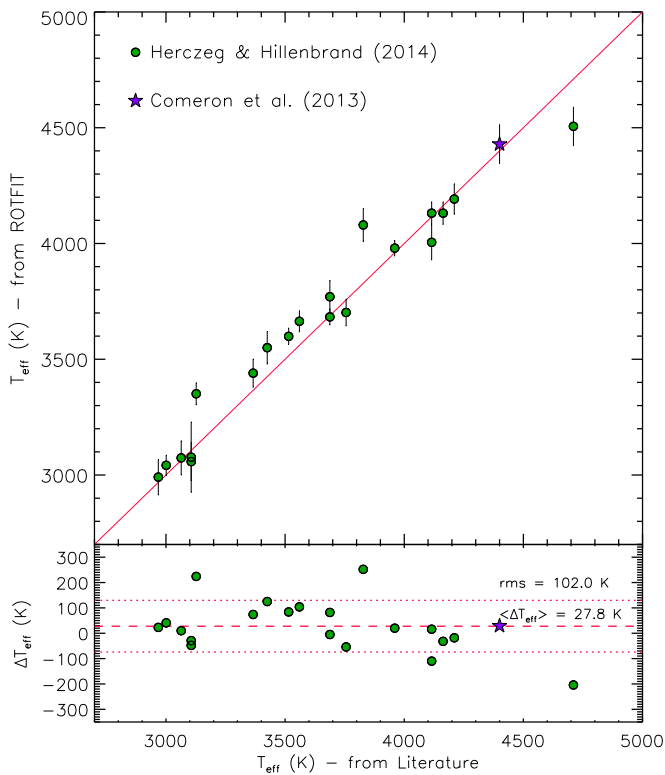}	
\caption{{\it Top panel)} Comparison between the effective temperature determined with ROTFIT and that derived by \citet{Herczeg2014}
 and \citet{Comeron2013}. The solid line in the top panel represents the 1:1 relationship.
The \teff\ differences ({\it Bottom panel}) show a mean value of about $+28$\,K (dashed line) and a standard 
deviation of about 100~K (dotted lines).}
\label{Fig:Teff_comp_literature}
\end{figure}

As a further check for the accuracy of our determinations we have searched for spectroscopic \teff\ measurements in the literature, not
based on X-Shooter data. 
We compared our results with the values reported by \citet{Comeron2013} and \citet{Herczeg2014}. Both works are based on low resolution ($R=500$--1000) 
optical spectra.
We have only one star in common with \citet{Comeron2013},
namely \object{SSTc2d160836.2-392302}, for which the \teff\ agreement is excellent (see Fig.~\ref{Fig:Teff_comp_literature}). 
Twenty of our targets have been observed by \citet{Herczeg2014}, who list the spectral type 
derived from suitable spectral indices. 
We converted their SpT into \teff\ by using the calibration put forward by them. The comparison is 
displayed in Fig.~\ref{Fig:Teff_comp_literature}.
The agreement is evident, with an average offset of +28\,K and a scatter of about 100\,K, as in Fig.~\ref{Fig:Teff_comp}. The latter 
can be considered as the typical external accuracy of our \teff\ values.
We note that, despite the lower number of points, the agreement in the M-type domain looks better than the one shown in Fig.~\ref{Fig:Teff_comp}. 
This is likely owing to the \teff\ scale, based on BT-Settl spectra, adopted both in the present work and in \citet{Herczeg2014}. 

Concerning \logg, the errors are in the range 0.1--0.5\,dex, with a median value of 0.21 dex. Only four objects have larger errors 
(up to 0.9\,dex). We did not find other estimates of gravity in the literature, except for the value of \logg=4.0 reported by 
\citet{Comeron2013} for \object{SSTc2d160836.2-392302}, which is consistent with ours (4.04$\pm$0.13).

\subsection{Projected rotational velocity}
\label{subsec:accuracy:vsini}
The intermediate resolution of X-Shooter allows us to measure \vsini\  only for relatively fast rotators. 
To check the minimum \vsini\  that can be measured in these spectra and to estimate the \vsini\  errors, 
we have run Monte Carlo simulations on synthetic spectra with the same resolution and sampling of X-Shooter, following the 
prescriptions given in \citet{Frasca2015}.
We found that the minimum detectable \vsini\ with the 0$\farcs$9 slit  is about 8\,\kms. 
Therefore, we considered as upper limits all \vsini\  values lower than 8\,\kms. We also evaluated the upper limit on \vsini\ 
for the spectra acquired with the  0$\farcs$4 slit, finding a value of 6\,\kms. Details about the Monte Carlo 
simulations can be found in Appendix\,\ref{sec:MonteCarlo}.  The errors of \vsini\ range from 1 to 15 \kms (Table~\ref{Tab:param}), 
with a median value of about 4\,\kms.

Determinations of \vsini\ can be found in the literature only for a handful of the brightest sources in our sample. For Sz\,68, \cite{Torres2006} 
report a value of \vsini$=38.9\pm0.6$\,\kms, which is compatible with our determination of 39.6$\pm1.2$\,\kms. \citet{Dubath1996} 
give only a lower limit of \vsini$>60$\,\kms\ for Sz\,121, which is one of the most rapidly rotating stars in our sample (\vsini$=87$\,\kms). 
The same authors report \vsini$=10.2\pm1.6$\,\kms\ for Sz\,108, for which we find an upper limit of 8\,\kms. The value of 22.4\,\kms\ 
quoted in the Catalog of Stellar Rotational Velocities \citep{Glebocki2005} for RY~Lup is marginally consistent with ours 
(\vsini$=16.3\pm5.3$\,\kms). 
Finally, in the same catalog, \vsini$=14.2$\,\kms\ is reported for Sz\,98, while we find \vsini$<8$\,\kms.

\subsection{Veiling}
\label{subsec:accuracy:veiling}
After several tests made with our code on Class\,III and X-Shooter archive spectra of non-accreting stars, we realized that values 
of veiling as large as 0.2 can be found, although no veiling is expected in such cases. This is probably a combined effect of the intermediate 
spectral resolution, small differences in the continuum setting between target and template spectra, and some  trade-off between parameters.
For this reason, all values $r\leq 0.2$ are considered as not significant and have been replaced with an upper limit of 0.2 in 
Table~\ref{Tab:param}.  We measured significant values of veiling ($r> 0.2$) for 12, 19, and 44 YSOs at 970\,nm, 710\,nm, and 620\,nm, 
respectively. This trend is in line with the enhancement of $r$ with decreasing wavelength (see Fig.~\ref{fig:veil}).

\citet{Herczeg2014} report values of the veiling at 750\,nm for sixteen of our targets. For about a half of them we derived only upper limits, 
$r_{710}\leq 0.2$, while for the remaining YSOs the veiling is always rather small. However, the comparison of these data  
shows that we can measure the veiling at 710\,nm whenever \citet{Herczeg2014} quote $r_{750}> 0.2$.

\begin{figure}[th]
\hspace{0cm}
\includegraphics[width=8cm]{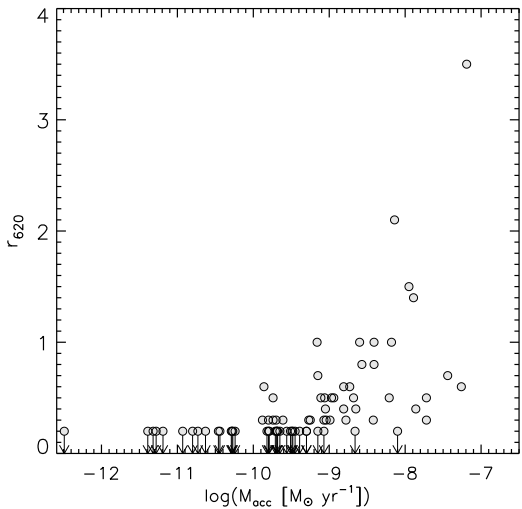}	
\caption{Veiling at 620\,nm as a function of the mass accretion rate measured by \citet{Alcala2017} on the same spectra.}
\label{Fig:veil_macc}
\end{figure}

 Using the mass accretion rates derived by \citet{Alcala2017} on the same data set, we found that high values of veiling ($r_{620}\ge 1.0$) are measured only for 
the strongest accretors (\Macc$>10^{-9}$, see Fig.~\ref{Fig:veil_macc}).

\section{Results}
\label{Sec:Results}

\subsection{Membership}\label{subsec:membership}

The atmospheric parameters, the RV, and the equivalent width of the lithium line at 6707.8\,\AA\  ($EW_{\rm Li}$) 
were used to verify the membership of our targets to the Lupus SFR. Here we used the $EW_{\rm Li}$ values derived by
\citet{Biazzo2016}.

\begin{figure}[th]
\hspace{-1cm}
\includegraphics[width=9.5cm]{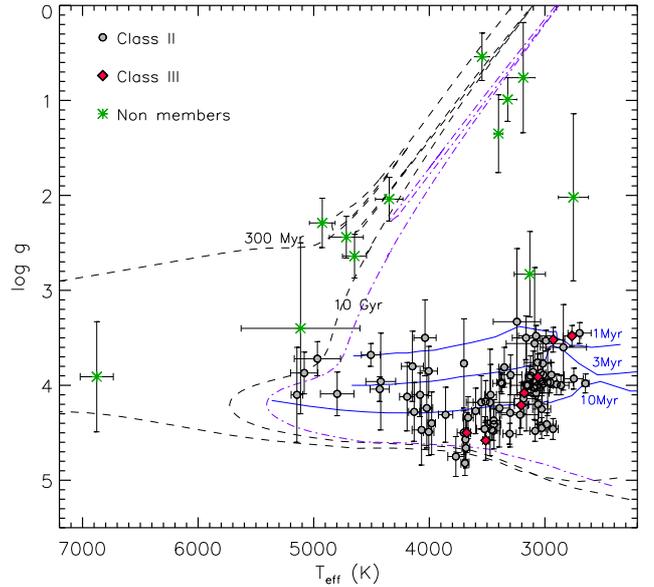}	
\caption{$T_{\rm eff}$--$\log g$ diagram for all the objects for which a \logg\  value has been derived with ROTFIT. 
Grey dots represent Class~II sources, red diamonds denote Class~III ones, while the non-members are indicated by green 
asterisks. The full blue lines are the PMS isochrones from \citet{Baraffe2015} at 1, 3, and 10\,Myr and solar metallicity.
The  black dashed lines are post-MS isochrones from the PARSEC database \citep{Bressan2012} for a 
solar metallicity at 300\,Myr and 10\,Gyr. The magenta dot-dashed line is the PARSEC isochrone at 
10\,Gyr and Z=0.060.}
\label{Fig:LoggTeff}
\end{figure}

The  \logg\  values derived with ROTFIT allow us to identify in our sample contaminants whose IR colors mimic
those of cool very young stars.  
In Fig.~\ref{Fig:LoggTeff} we show the \logg\--\teff\ 
diagram of our targets, where the stars suspected to be non-members have been highlighted with green asterisks.
We also plot the pre-main sequence (PMS) isochrones of \citet{Baraffe2015} and the post-main sequence isochrones from the PARSEC
database \citep{Bressan2012}.

 All green asterisks at \teff$<5000$\,K have $\log g<3.0$ and correspond most likely to giants. Four of these are indeed in the red 
giant branch (RGB) and possibly close to the red giant clump; a few others, with lower gravity, could be in the AGB phase.
The star with the highest temperature,  SSTc2dJ161148.7-381758, could be a MS star unrelated to the Lupus SFR, as indicated by its 
RV=$-47.6$\,\kms.  The green asterisk with the largest errors located at \teff=5114\,K and \logg=3.4 corresponds to SSTc2dJ161045.4-385455.
Its position on the \logg\--\teff\ plane may be compatible with the PMS locus, taking into account the errors, but the value of 
RV=$-115.1$\,\kms\ strongly suggests that it is not a member of the Lupus SFR.

\begin{figure}[th]
\hspace{-0.1cm}
\vspace{-.2cm}
\includegraphics[width=8.8cm]{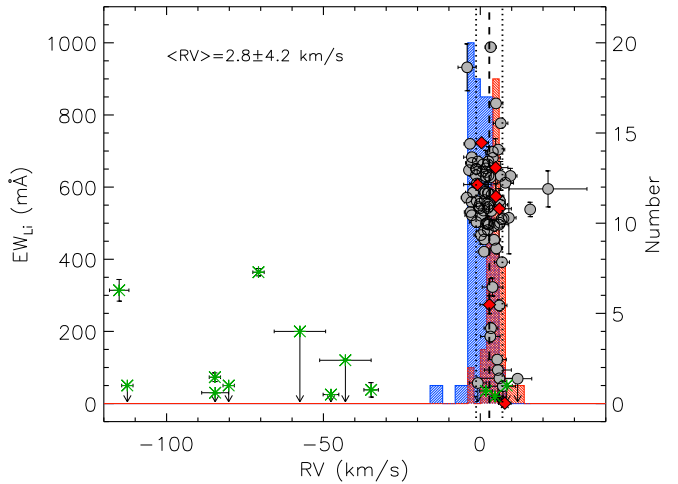}	
\caption{Equivalent width of the \ion{Li}{i} $\lambda$6708\,\AA\  line from \citet{Biazzo2016} versus radial velocity measured in the 
present paper. The meaning of the symbols 
is as in Fig.~\ref{Fig:LoggTeff}. The average RV of the Lupus members (2.8\,\kms) is marked by the vertical dashed line,
while the dotted lines delimit the 1$\sigma$ confidence region. The blue and red histograms represent the RV distributions
for the on-cloud and off-cloud Lupus stars found by \citet{Galli2013}.
 }
\label{Fig:RVEWLi}
\end{figure}

The other two indicators of membership, namely the RV and $EW_{\rm Li}$, are plotted in Fig.~\ref{Fig:RVEWLi} with the same symbols as in 
Fig.~\ref{Fig:LoggTeff}. 
It is worth noting that, except for two objects (SSTc2dJ160708.6-394723 and SSTc2dJ161045.4-385455), all the stars classified as non-members 
based on the \logg\--\teff\ diagram and/or discrepant RV, have undetectable \ion{Li}{i} lines. 
The non-member with the highest $EW_{\rm Li}$ (364\,m\AA) is SSTc2dJ160708.6-394723. For this object we estimate \teff$\simeq\,4650$\,K and 
\logg$\simeq 2.6$, thus it lies in the RGB region of the \logg\--\teff\ diagram (Fig.~\ref{Fig:LoggTeff}). Its RV is inconsistent with the Lupus SFR. 
 Likewise, SSTc2dJ161045.4-385455 has a rather large $EW_{\rm Li}$ (about 300\,m\AA), but, as already mentioned, its 
RV\,=\,$-115.1$\,\kms\ and the absence of emission lines rule out its membership to the Lupus SFR. Although with large errors, its atmospheric 
parameters suggest a G-type giant or subgiant.  These two stars are most likely lithium-rich giants. More details are given in Appendix\,\ref{sec:peculiar}.
 
Three other non-members based on the \logg\--\teff\ diagram (Sz\,78, Sz\,105, and SSTc2dJ161222.7-371328) have instead RVs compatible with Lupus, but 
no Lithium.		
These stars have all very low \logg\  values of 2.04, 0.76, and 0.99, respectively, thus ruling out a PMS nature. 

In conclusion, by using the three indicators, \logg, RV, and $EW_{\rm Li}$, we were able to reject 13 stars as Lupus members.
All the remaining 89 targets were considered as members, although a few of them have an RV slightly  outside the SFR average 
derived by us ($<RV>=2.8\pm4.2$\,\kms) and external to the RV distributions for on-cloud and off-cloud sources defined by \citet{Galli2013}.
Nonetheless, they display a large $EW_{\rm Li}$ that reinforces their membership; they could be spectroscopic binaries. 

A small group of six Class\,II sources, namely Sz\,69, Lup706, Par-Lup3-4, 2MASS\,J16085953-3856275, 2MASS\,J16085373-3914367, and 
SSTc2dJ154508.9-341734, has instead an RV well consistent with the Lupus SFR but a quite low lithium content ($EW_{\rm Li}<150$\,m\AA). 
Nevertheless, the spectra of these objects display other features that indicate their membership (c.f. strong and wide emission lines, 
veiling, etc., \citealt{Alcala2014,Alcala2017}). 
These cases will be discussed in more detail in \citet{Biazzo2016}.
 Finally, the star Sz\,94, classified as a Class\,III source with proper motions and RV compatible with Lupus, shows $EW_{\rm Li}\approx 0$, 
making its membership dubious.

\subsection{Hertzsprung-Russell diagram}\label{subsec:HR}

\begin{figure}  
\begin{center}
\includegraphics[width=8.8cm]{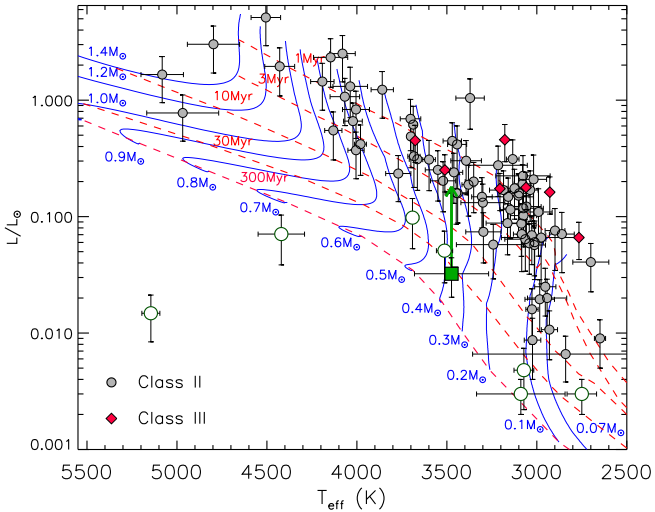}	
\caption{Hertzsprung-Russell diagram of the Lupus members. The evolutionary 
tracks of \citet{Baraffe2015} are shown as solid lines with the labels representing their masses. The isochrones at 1, 3, 10, 30 and 300~Myr 
by the same authors are shown with dashed lines. Big open circles denote subluminous objects.  The green arrowhead marks the position of
the flat source SSTc2dJ160708.6-391408  (green filled square) when assuming the bolometric luminosity of 0.18\,$L_{\sun}$ \citep[][]{Evans2009}.}
\label{fig:HR}
 \end{center}
\end{figure}

In \citet{Alcala2017} we derived masses using four different evolutionary tracks, but only for the Class\,II sources. Here we derive masses and ages for both 
the Class\,III and Class\,II sources homogeneously, and using the \teff\ values resulting from the ROTFIT analysis. 

In Fig.~\ref{fig:HR}, we report the position of our targets in the Hertzsprung-Russell (HR) diagram, where we used the \teff\  values derived 
in the present work and the luminosities calculated by \citet{Manara2013} and \citet{Alcala2014,Alcala2017},  with the same method, for 
the Class~III and Class~II sources, respectively. 
In the same figure we overplot the PMS evolutionary tracks and isochrones by \citet{Baraffe2015}. 

Most of the targets are located between the isochrones at 1 and 10 Myr, while a few of them
(Lup706, Sz106, Par-Lup3-4, Sz123B, Sz133, SSTc2dJ160703.9-391112, and Sz102), displayed with
larger open circles in Fig.~\ref{fig:HR}, appear to be subluminous. These are likely objects with rotation
axes perpendicular to the line of sight where the edge-on disks reduce the stellar luminosity by extinction 
and light scattering \citep{Alcala2014,Alcala2017}.  Several of these objects show in their optical spectra an enhancement
of outflow tracers like the low-velocity component of the \ion{O}{i} $\lambda$6300\,\AA ~line
\citep{Bacciotti2011,Natta2014,Whelan2014}. This line originates in a more extended volume with respect to the
accretion tracers that arise from a much closer region to the stellar surface, and are suppressed
by the optically thick edge-on disk. This effect is particularly important in Par-Lup3-4, Sz102
and Sz133 (\citealt{Bacciotti2011}, \citealt{Natta2014}, Nisini et al. in prep.).
Another object, namely SSTc2dJ160708.6-391408, classified as a flat IR source \citep{Merin2008}
and shown with a green filled square in Fig.~\ref{fig:HR}, appears subluminous because the derived stellar luminosity
does not account for reprocessed stellar radiation in the infalling envelope of gas and dust
(see Appendix~C in \citealt{Alcala2017}). This object is further discussed in Appendix~\ref{sec:peculiar}.

The evolutionary tracks and isochrones can be used to estimate the masses and ages of the targets from their location in the HR diagram
by minimizing the quantity:
\begin{equation}
\label{eq:HR_mass}
\chi^2 = \frac{(T_{\rm eff}-T_{\rm mod})^{2}}{\sigma_{T_{\rm eff}}^2} +\frac{(L-L_{\rm mod})^{2}}{\sigma_{L}^2} \,,
\end{equation}
where $T_{\rm eff}$ and $L$ are the stellar effective temperature and bolometric luminosity, respectively, with errors 
$\sigma_{T_{\rm eff}}$ and $\sigma_{L}$, respectively. The effective temperature and stellar luminosity of 
the evolutionary tracks are denoted with $T_{\rm mod}$ and $L_{\rm mod}$, respectively.
We assigned to each YSO the mass and age corresponding to the closest track and isochrone, which minimize $\chi^2$.

The same procedure was used for evaluating masses and ages with the  \citet{Siess2000} evolutionary models that do not cover the 
very low-mass regime but extend to higher masses. For the stars in the mass range 0.1--1.4\,$M_{\sun}$ we found an excellent agreement 
between the masses derived with the two sets of tracks, with differences within 0.05\,$M_{\sun}$ rms.
Masses and ages, with the exception of the underluminous sources, are reported in 
Table~\ref{Tab:param}. For the most massive star, SSTc2dJ160830.7-382827, whose position in the HR diagram is outside the mass range
covered by the  \citet{Baraffe2015} tracks, we adopted the \citet{Siess2000}  models and we found \Mstar=1.8\Msun\ and	$Age=2.9$\,Myr.

The mass determinations are in very close agreement with those of \citet{Alcala2017}. 
 As stressed in previous works, the age determination suffers from uncertainties due to the error on the distance of the clouds, 
source variability, extinction and depends on the adopted set of evolutionary tracks \citep[see, e.g.,][and references therein]{Comeron2008}. 
Therefore, the ages of the individual sources are affected by large uncertainties.
However, the median age of the full sample analyzed in the present work is 2\,Myr and the average value is 2.5\,Myr, in good agreement with 
previous determinations \citep[e.g.,][and references therein]{Alcala2014,Alcala2017}.

\subsection{Activity/accretion diagnostics}
\label{subsec:fluxes}

The emission lines  comprised in the X-Shooter spectra have been used as diagnostics of chromospheric 
activity and accretion in previous works of our group \citep[e.g.,][]{Stelzer2013,Alcala2014}.
We have shown how the exceptional richness of information of these spectra allows us to probe the different layers 
of the atmospheres of PMS stars  and to define the ``noise'' that chromospheric emission introduces into mass accretion estimates 
\citep{Manara2013}.  These spectra allowed us to considerably improve the accuracy of the determination of mass accretion with 
important implications on the \Macc--\Mstar\ relationship \citep{Alcala2014,Alcala2017}.  

 Here we give a closer look at the H$\alpha$, 	
\ion{Ca}{ii}\,K, \ion{Ca}{ii}~infrared triplet (IRT), and  \ion{Na}{i}\,D$_{1,2}$. These lines, particularly H$\alpha$ and \ion{Ca}{ii} lines, have 
been widely used as diagnostics for both chromospheric activity \citep[e.g.,][and reference therein]{Strassmeier2000,montesetal2001,MartinezArnaiz2011,Frasca2016} 
and accretion \citep[e.g.,][]{Muzerolle2003,Mohanty2005,Costigan2012,Alcala2014} due to their intensity and  the high sensitivity of
the CCD detectors at these wavelengths.
The cores of the \ion{Na}{i}\,D$_{1,2}$ lines are good diagnostics of chromospheric activity (especially in late-K and M-type stars, e.g.,  
\citealt{Houdebine2009,Gomes2011}), mass accretion \citep[e.g.,][]{Rigliaco2012,Alcala2014,Alcala2017} and winds \citep[e.g.,][]{NattaGiova1990,Facchini2016}.

\begin{figure}  
\begin{center}
\includegraphics[width=8.8cm]{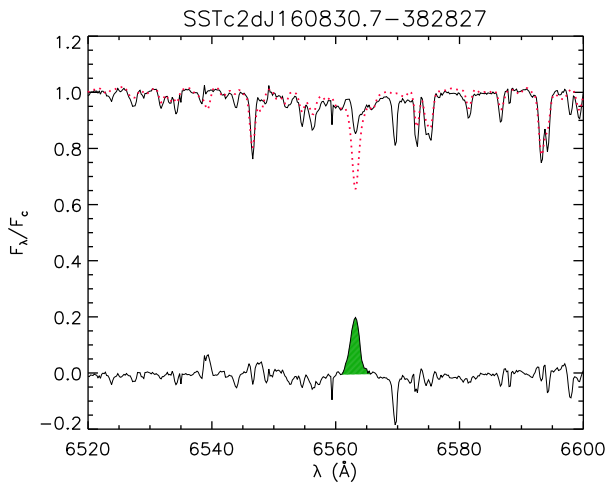}	
\vspace{-.5cm}
\caption{Example of the spectral subtraction for the H$\alpha$ line of SSTc2dJ160830.7-382827. 
The target spectrum normalized to the local continuum is represented by a solid black line, while the best-fitting BT-Settl spectrum 
is overplotted with a dotted red line. 
The difference spectrum is displayed in the bottom of the panel,  where the residual H$\alpha$ emission is highlighted by the hatched green area.} 
\label{fig:spectral_sub_Ha}
 \end{center}
\end{figure}

\begin{figure}  
\begin{center}
\includegraphics[width=8.8cm]{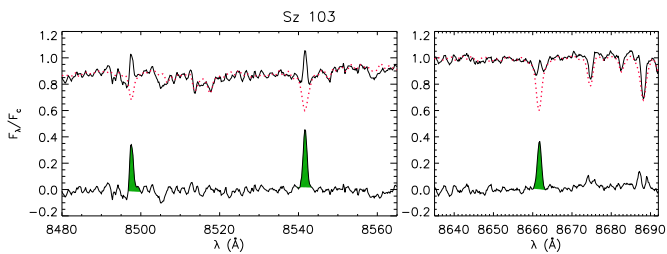}	
\vspace{0cm}
\caption{Example of the spectral subtraction for the \ion{Ca}{ii} IRT lines of Sz\,103.  
 Observed spectrum and inactive template are traced with a solid black line and a dotted red line, respectively.
The difference between observed and template spectrum is displayed in the bottom of each panel, along with the residual emission in the
\ion{Ca}{ii} IRT line cores (hatched green areas). 
}
\label{fig:spectral_sub_CaIRT}
 \end{center}
\end{figure}

\begin{figure}  
\begin{center}
\includegraphics[width=9.2cm]{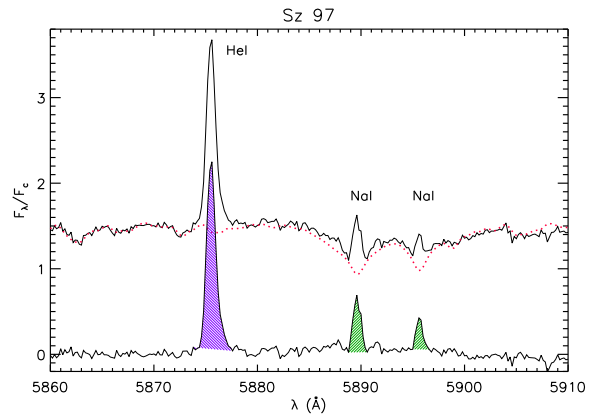}	
\vspace{0cm}
\caption{Example of the spectral subtraction for the \ion{Na}{i}\,D$_{1,2}$  and \ion{He}{i}\,D$_{3}$ lines of Sz\,97.  
 Solid and dotted red lines have the same meaning as in Fig.~\ref{fig:spectral_sub_Ha}.
An offset of 0.5 has been applied to both the observed and synthetic spectra for visualization purposes. 
The hatched green areas in the `difference' spectrum represent the \ion{Na}{i}\,D$_{1,2}$ excess emissions, while the 
\ion{He}{i}\,D$_{3}$ residual emission is filled with a violet hatching. 
}
\label{fig:spectral_sub_NaID2}
 \end{center}
\end{figure}

We calculated the equivalent widths (EWs) and  line fluxes by using the spectral subtraction method (see, e.g., 
\citealt{frascacatalano1994, montesetal1995}) to remove the photospheric flux and to evidence the core emission.
Although this correction is negligible for objects with high accretion rates and very strong emission lines, it is absolutely 
needed in cases where the photospheric profile is only filled-in with emission or the latter is very  weak. 

Figure~\ref{fig:spectral_sub_Ha} shows the result of the spectral subtraction on the H$\alpha$ line of SSTc2dJ160830.7-382827,
which is a target with a transitional disk and a weak accretor \citep{Alcala2017}. In this case the H$\alpha$ line displays an
absorption profile filled-in with emission, which is only revealed by subtracting the inactive template.
The latter has been obtained by the interpolation of the best-fitting BT-Settl spectra at the \teff\ and \logg\ of the target that 
are quoted in Table~\ref{Tab:param}.

The inactive templates are rotationally broadened, whenever \vsini\ is larger than the  minimum value that
can be resolved by the observations  (see Sect.~\ref{subsec:accuracy:vsini}), 
Doppler-shifted, and resampled on the spectral points of the target spectra before subtraction. 
The residual H$\alpha$ profile integrated over wavelength (hatched green area in Fig.~\ref{fig:spectral_sub_Ha}) provides us with the net 
H$\alpha$ EW.	

An example of the application of the spectral subtraction method to the \ion{Ca}{ii}\,IRT lines is shown in 
Fig.~\ref{fig:spectral_sub_CaIRT} for the star Sz\,103. 
We note that the subtraction of the inactive template is necessary for a correct measurement of the \ion{Ca}{ii}\,IRT EWs. 
In particular, the \ion{Ca}{ii}\,$\lambda$\,8662 line shows an emission in its core that does not reach the continuum.

The result of the spectral subtraction in the region of \ion{Na}{i}\,D$_{1,2}$  and \ion{He}{i}\,D$_{3}$ lines for Sz\,97 is shown in 
Fig.~\ref{fig:spectral_sub_NaID2}. It is clear that, although the underlying photospheric spectrum barely affects the strong \ion{He}{i}\,D$_{3}$ line,
it must be subtracted from the target spectrum to get the emission EWs in the \ion{Na}{i}\,D$_{1,2}$ line cores.
 We stress that the X-Shooter spectra were corrected for telluric absorption  lines also at these wavelengths, as described 
by \citet{Alcala2014}, and the interstellar \ion{Na}{i} absorption affects significantly only the more distant non-members (giants).

We converted the observed EWs into line fluxes per unit surface by multiplying them for the continuum flux of the 
synthetic spectrum corresponding to the \teff\  and \logg\  of the target. To get a more accurate value of the continuum flux, we interpolated 
in \teff\  and \logg\  within the grid of BT-Settl spectra. 
The fluxes of H$\alpha$, H$\beta$, \ion{Ca}{ii}~K, \ion{Ca}{ii}~IRT, and \ion{Na}{i}\,D lines are quoted in Table~\ref{Tab:fluxes}.
 In this table we do not report any flux value whenever the spectral subtraction does not give an emission residual profile.
We treat as upper limits the values for which the error is larger than the flux. 

We have already shown in \citet{Stelzer2013} the agreement between  line fluxes per unit surface calculated on the basis of model spectra, as in the
present paper, and those derived from the observed flux at Earth and the dilution factor, $(R_*/d)^2$, where $R_*$ and $d$ are the stellar radius and distance, 
respectively. 

We also calculated the ratio of the line and bolometric flux, $R'_{\rm line}  = F_{\rm line}/(\sigma T_{\rm eff}^4)$ as a further
chromospheric/accretion diagnostic\footnote{The prime in $R'_{\rm line}$ indicates that the photospheric contribution was 
subtracted, as usual in the definition of activity indices.}.

\begin{figure*}[htb]  
\begin{center}
\includegraphics[width=8.8cm]{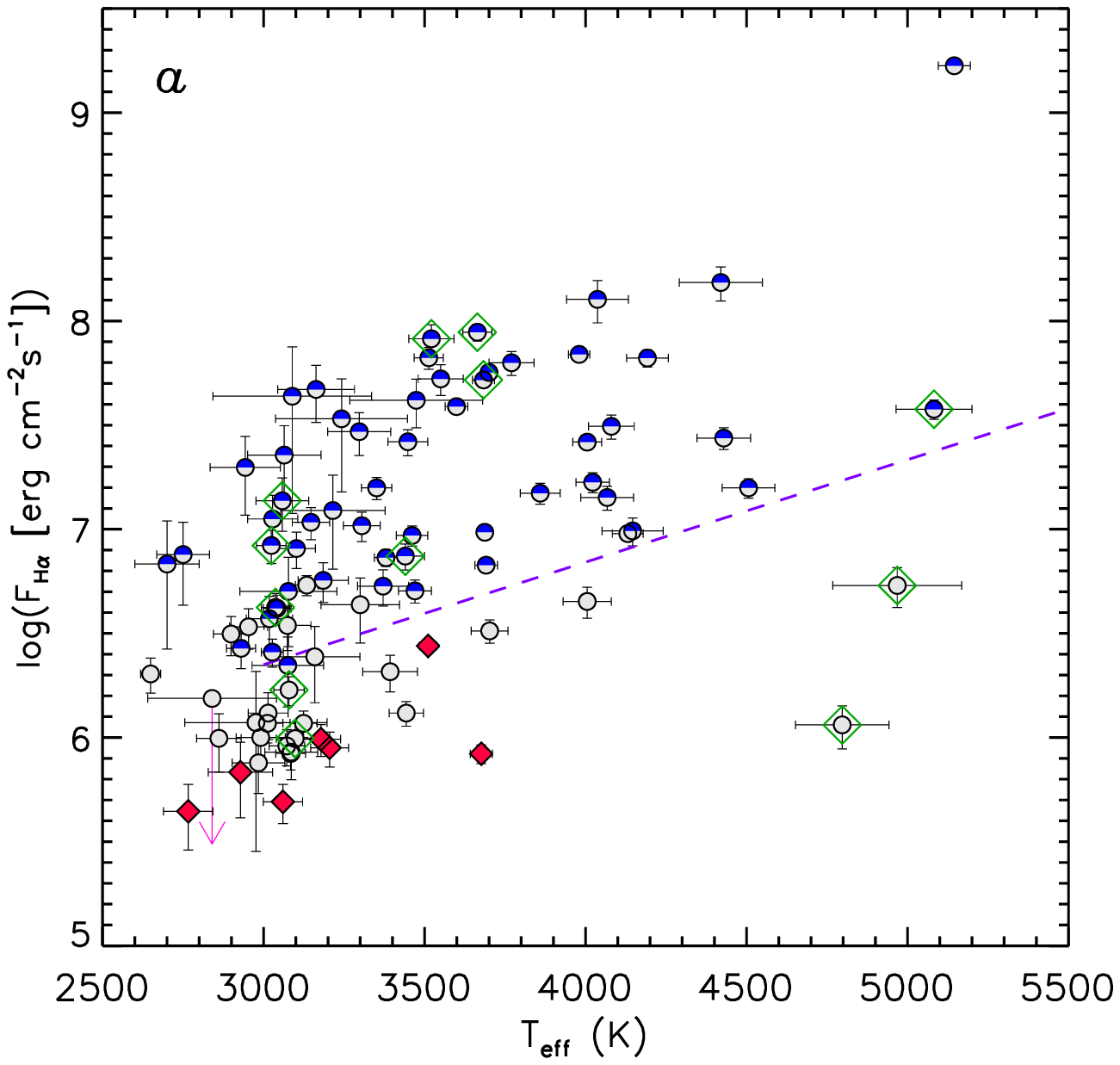}	
\includegraphics[width=8.8cm]{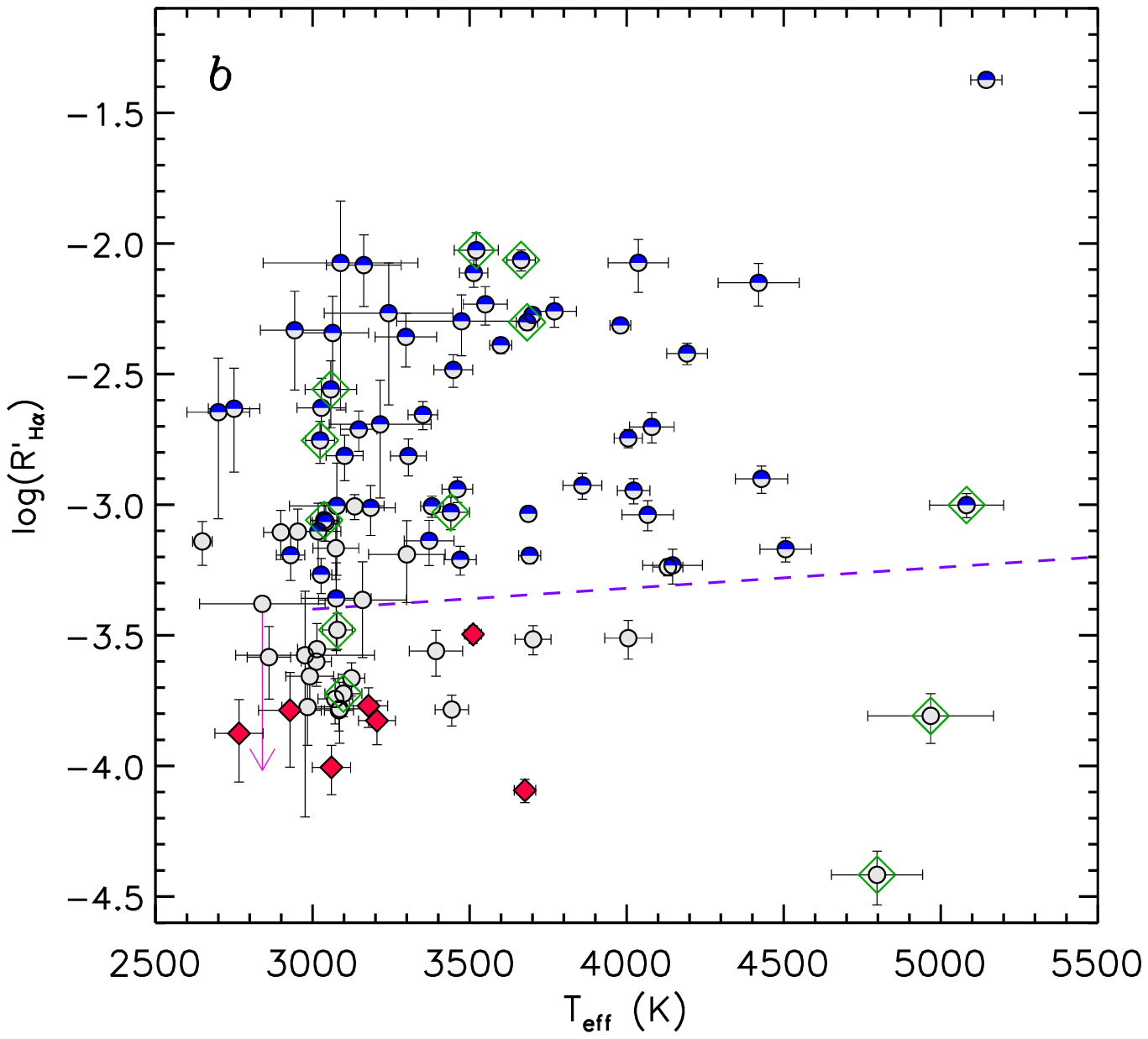}	
\vspace{-.3cm}
\caption{{\it Left panel}) H$\alpha$ flux versus $T_{\rm eff}$. {\it Right panel}) $R'_{\rm H\alpha}$ versus $T_{\rm eff}$.
In both panels the Class~II and Class~III objects are distinguished as in previous figures. The objects that can be considered 
accretors according to the \citet{WhiteBasri2003} criterion  are highlighted with a blue color in the upper half of the symbol. 
 Data for YSOs with transitional disks (TD) are enclosed into open green diamonds. In both panels, 
the dashed straight line is the boundary between chromospheric emission and accretion as derived by \citet{Frasca2015} for the 
Gamma Vel and Cha\,I SFRs.  The downward arrow in each panel indicates the only upper limit (2MASSJ16085373-3914367).
}
\label{Fig:Flux_Teff}
 \end{center}
\end{figure*}

\begin{figure*}[htb]  
\begin{center}
\includegraphics[width=6.3cm]{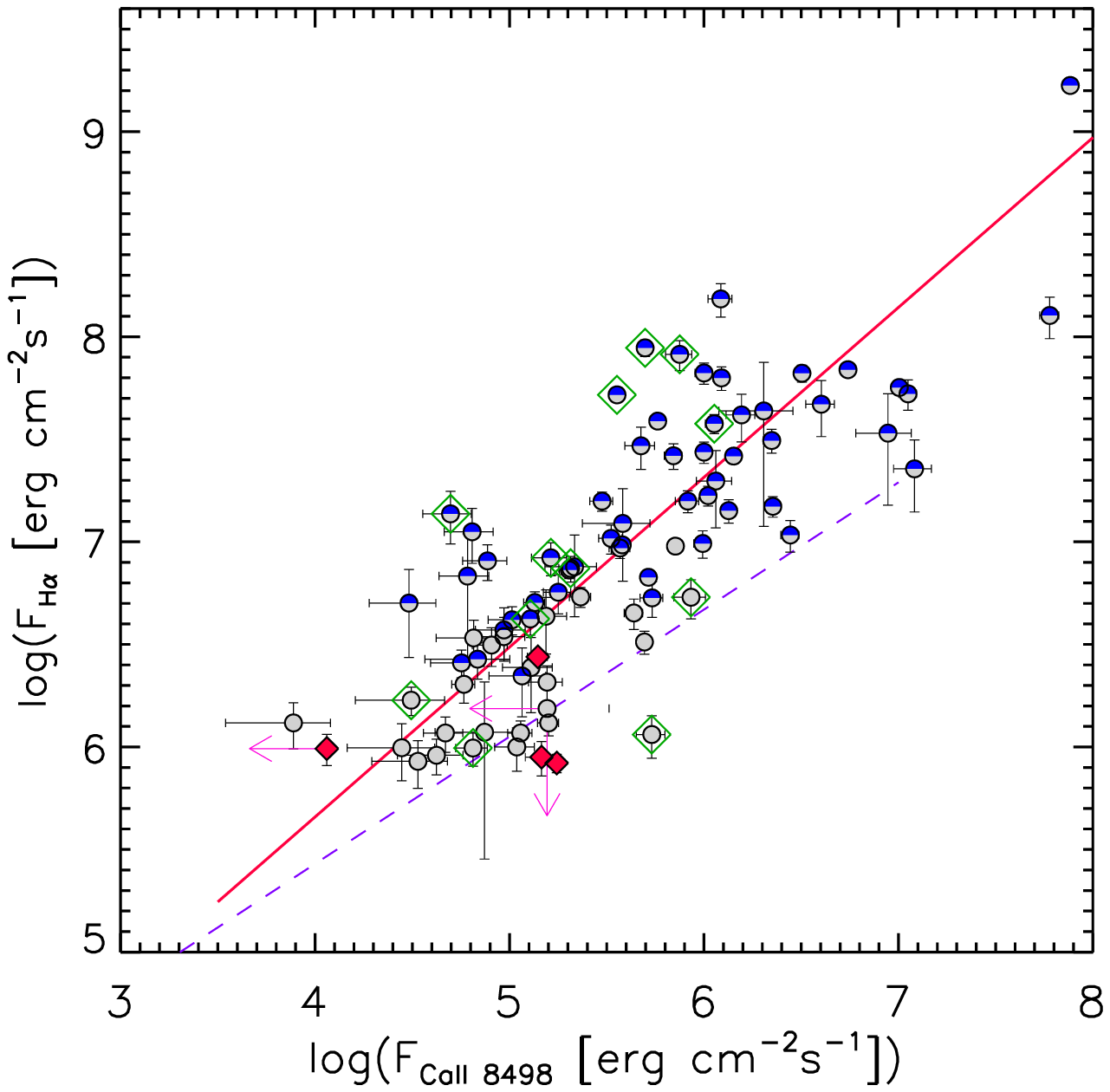}	
\hspace{-0.7cm}
\includegraphics[width=6.3cm]{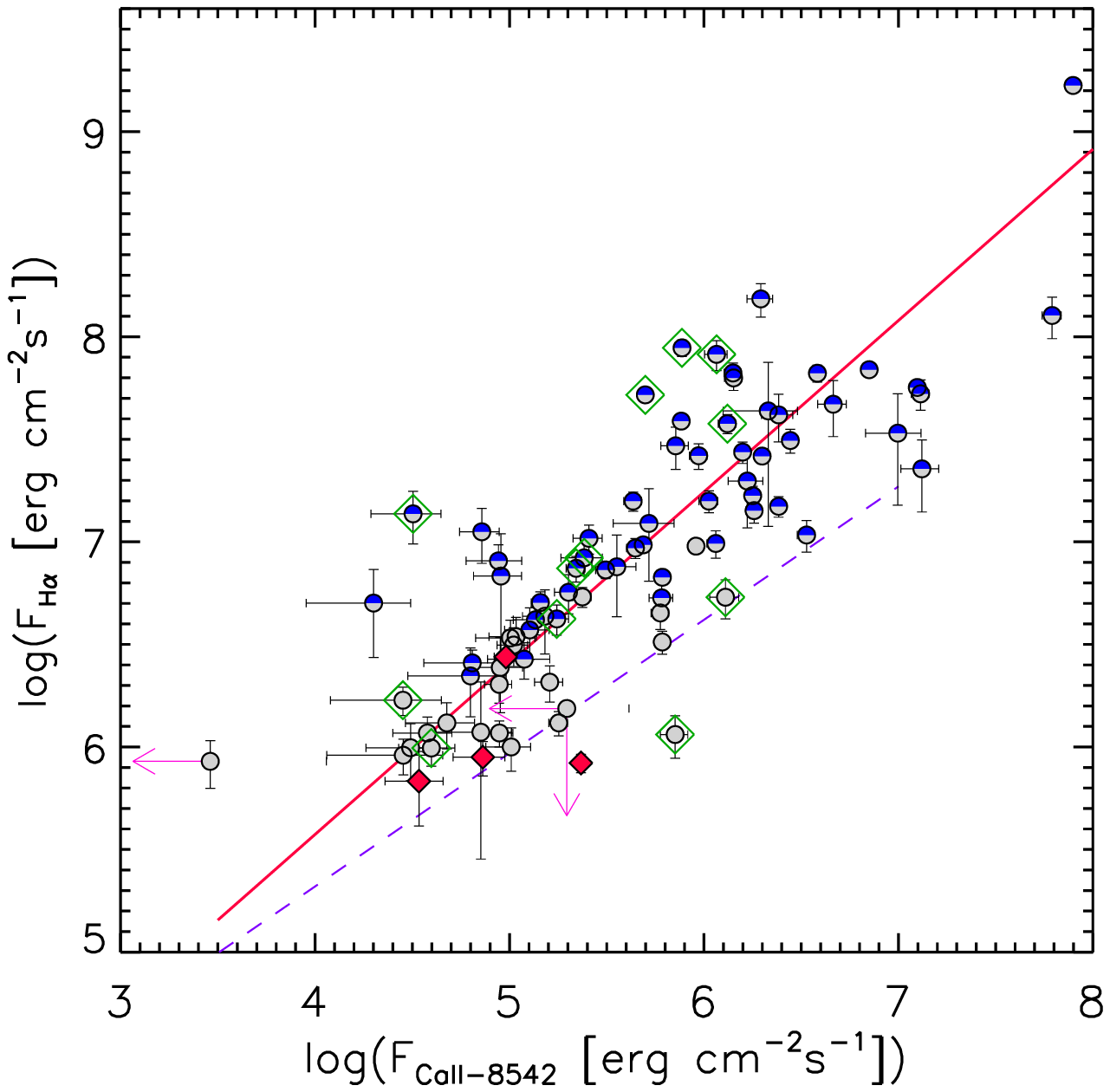}	
\hspace{-0.7cm}
\includegraphics[width=6.3cm]{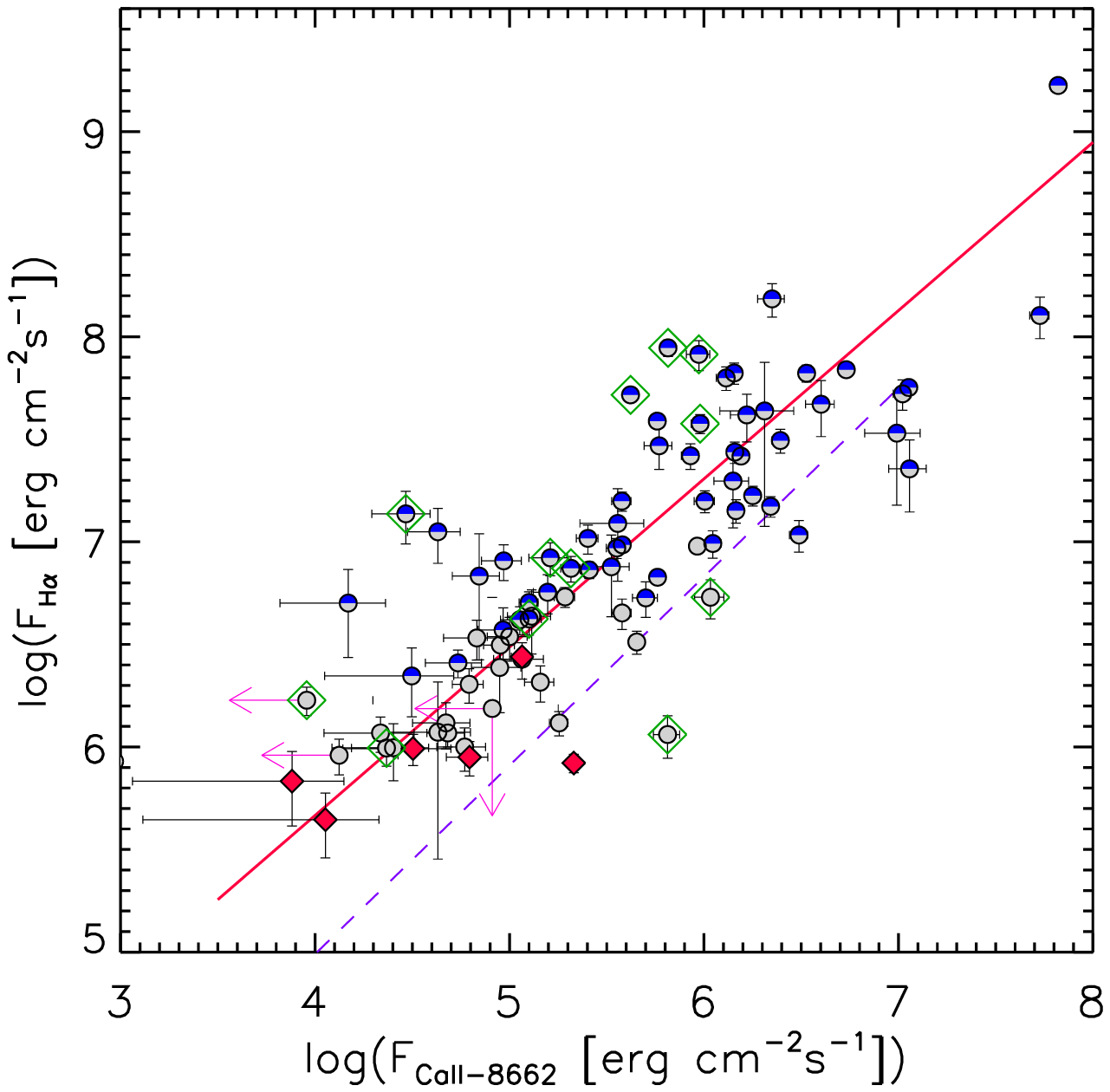}	
\vspace{0cm}
\caption{Flux--flux relations between H$\alpha$ and \ion{Ca}{ii}~IRT lines ($\lambda$\,8498, $\lambda$\,8542, and $\lambda$\,8662\,\AA, 
from the left to the right, respectively). The meaning of the symbols is as in Fig.~\ref{Fig:Flux_Teff}. 
The full red lines are the least-squares regressions, whose coefficients are reported in Eq.~\ref{Eq:FHa-FCaIRT}. 
The dashed lines represent the best fits of flux-flux relations found by \citet{Stelzer2013} for Class\,III objects. 
 Downward and leftward arrows indicate upper limits on H$\alpha$ and \ion{Ca}{ii} fluxes, respectively.
}
\label{Fig:Fha_FCa}
\end{center}
\end{figure*}

For the stars with H$\alpha$ line in emission above the local continuum, we also measured the full width at 10\% of
the line peak ($10\%W_{\rm H\alpha}$).		
This quantity is helpful for discriminating between accreting stars and those with pure chromospheric emission.
In particular, we defined ``candidate accretors'' the objects that fulfill the criterion of \citet[][]{WhiteBasri2003} 
that is based on both a $10\%W_{\rm H\alpha} \ge 270$ km\,s$^{-1}$ and  $EW_{H\alpha}$ larger than given thresholds depending on 
spectral type, and distinguished them in the following plots  with half-filled symbols.

In Fig.~\ref{Fig:Flux_Teff} we plot the H$\alpha$  flux, $F_{\rm H\alpha}$, and the ratio $R'_{\rm H\alpha}$ as a function 
of the effective temperature. 
In this figure, the boundary between the accreting objects and the 
chromospherically active stars, as defined by \citet{Frasca2015}  with the same criteria for the members of the Gamma~Vel cluster 
and Cha\,I SFR, is also marked.	 
 We note that all Class~III stars lie below the aforementioned boundary and all ``candidate accretors'' (half-filled symbols) are located above it, while
the other Class~II sources are scattered around, and mostly below, this dividing line. In particular, 17 Class\,II sources 
are located below the dividing line. However, this border cannot be considered as a knife edge that separates accretors from chromospheric sources;
therefore only the objects lying well below it are noteworthy. For instance, if we select only the YSOs with $\log(R'_{\rm H\alpha})\leq -3.7$, where most 
of the Class\,III sources lie, we end up with only eight Class\,II sources, three of which (Lup\,607, MY\,Lup, and SSTc2dJ160830.7-382827) are those considered as 
``weak accretors'' by \citet{Alcala2017}.
The other five objects, namely SSTc2dJ160000.6-422158, SSTc2dJ155925.2-423507, Sz\,95, SSTc2dJ161029.6-392215, and SSTc2dJ160703.9-391112 have 
rather low accretion rates according to \citet{Alcala2017}.       
The low H$\alpha$ flux could indicate that these objects have still rather dense dusty disks that give rise to the IR excess, but a low or moderate accretion rate
for which the H$\alpha$ flux is comparable with that of chromospheric sources.

The presence above the dividing line of some Class\,II object with a substantial accretion rate \citep{Alcala2017} and not 
satisfying the White \& Basri criterion shows that the latter is not 100\,\% reliable in selecting accretors.

 We note that the objects with transitional disks (TD), shown as green open diamonds in Fig.~\ref{Fig:Flux_Teff}, are spread all over the diagram and no clear 
behavior appears, in agreement with the finding by \citet{Alcala2017} for the mass accretion rate.

The H$\alpha$ and \ion{Ca}{ii}~IRT fluxes are well correlated, as apparent in Fig.~\ref{Fig:Fha_FCa} and suggested by the high value 
 of Spearman's rank correlation coefficients, which range from $\rho=0.80$ to $\rho=0.83$ with a significance $\sigma$ in the range 
$10^{-22}$--$10^{-19}$ \citep{Pressetal1992}, for the three \ion{Ca}{ii} lines. 
Least-squares regressions provide the following relations:

\begin{eqnarray}
\log F_{\rm H\alpha} & = & 2.35(\pm 0.32) + 0.83(\pm 0.07)\cdot\log F_{\rm CaII 8498}\nonumber  \\
\log F_{\rm H\alpha} & = & 2.23(\pm 0.29) + 0.84(\pm 0.06)\cdot\log F_{\rm CaII 8542}\nonumber  \\
\log F_{\rm H\alpha} & = & 2.38(\pm 0.27) + 0.82(\pm 0.07)\cdot\log F_{\rm CaII 8662}.
\label{Eq:FHa-FCaIRT}
\end{eqnarray}

 It is worth noting that most of the sources display H$\alpha$ fluxes in excess with respect to the mean flux-flux relations found by 
\citet{Stelzer2013} for Class\,III objects in Lupus, TWA, and $\sigma$\,Ori SFRs. This is particularly evident for the  ``candidate accretors''
(half-filled symbols) and suggests that the accretion produces a larger H$\alpha$ luminosity, compared to \ion{Ca}{ii}\,IRT 
lines, than that originating from chromospheres.
 This trend can be explained by the fact that strong accretors generally have also outflows or stronger winds, which 
contribute to the H$\alpha$ emission but do not significantly affect \ion{Ca}{ii} lines.

\begin{figure}[htb]  
\begin{center}
\includegraphics[width=8.cm]{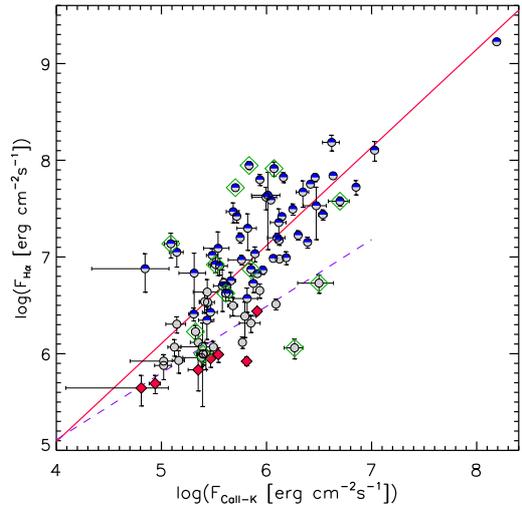}	
\caption{Flux--flux relationship between H$\alpha$ and \ion{Ca}{ii}~K line ($\lambda$\,3934\,\AA). The meaning of the symbols and lines 
is as in Fig.~\ref{Fig:Flux_Teff}. 
}
\label{Fig:Fha_FCaK}
\end{center}
\end{figure}

A similar behavior is found when plotting the H$\alpha$ versus the \ion{Ca}{ii}~K flux, as displayed in Fig.~\ref{Fig:Fha_FCaK}.
These data can be fitted by the following linear relation:
\begin{eqnarray}
\log F_{\rm H\alpha} & = & 1.04(\pm 0.41) + 1.01(\pm 0.07)\cdot\log F_{\rm CaII K}
\label{Eq:FHa-FCaK}
\end{eqnarray}

\begin{figure}[htb]   
\begin{center}
\includegraphics[width=8.cm]{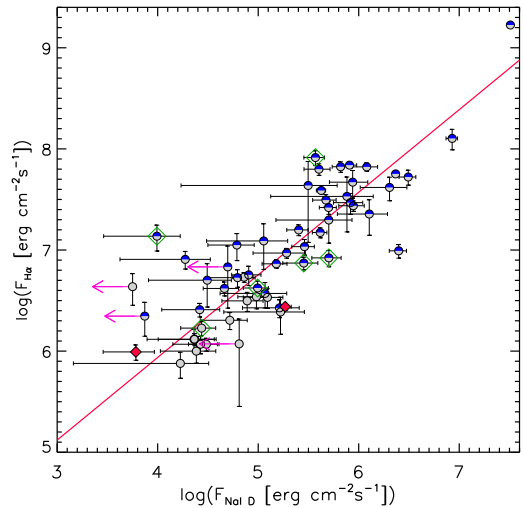}	
\vspace{-0.5cm}
\caption{Flux--flux relationship between H$\alpha$ and \ion{Na}{i}~D$_1$+D$_2$. The meaning of the symbols is as in Fig.~\ref{Fig:Flux_Teff}. 
The full red line is the least-squares regression.  Leftward arrows indicate upper limits on \ion{Na}{i} fluxes.
}
\label{Fig:Fha_FNaI}
\end{center}
\end{figure}

The H$\alpha$ and \ion{Na}{i}\,D$_{1,2}$ fluxes display a correlation ($\rho=0.81$, $\sigma=7\cdot 10^{-15}$), 
as shown in Fig.~\ref{Fig:Fha_FNaI}.	  
We remark that $F_{\rm NaI D}$ is the sum of the fluxes in the two sodium D lines. 
A least-squares regression provides the following relation:
\begin{equation}
\log F_{\rm H\alpha}  =  2.67(\pm 0.38) + 0.82(\pm 0.09)\cdot\log F_{\rm NaI D}.
\label{Eq:Fha_FNaID}
\end{equation}

\begin{figure*}[thb]  
\begin{center}
\includegraphics[width=6.3cm]{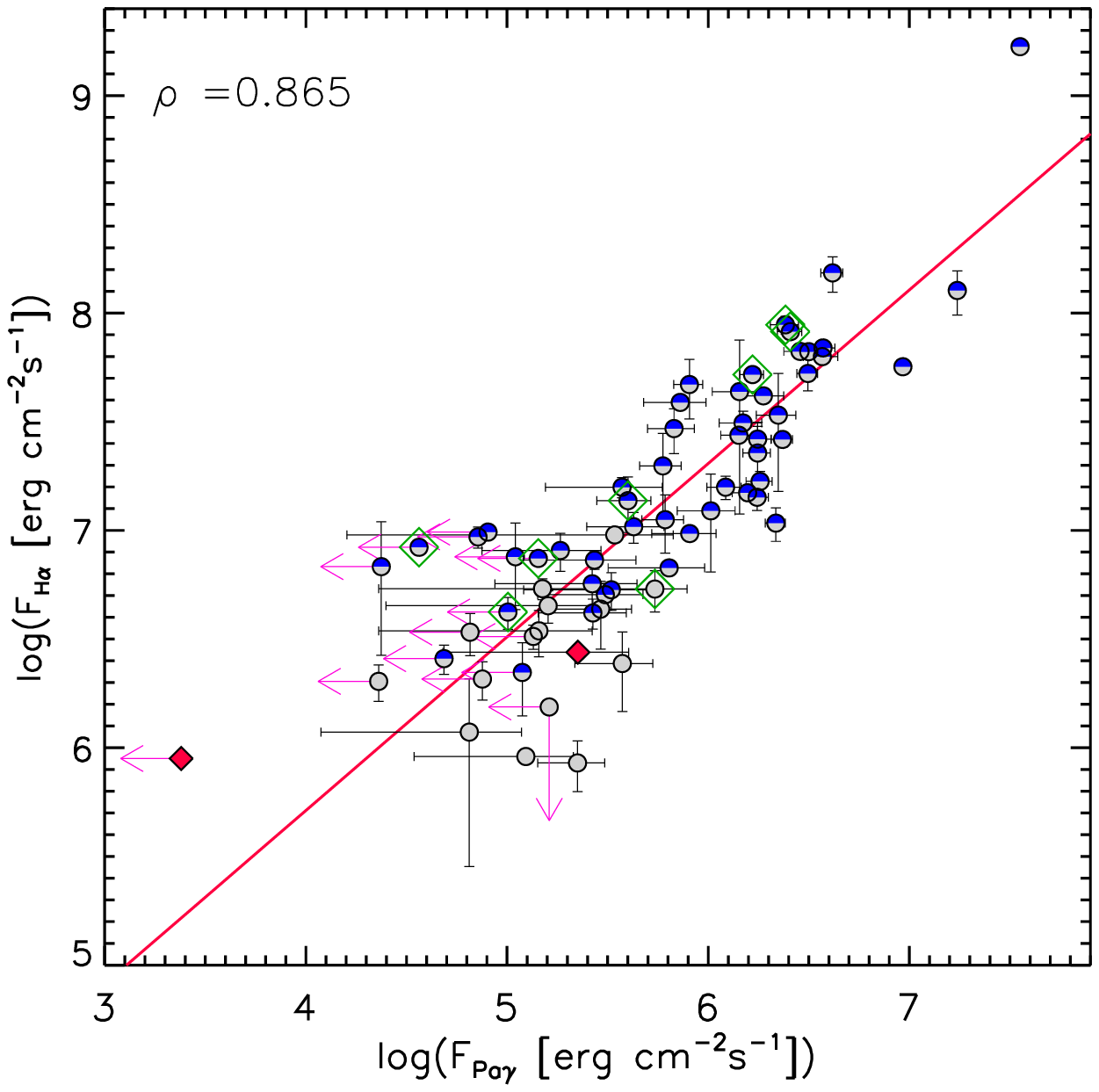}	
\hspace{-0.7cm}
\includegraphics[width=6.3cm]{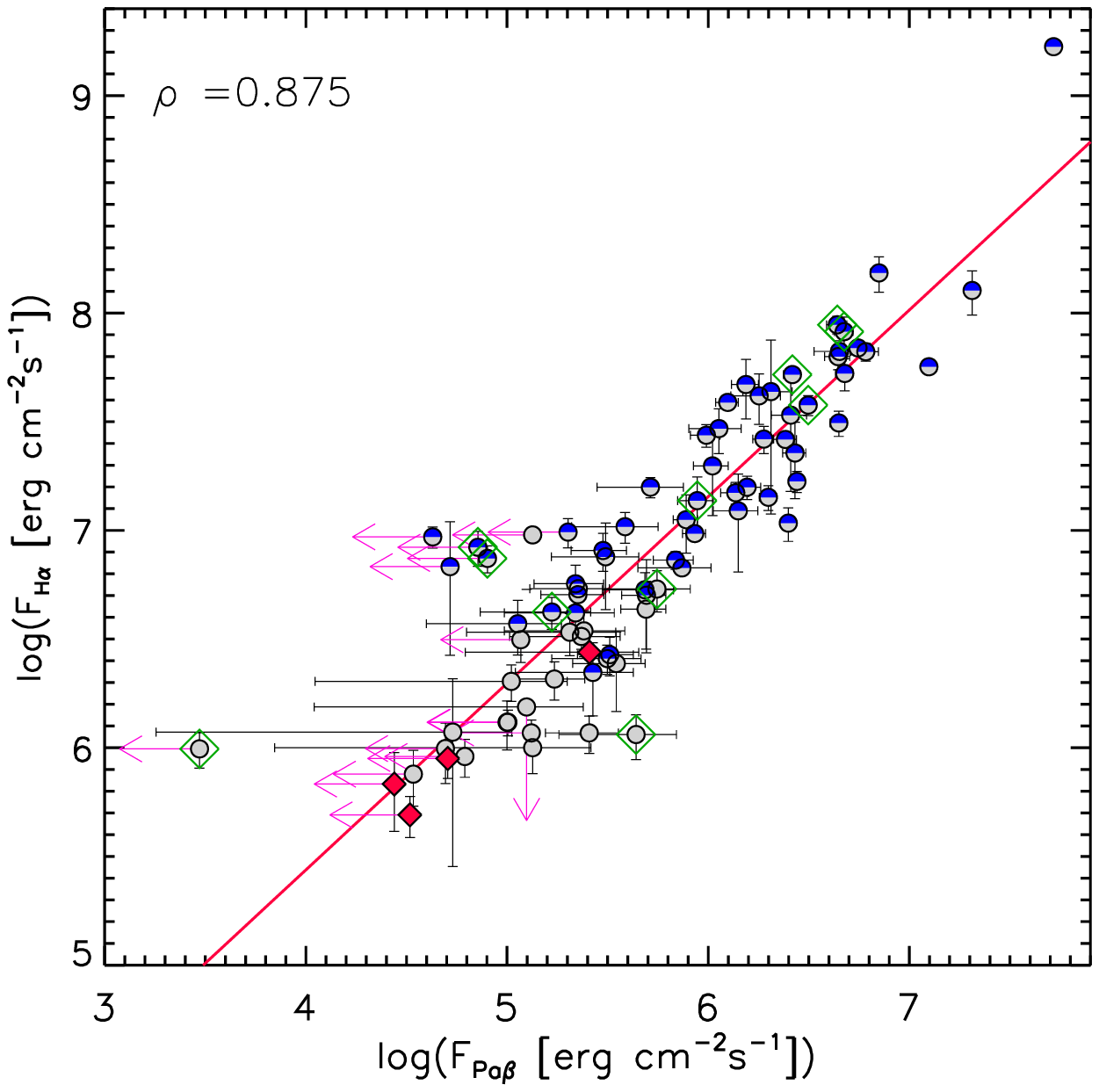}	
\hspace{-0.7cm}
\includegraphics[width=6.3cm]{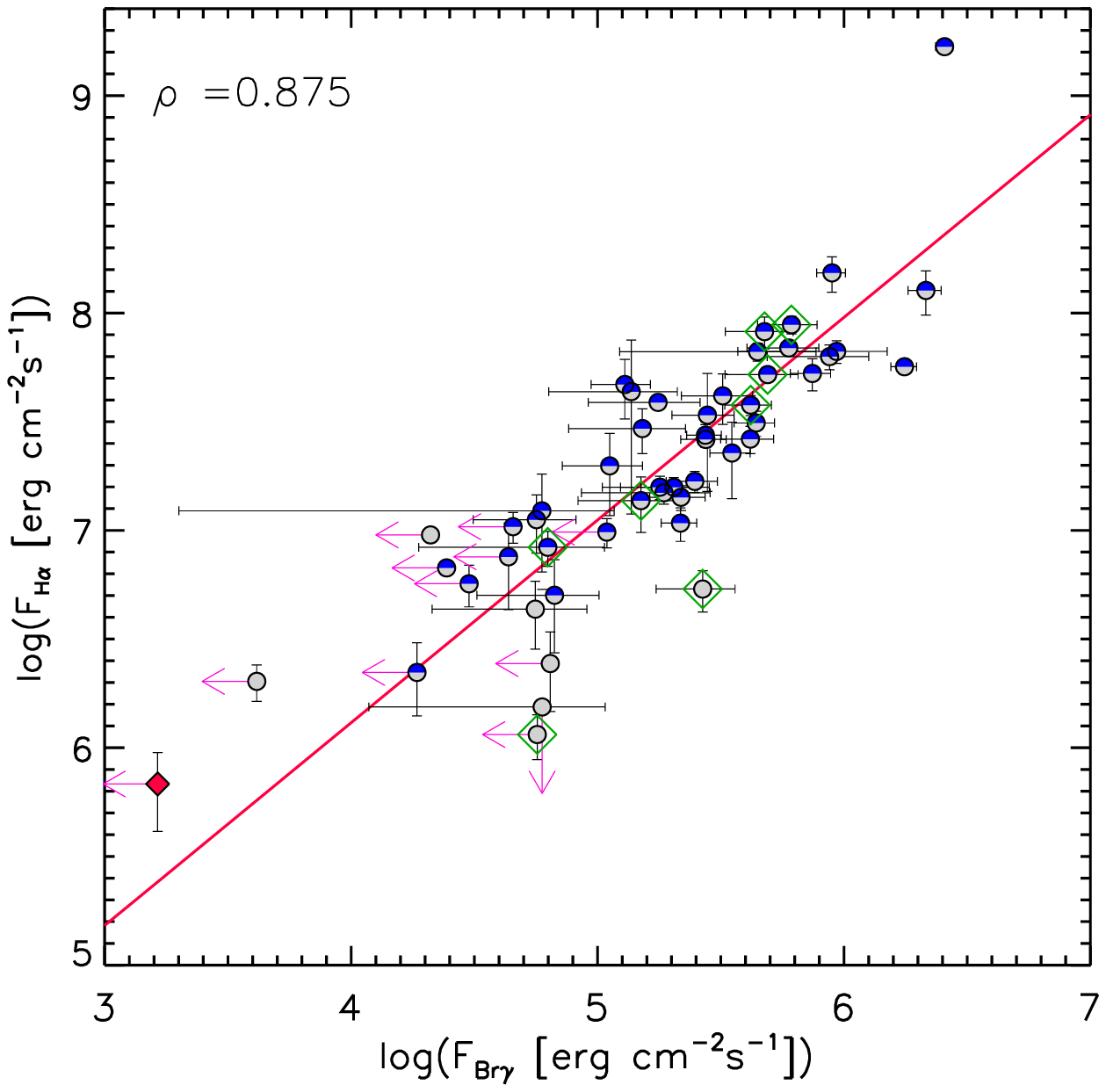}	
\vspace{0cm}
\caption{Flux--flux relations between H$\alpha$ and Pa$\gamma$, Pa$\beta$, and Br$\gamma$
(from the left to the right, respectively). The meaning of the symbols is as in Fig.~\ref{Fig:Flux_Teff}. 
The full red lines are the least-squares regressions, whose coefficients are reported in Eq.~\ref{Eq:FHa-FNIR}. 
Downward and leftward arrows indicate upper limits. The value of the rank-correlation coefficient $\rho$ is also reported in each box. 
}
\label{Fig:Fha_FNIR}
\end{center}
\end{figure*}

We have also applied the subtraction method to the strongest emission lines in the NIR spectra that are not severely 
affected by telluric absorption, namely the \ion{He}{i}\,$\lambda$10830\,\AA, Paschen (Pa$\gamma$ and Pa$\beta$) and Brackett (Br$\gamma$) lines.
The \ion{He}{i}\,$\lambda$10830\,\AA\  profile often displays excess absorption components that are related to stellar and/or disk 
winds \citep[e.g.,][]{Kurosawa2011}. Therefore, in these cases the spectral subtraction gives rise to extra absorption. We have, therefore, 
discarded this line, whose study is deferred to a subsequent work.

The fluxes of the three NIR lines, namely the Paschen $\beta$ and $\gamma$ lines and the Brackett $\gamma$ line, have been measured for most of 
the targets (Table~\ref{Tab:NIR}). The fluxes of these lines are
also well correlated with the H$\alpha$ flux, as shown in Fig.~\ref{Fig:Fha_FNIR}.
Least-squares regressions provide the following relations:
\begin{eqnarray}
\log F_{\rm H\alpha} & = & 2.52(\pm 0.32) + 0.80(\pm 0.08)\cdot\log F_{\rm Pa\gamma}\nonumber  \\
\log F_{\rm H\alpha} & = & 2.00(\pm 0.27) + 0.86(\pm 0.07)\cdot\log F_{\rm Pa\beta}\nonumber  \\
\log F_{\rm H\alpha} & = & 2.38(\pm 0.39) + 0.93(\pm 0.10)\cdot\log F_{\rm Br\gamma}.
\label{Eq:FHa-FNIR}
\end{eqnarray}

\subsection{Line fluxes and accretion}
\label{Sec:flux_accr}

\begin{figure*}[htb]  
\begin{center}
\includegraphics[width=6.2cm]{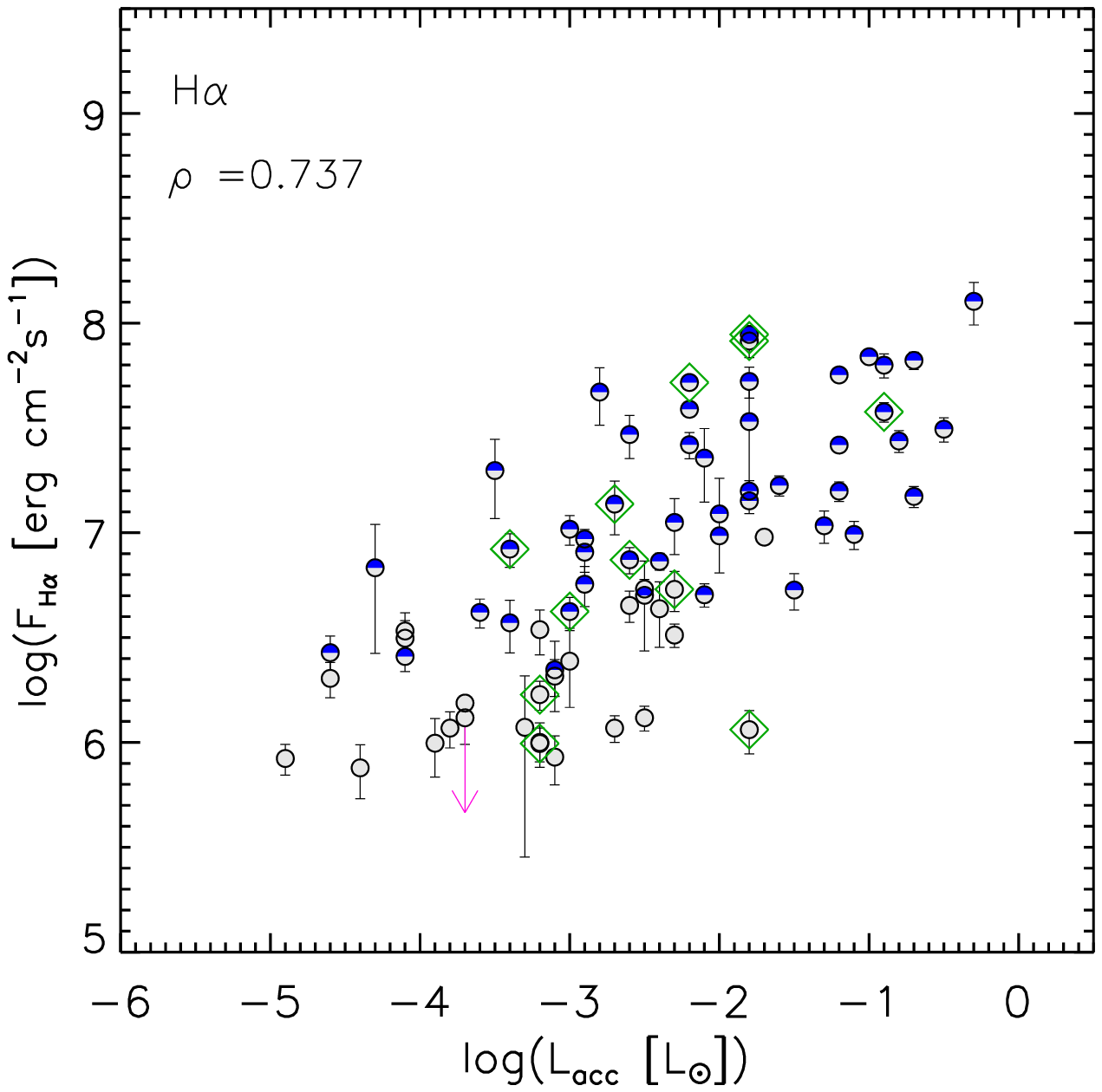}	
\hspace{-0.7cm}
\includegraphics[width=6.2cm]{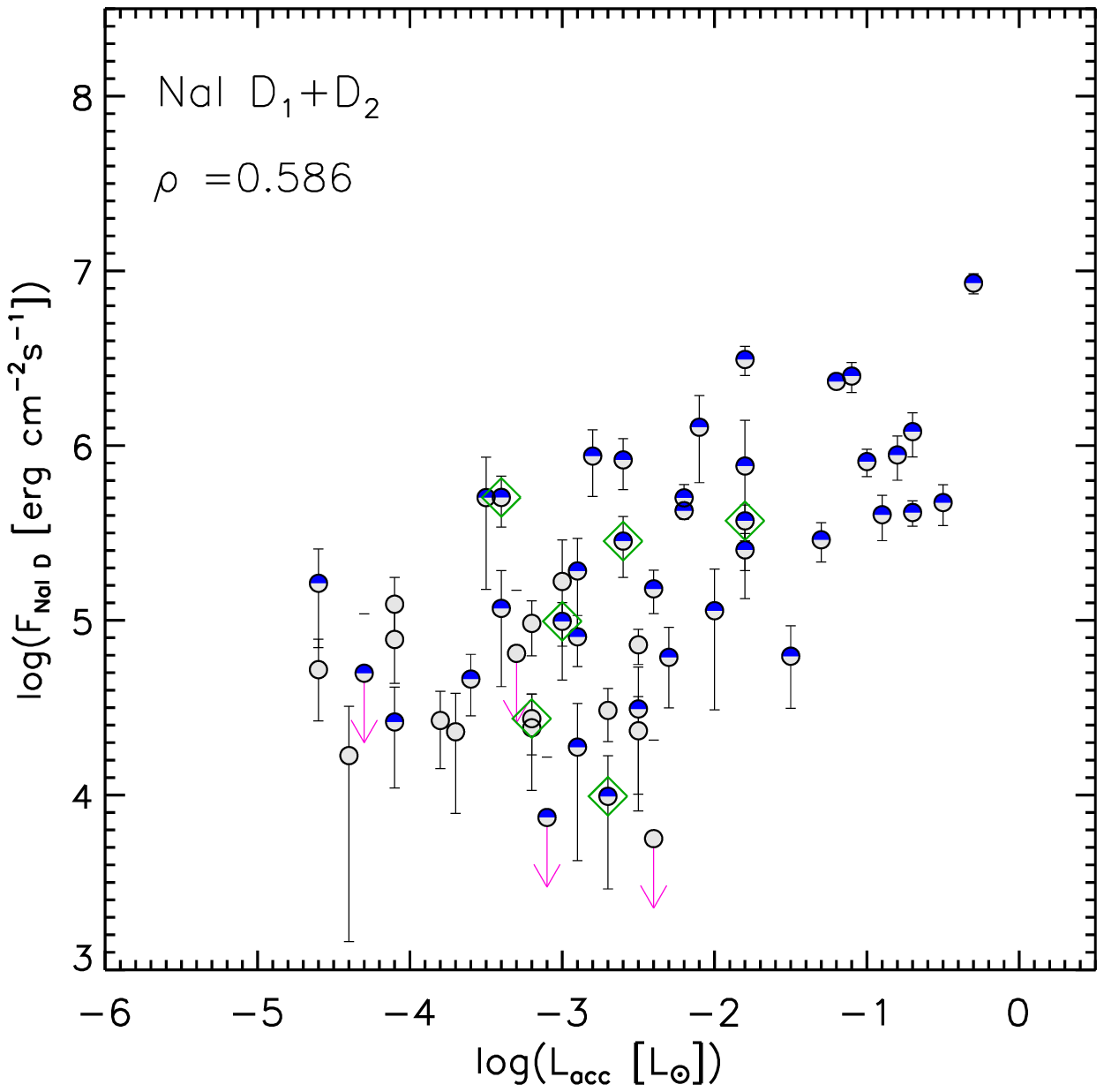}	
\hspace{-0.7cm}
\includegraphics[width=6.2cm]{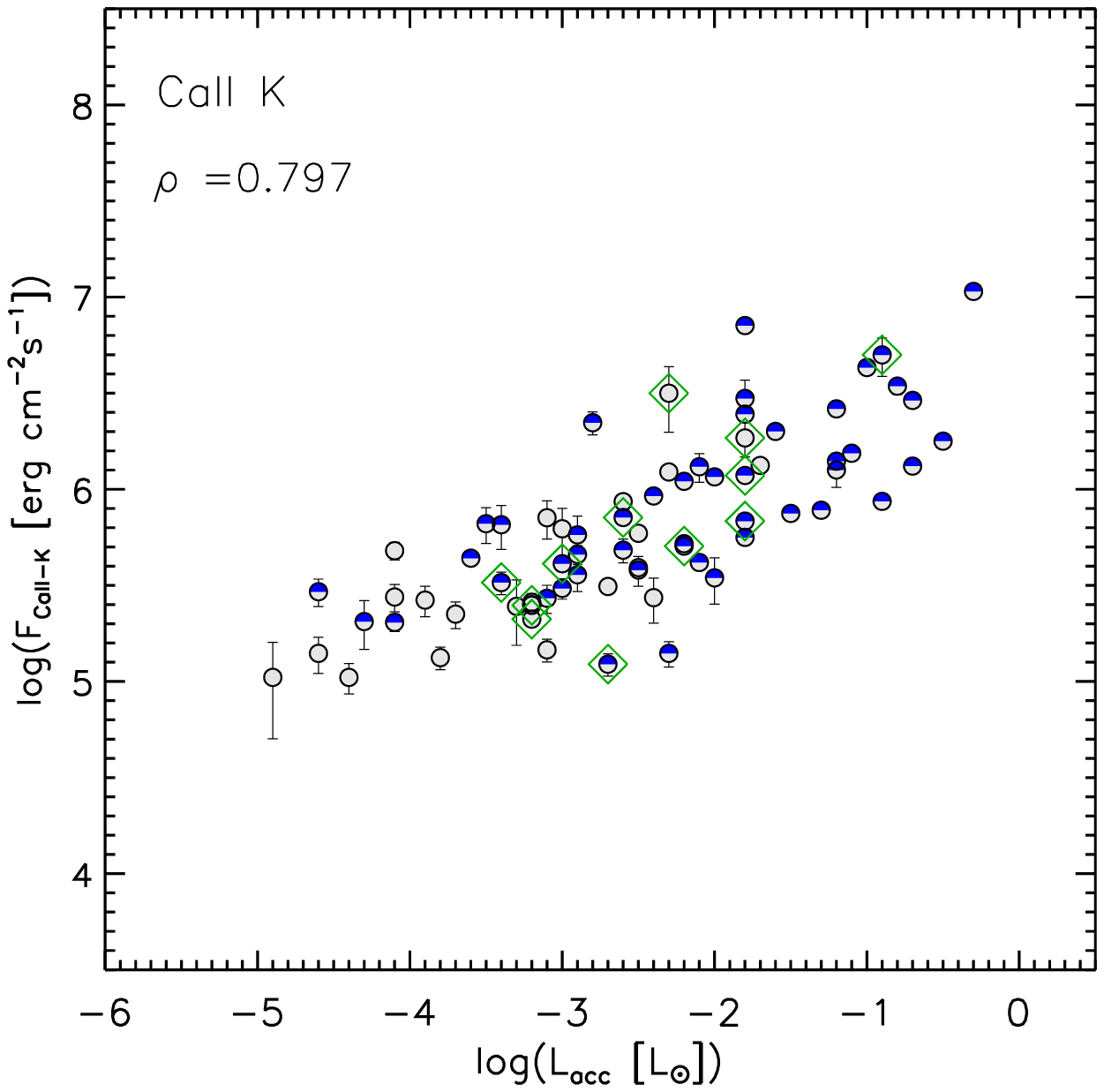}	
\vspace{0cm}
\includegraphics[width=6.2cm]{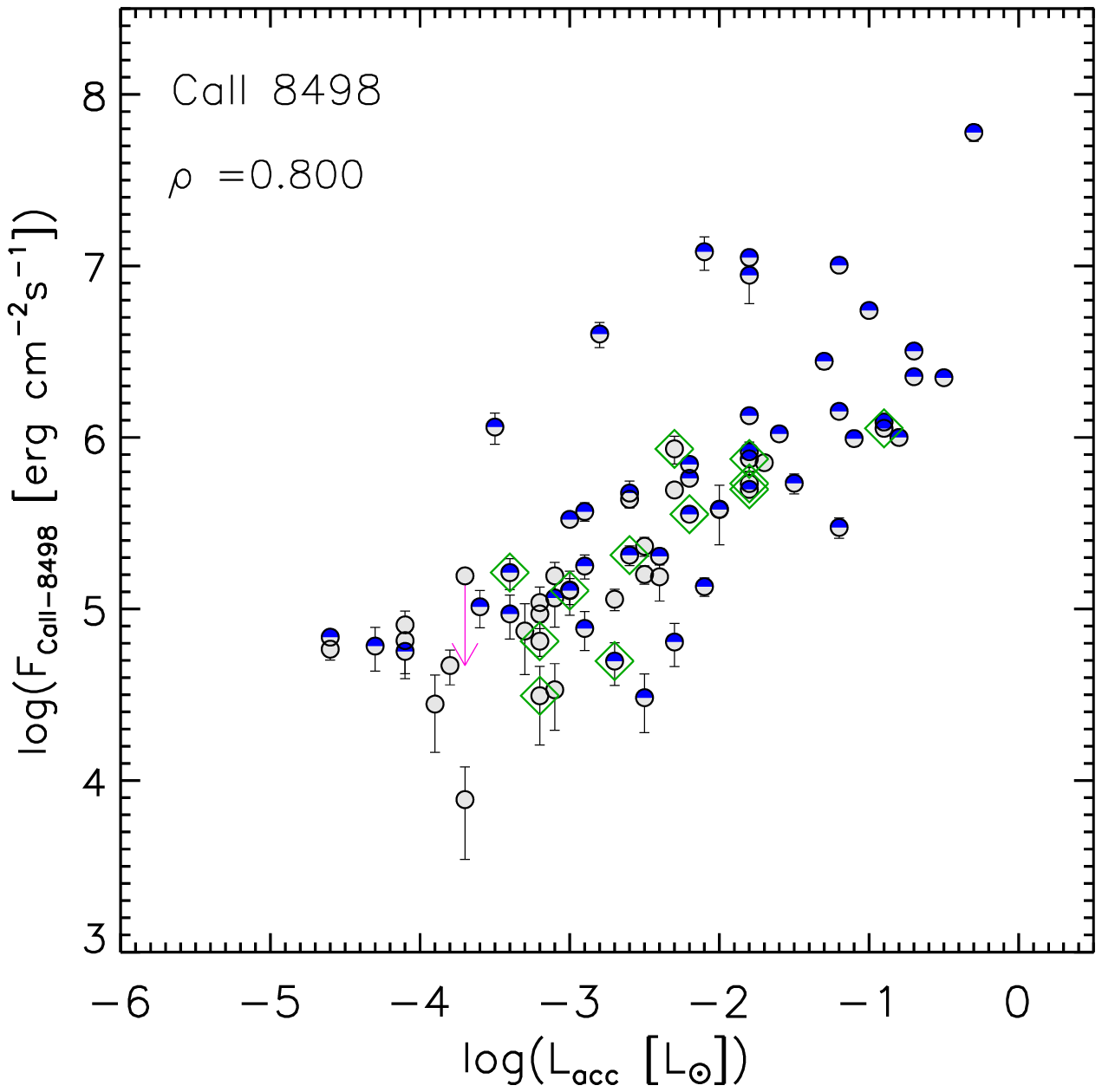}	
\hspace{-0.7cm}
\includegraphics[width=6.2cm]{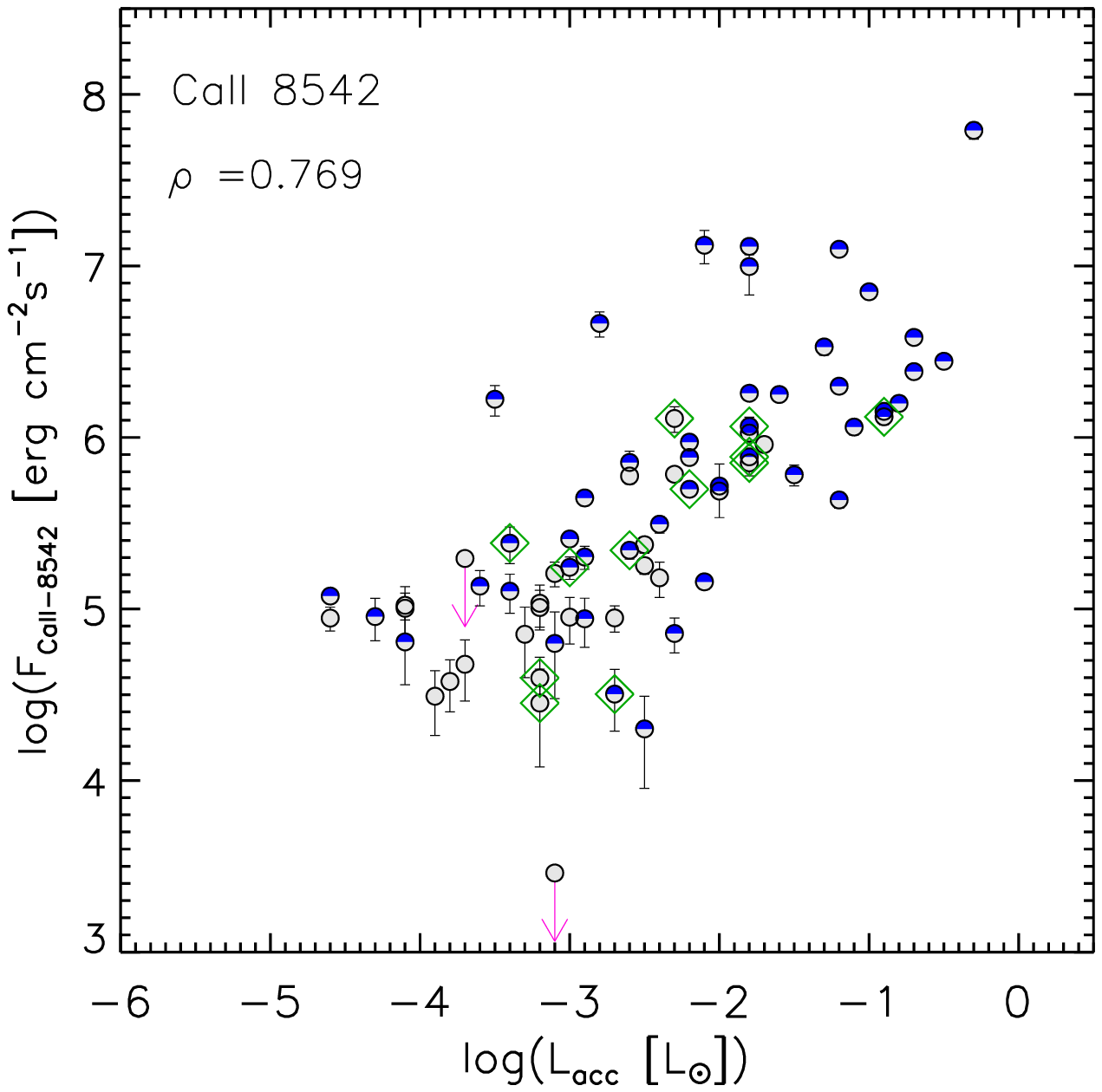}	
\hspace{-0.7cm}
\includegraphics[width=6.2cm]{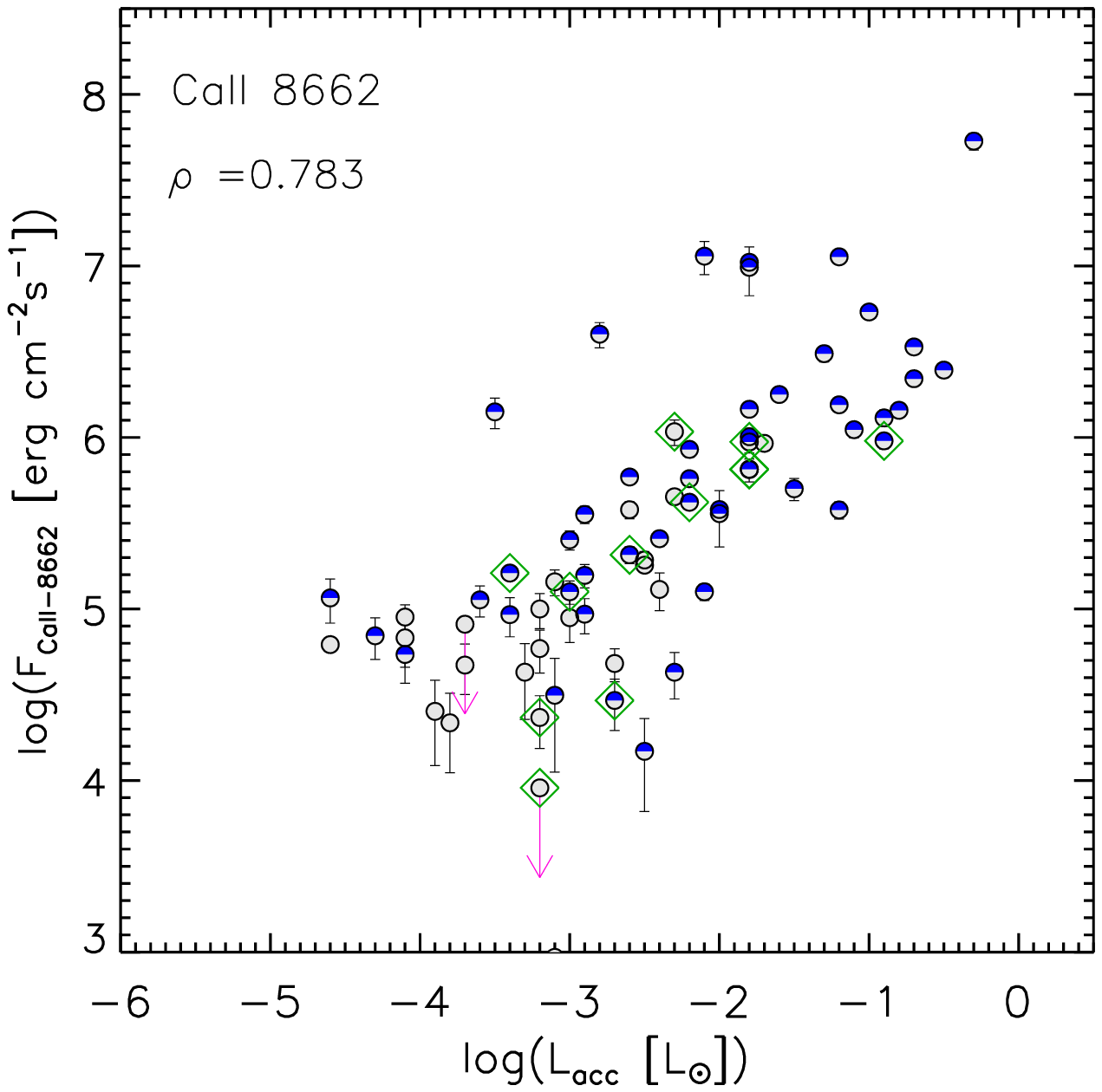}	
\vspace{0cm}
\caption{Line fluxes per unit surface as a function of the accretion luminosity.	
The meaning of the symbols is as in Fig.~\ref{Fig:Flux_Teff}. The value of the rank-correlation coefficient $\rho$ is also reported in each box. 
The subluminous objects have been excluded because $L_{\rm acc}$ is underestimated  \citep{Alcala2014,Alcala2017}, while the surface fluxes are not 
affected.  Downward arrows indicate upper limits.
}
\label{Fig:Lacc_Fline}
\end{center}
\end{figure*}

 Line fluxes are also very useful for evaluating the mass accretion rate and for comparing it with the one derived 
from primary indicators, like the Balmer continuum excess.
\citet{Alcala2017} calculated the accretion luminosity, $L_{\rm acc}$, by modelling the observed flux-calibrated spectra with the sum of a non-accreting
(Class\,III) template and the emission from a slab of hydrogen, which accounts for $L_{\rm acc}$. The only parameters needed to convert the flux at Earth into 
luminosity are the distance to the YSO and the extinction.

\citet{Mendigutia2015} suggested that the relations between accretion and line luminosity, $L_{\rm acc}-L_{\rm line}$, may be the result of the 
$L_{\rm acc}-L_{\rm *}$ and $L_{\rm line}-L_{\rm *}$ correlations. 
Indeed, both $L_{\rm acc}$ and $L_{\rm line}$ are obtained adopting the same distance and extinction. Therefore, in their hypothesis, the observed trends 
may be a ``calibration'' effect rather than physical correlations.

In the present work we have calculated the  line fluxes per unit surface based on the continuum fluxes of the model spectra at the \teff\ and \logg\  
of the target.
Thus, neither the distance nor the extinction enter in the determination of $F_{\rm line}$.

We show in Fig.~\ref{Fig:Lacc_Fline} that  line fluxes are highly correlated with the accretion luminosity reported by \citet{Alcala2017}, as also 
witnessed by the rank-correlation coefficients ranging from 0.59 to 0.80 with significances always smaller than $10^{-6}$.	
This excludes a calibration effect and strengthens the validity of the emission lines as accretion diagnostics.

\subsection{Flux ratios}
\label{Sec:fluxratio}

\begin{figure}  
\begin{center}
\includegraphics[width=8.8cm]{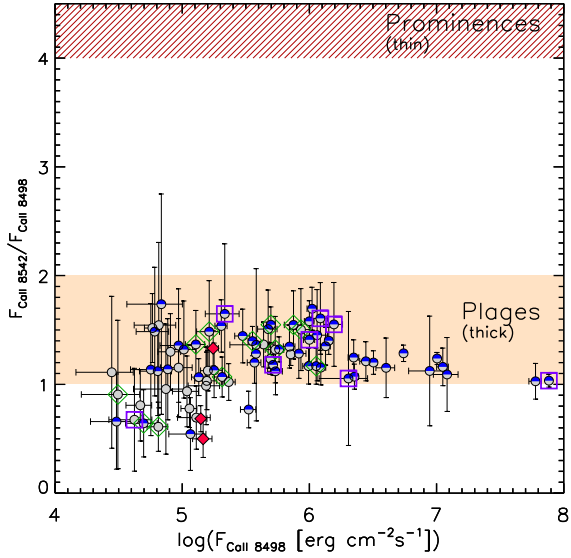}	
\vspace{-0.5cm}
\caption{Flux ratio $F_{\rm CaII 8542}/F_{\rm CaII 8498}$ versus $F_{\rm CaII 8498}$.
The range of values typical for solar plages and prominences are also shown by the shaded and hatched areas, respectively.
 Data for subluminous YSOs are enclosed into open violet squares.
}
\label{Fig:FCaIIratio}
\end{center}
\end{figure}

The ratio of  fluxes in two \ion{Ca}{ii}-IRT lines,  $F_{\rm CaII 8542}/F_{\rm CaII 8498}$, is sensitive to the conditions of the
emitting plasma, particularly to its optical depth.
It was shown that for optically thick sources, such as dense chromospheric solar or stellar plages, the ratio is typically
in the range 1--2; optically-thin emission sources, like solar prominences seen off-limb, have instead large values of this ratio,
$F_{\rm CaII 8542}/F_{\rm CaII 8498}\simeq$\,4--9 \citep[see, e.g.,][and references therein]{HerbigSoderblom1980,Landman1980}.	 
As apparent in Fig.~\ref{Fig:FCaIIratio}, the values of this flux ratio are low for all Lupus members in our sample, both Class~II and III. 
This suggests that the bulk of emission is originating in optically-thick regions, either chromospheric 
plages or the impact regions of accretion flows near the YSOs surface, as already suggested, e.g., by \citet{HerbigSoderblom1980} and  \citet{Alcala2014}. 

\begin{figure}  
\begin{center}
\includegraphics[width=8.6cm]{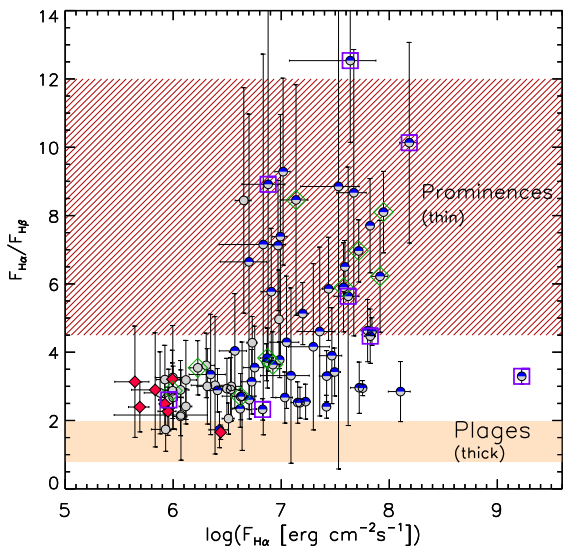}	
\vspace{-0.5cm}
\caption{Balmer decrement ($F_{\rm H\alpha}/F_{\rm H\beta}$) versus $F_{\rm H\alpha}$. 
The range of values typical for solar plages and prominences are also shown by the shaded and hatched areas, respectively.
 Data for subluminous YSOs are enclosed into open violet squares.
}
\label{Fig:Balmer_decr}
 \end{center}
\end{figure}

Another indicator of the physical conditions of the emitting matter is the Balmer decrement, which is here defined as the ratio of the H$\alpha$
and H$\beta$ fluxes. We calculated the H$\beta$ flux at the stellar surface in the same way as the H$\alpha$ flux.
The Balmer decrement is displayed in Fig.~\ref{Fig:Balmer_decr}. We note that all Class\,III sources have low values of $F_{\rm H\alpha}/F_{\rm H\beta}$,
which range from about 1.5 to 3. These values are, on average, slightly larger than those typical of solar plages or pre-flare active regions (1--2), but 
significantly lower than in optically thin regions such as solar prominences observed off limb \citep[e.g.,][]{tandberg1967,landman1979}.
This behavior was already observed in stars with a very high chromospheric activity level and explained as the result of different physical 
conditions in the chromospheres or of a combination of plage-like and prominence-like regions in the observed flux 
\citep[e.g.,][]{hallramsey1992,chesteretal1994,Stelzer2013}.

Several accretors share the same locus of the Class\,III sources in this diagram, but some others (about one third) display very high values of the Balmer decrement,
which indicates a prevalence of optically thin emission in their Balmer lines, similarly to  results from other studies of accretors
\citep[e.g.,][]{Whelan2014,Frasca2015,Antoniucci2017}.

This suggests that \ion{Ca}{ii} and Balmer lines carry informations about different effects/regions of the 
accretion process, the former mainly originating in optically thick regions, like, e.g., the impact shocks produced by the accretion flows near the 
stellar photosphere. The Balmer lines can be formed instead in different parts of the accretion funnels with a large range of 
optical depth.  

 We note that the subluminous YSOs, with the exception of SSTc2dJ160703.9-391112, Sz\,106 and Sz\,102, display large values of 
the Balmer decrement. This can be understood if we assume that an edge-on disk is the cause of the under-luminosity. In this framework, the densest, 
and thus optically thickest, regions of the accretion flows are suppressed by the optically thick edge-on disk. 
Moreover, \citet{Antoniucci2017} report that the whole Balmer series decrement shows a common shape (type 1 in their classification) in most subluminous objects, which is 
compatible with an edge-on geometry.
The rather low values of Balmer decrement for SSTc2dJ160703.9-391112, Sz\,106 and Sz\,102 may be the result of the geometry of the star/disk system 
and/or of the strong mass accretion rate, which gives rise to dense flows at rather large distances from the central star.

\section{Summary and conclusions}
\label{Sec:Conclusions}
We have presented the results of the analysis of X-Shooter@VLT spectra of 102 YSO candidates, mostly of infrared Class\,II, in 
the Lupus star forming region. 
The application of the code ROTFIT to specific spectral regions at optical and near-infrared wavelengths  has allowed us to measure the 
effective temperature (\teff), surface gravity (\logg), radial velocity (RV) and 
projected rotational velocity (\vsini) with median errors of $\sigma_{T_{\rm eff}}=70$\,K, $\sigma_{\log g}=0.21$\,dex, $\sigma_{\rm RV}=2.3$\,\kms, 
and $\sigma_{v\sin i}=4$\,\kms, respectively. By means of Monte Carlo simulations, we have estimated  the minimum detectable \vsini\ 
as 8\,\kms\ with the 0$\farcs$9 VIS slit and 6\,\kms\ with the 0$\farcs$4 VIS slit, respectively.
The comparison with the effective temperatures inferred from the spectral classification performed on the same spectra by \citet{Manara2013} and 
\citet{Alcala2014,Alcala2017} and with literature data stemming from low-resolution spectroscopy indicates an external accuracy of about 100\,K.
A similar statistical evaluation cannot be done for the accuracy of \logg\ and \vsini, because of the very few values of these parameters available 
in the literature, which are, however, in good agreement with our determinations.
We have also derived the veiling at different wavelengths and confirm  previous results that the strongest accretors possess the highest veiling values.

We have shown that 13  candidate members of the Lupus SFR must be definitely considered as non-members on the basis of their discrepant RV with respect 
to the Lupus SFR and/or the very low \logg\ values, which suggest that at least 11 of them are background giants in the RGB or AGB phase. 
Two of them turned out to be lithium-rich giants.

Thanks to the subtraction of inactive template spectra, we have measured the line fluxes corrected for the contribution of the photospheric
absorption, even in the objects with the weakest chromospheric or accretion emission.
We found that all the Class\,III sources lie, as expected, in the locus of the \teff-$F_{\rm H\alpha}$ diagram occupied by the objects where chromospheric 
emission dominates over accretion as defined by \citet{Frasca2015}. The Class\,II sources are mostly above the boundary between chromospheres and accretion, 
but some of them are also in the domain of the chromospheric sources. However, all the objects fulfilling the \citet{WhiteBasri2003} criterion, based on 
the 10\% width and EW of the H$\alpha$ emission line, lie, correctly, above the aforementioned boundary, i.e. in the domain of accretors according to 
\citet{Frasca2015}.
We do not see any particular trend for the YSOs with transitional disks in this diagram. They display both low and high fluxes similarly to the behavior 
of their mass accretion rate shown by \citet{Alcala2017}.

The strong correlation between our  line fluxes per unit surface and the accretion luminosity ($L_{\rm acc}$) derived by \citet{Alcala2017} from 
the Balmer continuum excess has allowed us to confirm the validity of emission lines as accretion diagnostics and to exclude that the relations between $L_{\rm acc}$ 
and line luminosities are only the result of calibration effects.

We have also investigated the relations between the H$\alpha$ flux and the  fluxes in different emission lines (\ion{Ca}{ii}\,IRT, \ion{Ca}{ii}\,K, 
\ion{Na}{i}\,D, Pa$\gamma$, Pa$\beta$, and Br$\gamma$).
We note that, with the exception of the Class\,III sources and some of the weak accretors, all the objects display an H$\alpha$ flux in excess to what it is expected 
from the relations found by \citet{Stelzer2013} for Class\,III sources. 
This behavior is likely the result of different physical conditions in the emitting regions of the accretors  with respect to chromospheric emitters, 
which seem to also affect the flux ratios. 
The latter have been calculated for two \ion{Ca}{ii}-IRT lines,  $F_{\rm CaII 8542}/F_{\rm CaII 8498}$, and for the first two members of the Balmer series,
$F_{\rm H\alpha}/F_{\rm H\beta}$, i.e. the Balmer decrement.
We found that the \ion{Ca}{ii}-IRT flux ratio takes always small values ($F_{\rm CaII 8542}/F_{\rm CaII 8498}<2$) that indicate an optically thick emission
source. The latter can be identified with the accretion shock near the stellar photosphere. 
The Balmer decrement takes instead, for several accretors, high values typical of optically thin emission, suggesting that the Balmer emission originates in different
parts of the accretion funnels with a smaller optical depth.

\begin{acknowledgements}
We are very grateful to the referee, Fernando Comer\'{o}n, for his useful comments and suggestions.
We thank the ESO staff, in particular Markus Wittkowski and Giacomo Beccari for their excellent
support during phase-2 proposal preparation, and the Paranal staff for their
support during the observations. We thank G. Cupani, V. D'Odorico, P. Goldoni and 
A. Modigliani for their help with the X-Shooter pipeline. Financial support from 
INAF is also  acknowledged. CMF acknowledges an ESA Research Fellowship funding.
This research made use of the SIMBAD database, operated at the CDS (Strasbourg, France).
\end{acknowledgements}

\bibliographystyle{aa}

{}

\setlength{\tabcolsep}{3pt}

\begin{table*}[htb]
\caption[]{Stellar parameters.}
\scriptsize
\begin{tabular}{lllrrrrrrrrrrrrrlr}
\hline
\hline
\noalign{\smallskip}
 Name	                  & 	     HJD & Memb/ & \teff & $\sigma_{T_{\rm eff}}$ & \logg & err & \vsini &  err &    RV   &  err & $r_{970}$ & $r_{710}$ & $r_{620}$ & $r_{540}$ & $r_{450}$ & $M_{\ast}$  & Age \\ 
                          &(-2\,450\,000)& class & \multicolumn{2}{c}{(K)} &  \multicolumn{2}{c}{(dex)} & \multicolumn{2}{c}{(\kms)} &  \multicolumn{2}{c}{(\kms)} &  &  &  &  & &{\tiny ($M_{\sun}$)}  & {\tiny (Myr)}  \\  
\hline
\noalign{\smallskip}
  Sz\,66  		  &  6035.7578 & Y/II &   3351  &    47   & 3.81  &  0.21  &  $\leq$8.0  &  ...    &	  2.4  &   1.8 &  $\leq$0.2 & $\leq$0.2 &	0.8 &	    1.5   &  	  2.0	&  0.30   &  3.9   \\	
  AKC2006\,19		  &  5674.7067 & Y/II &   3027  &    34   & 4.45  &  0.11  &  $\leq$8.0  &  ...    &	  9.6  &   2.1 &  $\leq$0.2 & $\leq$0.2 & $\leq$0.2 &	    ...   &  	  ...	&  0.09   &  8.0   \\	
  Sz\,69  		  &  5674.6058 & Y/II &   3163  &   119   & 3.50  &  0.23  &	    33.0 &   4.0   &	  5.4  &   2.9 &  $\leq$0.2 & $\leq$0.2 & $\leq$0.2 &	    ...   &  	  ...	&  0.20   &  2.6   \\	
  Sz\,71  		  &  6035.6010 & Y/II &   3599  &    35   & 4.27  &  0.24  &  $\leq$8.0  &  ...    &	 -3.3  &   1.9 &  $\leq$0.2 &	    0.3 &	0.5 &	    0.6   &  	  0.6	&  0.40   &  2.0   \\	
  Sz\,72  		  &  6035.6987 & Y/II &   3550  &    70   & 4.18  &  0.28  &  $\leq$8.0  &  ...    &	  6.9  &   2.4 &  $\leq$0.2 & $\leq$0.2 & $\leq$0.2 &	    ...   &  	  ...	&  0.40   &  2.9   \\	
  Sz\,73  		  &  6035.7983 & Y/II &   3980  &    33   & 4.40  &  0.33  &        31.0 &   3.0   & 	  5.0  &   2.2 &  $\leq$0.2 &	    0.5 &	1.0 &	    1.0   &  	  0.5	&  0.70   &  3.7   \\	
  Sz\,74  		  &  6035.6747 & Y/II &   3371  &    79   & 3.98  &  0.11  &        30.2 &   1.0   & 	  1.0  &   1.5 &  $\leq$0.2 & $\leq$0.2 &	0.4 &	    0.7   &  	  0.8	&  0.30   &  0.5   \\	
  Sz\,83  		  &  6035.6842 & Y/II &   4037  &    96   & 3.5   &  0.4   &         8.5 &   4.8   &	  3.3  &   1.8 &       1.4  &	    2.4 &       3.5 &       ...   &  	  ...	&  0.70   &  0.8   \\	
  Sz\,84  		  &  6035.6145 & Y/II &   3058  &    82   & 4.20  &  0.21  &	    21.3 &   4.1   &	 -3.1  &   2.0 &  $\leq$0.2 & $\leq$0.2 &       0.3 &       0.9   &  	  1.5	&  0.10   &  1.8   \\	
  Sz\,130 		  &  5293.8661 & Y/II &   3448  &    62   & 4.48  &  0.10  &  $\leq$8.0  &   ...   &	  3.6  &   2.6 &  $\leq$0.2 & $\leq$0.2 &	0.7 &	    ...   &  	  ...	&  0.30   &  2.3   \\	
  Sz\,88\,A		  &  6035.8400 & Y/II &   3700  &    19   & 3.77  &  0.47  &  $\leq$8.0  &   ...   &	  6.5  &   2.3 &       0.4  &	    1.4 &       2.1 &       ...   &  	  ...	&  0.50   &  4.0   \\	
  Sz\,88\,B		  &  6035.8400 & Y/II &   3075  &   111   & 4.08  &  0.12  &  $\leq$8.0  &   ...   &	  5.7  &   2.1 &  $\leq$0.2 &	    0.3 &       0.3 &       ...   &  	  ...	&  0.20   &  2.0   \\	
  Sz\,91  		  &  6035.8270 & Y/II &   3664  &    45   & 4.34  &  0.22  &  $\leq$8.0  &   ...   &	  4.0  &   2.4 &  $\leq$0.2 &       0.4 &       0.6 &       0.9   &  	  0.9	&  0.40   &  1.5   \\	
  Lup\,713		  &  5292.7036 & Y/II &   2943  &   109   & 3.89  &  0.13  &	    24.0 &   4.0   & 	  3.9  &   3.3 &  $\leq$0.2 & $\leq$0.2 &	... &	    ...   &  	  ...	&  0.08   &  4.7   \\	
  Lup\,604s		  &  5292.7434 & Y/II &   3014  &    62   & 3.89  &  0.12  &	    31.5 &   2.8   & 	  2.7  &   1.9 &  $\leq$0.2 & $\leq$0.2 & $\leq$0.2 &	    ...   &  	  ...	&  0.10   &  2.2   \\	
  Sz\,97  		  &  5674.6202 & Y/II &   3185  &    78   & 4.14  &  0.16  &	    25.1 &   1.5   &	  2.4  &   2.2 &  $\leq$0.2 & $\leq$0.2 & $\leq$0.2 &	    0.7   &  	  0.9	&  0.20   &  1.3   \\	
  Sz\,99  		  &  5293.8505 & Y/II &   3297  &    98   & 3.89  &  0.21  &	    38.0 &   3.0   &	  3.2  &   3.1 &  $\leq$0.2 & $\leq$0.2 & $\leq$0.2 &	    ...   &  	  ...	&  0.30   &  7.5   \\	
  Sz\,100 		  &  5674.6454 & Y/II &   3037  &    44   & 3.87  &  0.21  &	    16.6 &   5.4   & 	  2.7  &   2.5 &  $\leq$0.2 & $\leq$0.2 & $\leq$0.2 &	    ...   &  	  ...	&  0.10   &  4.7   \\	
  Sz\,103 		  &  5674.6604 & Y/II &   3380  &    36   & 3.97  &  0.10  &	    12.0 &   4.0   & 	  1.4  &   2.2 &  $\leq$0.2 & $\leq$0.2 &	0.3 &	    0.7   &  	  0.7	&  0.30   &  3.3   \\	
  Sz\,104 		  &  5292.7601 & Y/II &   3074  &    73   & 3.96  &  0.10  &  $\leq$8.0  &   ...   &	  2.3  &   2.3 &  $\leq$0.2 & $\leq$0.2 & $\leq$0.2 &	    ...   &  	  ...	&  0.20   &  2.4   \\	
  Lup\,706		  &  5292.7882 & Y/II &   2750  &    82   & 3.93  &  0.11  &	    25.0 &  15.0   & 	 11.9  &   4.5 &  $\leq$0.2 & $\leq$0.2 &	... &	    ...   &  	  ...	& ...$^{a}$ &  ...$^{a}$ \\   
  Sz\,106 		  &  6035.8117 & Y/II &   3691  &    35   & 4.82  &  0.13  &  $\leq$8.0  &   ...   &	  8.0  &   2.6 &  $\leq$0.2 & $\leq$0.2 &       0.5 &       0.5   &  	  0.5	& ...$^{a}$ &  ...$^{a}$ \\   
  Par-Lup3-3		  &  5292.8638 & Y/II &   3461  &    49   & 4.47  &  0.11  &  $\leq$8.0  &   ...   &	  3.5  &   2.5 &  $\leq$0.2 & $\leq$0.2 & $\leq$0.2 &	    ...   &  	  ...	&  0.30  &  1.5   \\  
  Par-Lup3-4		  &  5293.6849 & Y/II &   3089  &   246   & 3.56  &  0.80  &	    12.0 &  10.0   &	  5.6  &   4.3 &	0.3 &	    0.3 &	... &	    ...   &  	  ...	& ...$^{a}$ &  ...$^{a}$ \\   
  Sz\,110 		  &  5674.8476 & Y/II &   3215  &   162   & 4.31  &  0.21  &  $\leq$8.0  &   ...   &	  2.6  &   2.3 &	0.4 &	    0.4 &	1.0 &	    2.7   &  	  3.3	&  0.20   &  0.6   \\  
  Sz\,111 		  &  6035.6286 & Y/II &   3683  &    34   & 4.66  &  0.21  &  $\leq$8.0  &   ...   &	 -1.2  &   2.1 &  $\leq$0.2 &	    0.3 &       0.5 &       0.6   &  	  0.6	&  0.50   &  4.2   \\ 
  Sz\,112 		  &  6035.7704 & Y/II &   3079  &    47   & 3.96  &  0.10  &  $\leq$8.0  &   ...   &	  6.2  &   1.7 &  $\leq$0.2 & $\leq$0.2 & $\leq$0.2 &       ...   &  	  ...	&  0.10   & 11.3   \\ 
  Sz\,113 		  &  5674.8663 & Y/II &   3064  &   114   & 3.76  &  0.27  &  $\leq$8.0  &   ...   &	  6.1  &   2.4 &	0.3 &	    0.5 &	... &	    ...   &  	  ...	&  0.10   &  1.8   \\ 
 2MASS\,J16085953-3856275 &  5674.7905 & Y/II &   2649  &    31   & 3.98  &  0.10  &  $\leq$8.0  &   ...   &	  7.2  &   4.0 &  $\leq$0.2 & $\leq$0.2 & $\leq$0.2 &	    ...   &  	  ...	&  0.025  &  1.0   \\ 
  SSTc2d160901.4-392512   &  5674.7331 & Y/II &   3305  &    57   & 4.51  &  0.11  &  $\leq$8.0  &   ...   &	 15.9  &   0.7 &  $\leq$0.2 & $\leq$0.2 & $\leq$0.2 &	    0.7   &  	  0.6	&  0.30   &  6.1   \\ 
  Sz\,114 		  &  5674.8913 & Y/II &   3134  &    35   & 3.92  &  0.12  &  $\leq$8.0  &   ...   &	  4.0  &   2.4 &  $\leq$0.2 &	    0.3 &	0.3 &	    0.6   &  	  0.8	&  0.20   &  2.4   \\ 
  Sz\,115 		  &  6035.8529 & Y/II &   3124  &    42   & 3.90  &  0.21  &         9.2 &   6.3   &      6.4  &   2.3 &  $\leq$0.2 & $\leq$0.2 & $\leq$0.2 &       ...   &  	  ...	&  0.20   &  2.3   \\ 
  Lup\,818s		  &  5674.7539 & Y/II &   2953  &    59   & 3.97  &  0.11  &	    12.4 &   6.0   &      5.1  &   2.0 &  $\leq$0.2 & $\leq$0.2 & $\leq$0.2 &       ...   &  	  ...	&  0.08   &  4.2   \\ 
  Sz123\,A		  &  6035.7844 & Y/II &   3521  &    70   & 4.46  &  0.13  &	    12.3 &   3.0   &      1.8  &   1.8 &  $\leq$0.2 & $\leq$0.2 &       0.6 &       1.2   &  	  1.2	&  0.40   &  4.4   \\ 
  Sz123\,B		  &  6035.8971 & Y/II &   3513  &    45   & 4.17  &  0.22  &  $\leq$8.0  &   ...   &      7.4  &   2.3 &  $\leq$0.2 & $\leq$0.2 &       0.6 &       0.8   &  	  0.9	& ...$^{a}$ &  ...$^{a}$ \\  
  SST-Lup3-1		  &  5292.8841 & Y/II &   3042  &    43   & 3.95  &  0.11  &  $\leq$8.0  &   ...   &      6.3  &   2.6 &  $\leq$0.2 & $\leq$0.2 & $\leq$0.2 &	    ...   &  	  ...	&  0.10   &  3.2   \\
\noalign{\medskip}
  Sz\,65                  &  7177.5239 & Y/II &   4005  &    75   & 3.85  &  0.26  &  $\leq$8.0  &  ...    &	 -2.7  &   2.0 &  $\leq$0.2 & $\leq$0.2 &	0.3 &	    0.3   & $\leq$0.2   &  0.70   &   1.9   \\ 
  AKC2006\,18 	          &  7132.7794 & Y/II &   2930  &    45   & 4.46  &  0.11  &  $\leq$8.0  &  ...    &	  9.1  &   2.3 &  $\leq$0.2 & $\leq$0.2 & $\leq$0.2 &	    ...   &	  ...   &  0.07   &   8.3   \\ 
  SSTc2dJ154508.9-341734  &  7188.6599 & Y/II &   3242  &   205   & 3.33  &  0.77  &  $\leq$8.0  &  ...    &	 -0.8  &   2.7 &	0.6 &	    0.7 &	0.8 &	    ...   &	  ...   &  0.20   &   4.8   \\ 
  Sz\,68                  &  7160.7168 & Y/II &   4506  &    82   & 3.68  &  0.12  &	    39.6 &   1.2   &	 -4.3  &   1.8 &	0.3 &	    0.3 &	0.3 &	    0.3   &	  0.3   &  1.20   &   0.5   \\ 
  SSTc2dJ154518.5-342125  &  7199.4886 & Y/II &   2700  &   100   & 3.45  &  0.11  &	    13.0 &   6.0   &	  4.4  &   2.9 &  $\leq$0.2 & $\leq$0.2 & $\leq$0.2 &	    ...   &	  ...   &  0.07   &   0.5   \\ 
  Sz\,81\,A               &  7253.5752 & Y/II &   3077  &   151   & 3.48  &  0.11  &  $\leq$8.0  &  ...    & 	 -0.1  &   2.9 &  $\leq$0.2 & $\leq$0.2 &	0.3 &	    ...   &	  ...	&  0.20   &   1.0   \\ 
  Sz\,81\,B               &  7253.5752 & Y/II &   2991  &    76   & 3.53  &  0.11  &	   25.0  &   5.0   & 	  1.2  &   2.4 &  $\leq$0.2 & $\leq$0.2 &	0.3 &	    ...   &	  ...	&  0.10   &   1.7   \\ 
  Sz\,129                 &  7199.5644 & Y/II &   4005  &    45   & 4.49  &  0.25  &  $\leq$10.0 &  ...    & 	  3.2  &   2.5 &  $\leq$0.2 & $\leq$0.2 &	1.0 &	    1.4   &	  1.1	&  0.70   &   3.5   \\ 
  SSTc2dJ155925.2-423507  &  7200.5241 & Y/II &   2984  &    82   & 4.41  &  0.12  &  $\leq$8.0  &  ...    &	  6.5  &   2.5 &  $\leq$0.2 & $\leq$0.2 & $\leq$0.2 &	    ...   &	  ...   &  0.09   &   5.8   \\ 
  RY~Lup                  &  7205.7799 & Y/II &   5082  &   118   & 3.87  &  0.22  &	    16.3 &   5.3   &	  1.3  &   2.0 &	0.3 & $\leq$0.2 &	0.5 &	    0.4   & $\leq$0.2   &  1.40   &  10.2   \\ 
  SSTc2dJ160000.6-422158  &  7115.8505 & Y/II &   3086  &    82   & 4.03  &  0.11  &  $\leq$8.0  &  ...    &	  2.5  &   2.2 &  $\leq$0.2 & $\leq$0.2 & $\leq$0.2 &	    ...   &	  ...   &  0.20   &   2.6   \\ 
  SSTc2dJ160002.4-422216  &  7204.6729 & Y/II &   3159  &   140   & 4.00  &  0.48  &  $\leq$8.0  &  ...    &	  2.6  &   2.6 &  $\leq$0.2 & $\leq$0.2 & $\leq$0.2 &	    0.3   &	  0.3   &  0.20   &   1.5   \\ 
  SSTc2dJ160026.1-415356  &  7201.5176 & Y/II &   2976  &   221   & 3.97  &  0.35  &  $\leq$8.0  &  ...    &	 -1.1  &   2.3 &  $\leq$0.2 & $\leq$0.2 &	0.3 &	    ...   &	  ...   &  0.10   &   1.6   \\ 
  MY~Lup                  &  7199.5763 & Y/II &   4968  &   200   & 3.72  &  0.18  &	    29.1 &   2.0   &	  4.4  &   2.1 &  $\leq$0.2 & $\leq$0.2 & $\leq$0.2 & $\leq$0.2   & $\leq$0.2   &  1.10   &  16.6   \\ 
  Sz\,131	          &  7204.7286 & Y/II &   3300  &   122   & 4.29  &  0.36  &  $\leq$8.0  &  ...    &	  2.4  &   2.2 &  $\leq$0.2 & $\leq$0.2 &	0.3 &	    ...   &	  ...   &  0.20   &   1.6   \\ 
  Sz\,133                 &  7205.7011 & Y/II &   4420  &   129   & 3.96  &  0.51  &  $\leq$8.0  &  ...    &	  0.7  &   2.5 &	0.5 &	    0.5 &	0.7 &	    0.7   &	  0.7   & ...$^{a}$ &  ...$^{a}$ \\ 
SSTc2dJ160703.9-391112    &  7542.7095 & Y/II &   3072  &    55   & 4.01  &  0.11  &  $\leq$8.0  &  ...    &	  1.8  &   2.7 & $\leq$0.2 & $\leq$0.2 & $\leq$0.2  &	   ...    &	  ...   & ...$^{a}$ &  ...$^{a}$ \\    
SSTc2dJ160708.6-391408    &  7545.5860 & Y/II &   3474  &   206   & 4.18  &  0.56  &  $\leq$8.0  &  ...    &	 -4.2  &   2.9 &        0.7 &	   0.9 &       1.0  &	   ...    &	  ...   &  0.30$^{b}$ & 3.0$^{b}$   \\	
  Sz\,90                  &  7215.6018 & Y/II &   4022  &    52   & 4.24  &  0.42  &  $\leq$8.0  &  ...    &      1.6  &   2.3 &  $\leq$0.2 & $\leq$0.2 &	0.5 &	    0.5   &	  0.5	&  0.70   &   2.1   \\ 
  Sz\,95                  &  7215.6243 & Y/II &   3443  &    53   & 4.37  &  0.12  &  $\leq$8.0  &  ...    &     -2.8  &   2.3 &  $\leq$0.2 & $\leq$0.2 & $\leq$0.2 &	    ...   &	  ...	&  0.30   &   0.8   \\ 
  Sz\,96                  &  7206.7765 & Y/II &   3702  &    57   & 4.50  &  0.23  &  $\leq$8.0  &  ...    &     -2.7  &   2.6 &  $\leq$0.2 & $\leq$0.2 &	0.4 &	    0.3   &	  0.3	&  0.40   &   0.6   \\ 
  2MASSJ16081497-3857145  &  7132.8450 & Y/II &   3024  &    46   & 3.96  &  0.23  &        22.0 &   4.2   &	  6.0  &   2.9 &  $\leq$0.2 & $\leq$0.2 & $\leq$0.2 &	    ...   &	  ...   &  0.09   &  15.2   \\ 
  Sz\,98                  &  7205.6379 & Y/II &   4080  &    71   & 4.10  &  0.23  &  $\leq$8.0  &  ...    & 	 -1.4  &   2.1 &	0.3 &	    0.3 &	0.6 &	    0.5   &	  0.5	&  0.70   &   0.5   \\ 
  Lup\,607                &  7165.7248 & Y/II &   3084  &    46   & 4.48  &  0.11  &  $\leq$8.0  &  ...    & 	  6.8  &   2.4 &  $\leq$0.2 & $\leq$0.2 & $\leq$0.2 &	    ...   &	  ...	&  0.10   &   5.8   \\ 
  Sz\,102 	          &  7129.8096 & Y/II &   5145  &    50   & 4.10  &  0.50  &	    41.0 &   2.8   &	 21.6  &  12.4 &	1.0 &	    2.0 &	2.5 &	    ...   &	  ...   & ...$^{a}$ &  ...$^{a}$ \\  
  SSTc2dJ160830.7-382827  &  7205.6604 & Y/II &   4797  &   145   & 4.09  &  0.23  &  $\leq$8.0  &  ...    &	  1.2  &   1.9 &  $\leq$0.2 & $\leq$0.2 & $\leq$0.2 &	    0.1   & $\leq$0.2   &  1.80*  &   2.9*  \\ 
SSTc2dJ160836.2-392302    &  7522.7342 & Y/II &   4429  &   83    & 4.04  &  0.13  &  $\leq$8.0  &   ...   &      2.2  &   2.1 &  $\leq$0.2 & $\leq$0.2 &       0.3 &       0.3   & $\leq$0.2   &  1.10   &   1.3   \\   
  Sz\,108\,B              &  7191.6713 & Y/II &   3102  &    59   & 3.86  &  0.22  &  $\leq$8.0  &  ...    &	  0.2  &   2.2 &  $\leq$0.2 & $\leq$0.2 & $\leq$0.2 &	    ...   &	  ...   &  0.20   &   2.3   \\ 
  2MASSJ16085324-3914401  &  7215.6425 & Y/II &   3393  &    85   & 4.24  &  0.41  &  $\leq$8.0  &  ...    &	  0.9  &   2.2 &  $\leq$0.2 & $\leq$0.2 &	0.3 &	    ...   &	  ...   &  0.30   &   1.1   \\ 
  2MASSJ16085373-3914367  &  7165.7974 & Y/II &   2840  &   200   & 3.60  &  0.45  &  $\leq$8.0  &  ...    &	  6.1  &   6.9 &  $\leq$0.2 & $\leq$0.2 & $\leq$0.2 &	    ...   &	  ...   &  0.07   &  13.8   \\ 
  2MASSJ16085529-3848481  &  7215.6660 & Y/II &   2899  &    55   & 3.99  &  0.11  &  $\leq$8.0  &  ...    &	 -1.1  &   2.4 &  $\leq$0.2 & $\leq$0.2 & $\leq$0.2 &	    ...   &	  ...   &  0.09   &   0.7   \\ 
  SSTc2dJ160927.0-383628  &  7216.5318 & Y/II &   3147  &    58   & 4.00  &  0.10  &  $\leq$8.0  &  ...    &	  3.4  &   2.2 &	0.5 &	    0.5 &	1.5 &	    ...   &	  ...   &  0.20   &   2.2   \\ 
  Sz\,117                 &  7216.5747 & Y/II &   3470  &    51   & 4.10  &  0.32  &  $\leq$8.0  &  ...    &     -1.4  &   2.2 &  $\leq$0.2 & $\leq$0.2 &	0.4 &	    0.7   &	  0.7	&  0.30   &   0.7   \\ 
  Sz\,118                 &  7216.6035 & Y/II &   4067  &    82   & 4.47  &  0.37  &  $\leq$8.0  &  ...    &     -0.8  &   2.3 &  $\leq$0.2 & $\leq$0.2 &	0.5 &	    0.5   &	  0.5	&  0.70   &   1.0   \\ 
  2MASSJ16100133-3906449  &  7216.6318 & Y/II &   3018  &    72   & 3.77  &  0.23  &	    15.9 &   5.4   &	  0.1  &   2.6 &  $\leq$0.2 & $\leq$0.2 & $\leq$0.2 &	    ...   &	  ...   &  0.10   &   1.8   \\ 
  SSTc2dJ161018.6-383613  &  7252.5473 & Y/II &   3012  &    48   & 3.98  &  0.10  &  $\leq$8.0  &  ...    &	 -0.2  &   2.5 &  $\leq$0.2 & $\leq$0.2 & $\leq$0.2 &	    ...   &	  ...   &  0.10   &   2.6   \\ 
  SSTc2dJ161019.8-383607  &  7242.5936 & Y/II &   2861  &    69   & 4.00  &  0.10  &	    25.0 &   7.0   &	  1.8  &   2.5 &  $\leq$0.2 & $\leq$0.2 & $\leq$0.2 &	    ...   &	  ...   &  0.09   &   0.5   \\ 
  SSTc2dJ161029.6-392215  &  7247.5973 & Y/II &   3098  &    59   & 4.02  &  0.10  &	    13.0 &   4.0   &	 -2.7  &   2.2 &  $\leq$0.2 & $\leq$0.2 & $\leq$0.2 &	    ...   &	  ...   &  0.20   &   2.0   \\ 
  SSTc2dJ161243.8-381503  &  7213.6804 & Y/II &   3687  &    22   & 4.57  &  0.21  &  $\leq$8.0  &  ...    &	 -2.3  &   2.4 &  $\leq$0.2 & $\leq$0.2 &	0.3 &	    ...   &	  ...   &  0.40   &   0.5   \\ 
  SSTc2dJ161344.1-373646  &  7200.4846 & Y/II &   3028  &    78   & 4.25  &  0.14  &  $\leq$8.0  &  ...    &	 -1.2  &   2.3 &  $\leq$0.2 &	    0.3 &	0.5 &	    ...   &	  ...   &  0.10   &   1.8   \\ 
\hline											       
\end{tabular}									       
\normalsize
\label{Tab:param}
\end{table*}

\addtocounter{table}{-1}

\begin{table*}[htb]
\caption[]{{\it Continued.}}
\scriptsize
\begin{tabular}{lllrrrrrrrrrrrrrlr}
\hline
\hline
\noalign{\smallskip}
 Name	                  & 	     HJD & Memb/ & \teff & $\sigma_{T_{\rm eff}}$ & \logg & err & \vsini &  err &    RV   &  err & $r_{970}$ & $r_{710}$ & $r_{620}$ & $r_{540}$ & $r_{450}$  & $M_{\ast}$  & Age \\ 
                          &(-2\,450\,000)& Class & \multicolumn{2}{c}{(K)} &  \multicolumn{2}{c}{(dex)} & \multicolumn{2}{c}{(\kms)} &  \multicolumn{2}{c}{(\kms)} &  &  &  &  & & ($M_{\sun}$)  & (Myr) \\ 
\hline
\noalign{\smallskip}
  GQ~Lup                  &  5321.7673 & Y/II &   4192  &   65    & 4.12  &  0.36  &  $\leq$6.0  &    ...  & 	 -3.6  &   1.3 & 	0.3 &	    0.3 &	0.5 &	    0.8   &       0.6	&  0.80   &   0.9   \\
  Sz\,76                  &  6775.7518 & Y/II &   3440  &   60    & 4.41  &  0.26  &  $\leq$6.0  &    ...  & 	  1.4  &   1.0 &  $\leq$0.2 & $\leq$0.2 & $\leq$0.2 &	    0.7   &       0.6	&  0.30   &   2.3   \\
  Sz\,77                  &  5321.8825 & Y/II &   4131  &   48    & 4.28  &  0.31  &       6.7   &    1.5  & 	  2.4  &   1.5 &  $\leq$0.2 & $\leq$0.2 &	0.4 &	    0.5   &       0.3	&  0.80   &   3.0   \\
  RXJ1556.1-3655          &  6775.7703 & Y/II &   3770  &   70    & 4.75  &  0.21  &      12.7	 &    1.3  & 	  2.6  &   1.2 &  $\leq$0.2 &	    0.3 &	1.4 &	    ...   &       ...	&  0.60   &   7.8   \\
  IM~Lup                  &  5320.5707 & Y/II &   4146  &   95    & 3.80  &  0.37  &      17.1	 &    1.4  & 	 -0.5  &   1.3 &  $\leq$0.2 & $\leq$0.2 & $\leq$0.2 & $\leq$0.2   & $\leq$0.2   &  0.70   &   0.5   \\
  EX~Lup                  &  5320.6616 & Y/II &   3859  &   62    & 4.31  &  0.29  &       7.5	 &    1.9  & 	  1.9  &   1.4 & 	0.4 & $\leq$0.2 &	0.7 &	    1.0   &       0.7   &  0.50   &   0.5   \\
\noalign{\medskip}
  Sz\,94  		  &  5292.7286 & Y/III&   3205  &    59   & 4.21  &  0.24  &        38.0 &    3.0  &      7.8  &   2.0 &  $\leq$0.2 & $\leq$0.2 & $\leq$0.2 &       ...   &  	  ...   &  0.20   &   1.3   \\  
  Par\,Lup3\,1		  &  5293.8248 & Y/III&   2766  &    77   & 3.48  &  0.11  &        31.0 &    7.0  &      4.9  &   3.9 &  $\leq$0.2 & $\leq$0.2 & $\leq$0.2 &       ...   &  	  ...   &  0.08   &   0.5   \\  
  Par\,Lup3\,2		  &  5292.8383 & Y/III&   3060  &    61   & 3.91  &  0.24  &        33.0 &    5.0  &      5.0  &   2.5 &  $\leq$0.2 & $\leq$0.2 & $\leq$0.2 &       ...   &  	  ...   &  0.10   &   1.8   \\  
  Sz\,107 		  &  5674.6759 & Y/III&   2928  &   100   & 3.52  &  0.13  &        66.6 &    4.1  &      6.1  &   3.8 &  $\leq$0.2 & $\leq$0.2 & $\leq$0.2 &       ...   &  	  ...   &  0.10   &   0.5   \\  
  Sz\,108\,A		  &  7191.6832 & Y/III&   3676  &    34   & 4.50  &  0.21  &  $\leq$8.0  &    ...  &	  0.4  &   2.2 &  $\leq$0.2 & $\leq$0.2 &	0.4 &       0.5   &  	  0.3   &  0.40   &   1.0   \\  
  Sz\,121 		  &  6035.6419 & Y/III&   3178  &    61   & 4.08  &  0.12  &        87.0 &    8.0  &     -0.9  &   4.4 &  $\leq$0.2 & $\leq$0.2 & $\leq$0.2 &       ...   &  	  ...   &  0.20   &   0.5   \\  
  Sz\,122 		  &  6035.7433 & Y/III&   3511  &    27   & 4.58  &  0.21  &       149.2 &    1.0  &      2.9  &   3.0 &  $\leq$0.2 & $\leq$0.2 & $\leq$0.2 &       ...   &  	  ...   &  0.40   &   6.8   \\  
\noalign{\medskip}
  Sz\,78  		  &  7199.5386 & N$^{c}$ & 4347 &   119   & 2.04  &  0.23  &  $\leq$8.0  &  ...    &      1.8  &   2.0 &  $\leq$0.2 & $\leq$0.2 & $\leq$0.2 & $\leq$0.2   & $\leq$0.2   &   ...   &   ...   \\ 
  Sz\,79  		  &  7199.5523 & N$^{c}$ & 4718 &   147   & 2.44  &  0.22  &	    19.3 &   2.1   &    -84.6  &   1.8 &  $\leq$0.2 & $\leq$0.2 & $\leq$0.2 &       ...   &	  ...   &   ...   &   ...   \\ 
  IRAS15567-4141	  &  7205.6767 & N$^{c}$ & 3130 &   135   & 2.83  &  0.45  &	    30.0 &  14.0   &	-43.0  &   8.2 &  $\leq$0.2 & $\leq$0.2 & $\leq$0.2 &	    ...   &	  ...   &   ...   &   ...   \\ 
  SSTc2dJ160034.4-422540  &  7204.7102 & N$^{c}$ & 2050 &   150   & ...   &  ...   &	    ...  &  ...    &	-84.5  &   4.3 &        ... &      ...  &       ... &       ...   &	  ...   &   ...   &   ...   \\ 
  SSTc2dJ160708.6-394723  &  7206.7441 & N$^{c}$ & 4647 &   104   & 2.64  &  0.23  &	    10.1 &   1.0   &    -70.7  &   1.8 &  $\leq$0.2 & $\leq$0.2 & $\leq$0.2 & $\leq$0.2   & $\leq$0.2   &   ...   &   ...   \\ 
  Sz\,92  		  &  7205.7363 & N$^{d}$ & 4926 &   111   & 2.29  &  0.26  &  $\leq$8.0  &  ...    &    -34.8  &   2.3 &  $\leq$0.2 & $\leq$0.2 & $\leq$0.2 & $\leq$0.2   & $\leq$0.2   &   ...   &   ...   \\ 
  2MASSJ16080618-3912225  &  7132.9002 & N$^{c}$ & 2753 &   130   & 2.02  &  0.88  &	    13.0 &   6.0   &    -57.5  &   8.2 &  $\leq$0.2 & $\leq$0.2 & $\leq$0.2 &	    ...   &	  ...   &   ...   &   ...   \\ 
  Sz\,105 		  &  6035.6593 & N$^{c}$ & 3187 &   102   & 0.76  &  0.58  &	    14.3 &   7.8   &      4.7  &   1.1 &        0.4 &       0.4 &       0.5 &       ...   &	  ...   &   ...   &   ...   \\ 
  SSTc2dJ161045.4-385455  &  7545.5166 & N$^{c}$ & 5114 &   514   & 3.40  &  0.90  &        42.6 &   2.3   &   -115.1  &   3.1 &  $\leq$0.2 & $\leq$0.2 & $\leq$0.2 & $\leq$0.2   & $\leq$0.2   &   ...   &   ...   \\ 
  SSTc2dJ161148.7-381758  &  7116.8337 & N$^{c}$ & 6877 &   143   & 3.91  &  0.58  &	    23.1 &   2.4   &	-47.6  &   2.5 &  $\leq$0.2 & $\leq$0.2 & $\leq$0.2 & $\leq$0.2   & $\leq$0.2   &   ...   &   ...   \\ 
  SSTc2dJ161211.2-383220  &  7520.7096 & N$^{c}$ & 3401 &    37   & 1.35  &  0.41  &	    19.0 &   15.0  &	-80.2  &   1.8 &  $\leq$0.2 & $\leq$0.2 & $\leq$0.2 &	    ...   &       ...   &   ...   &   ...   \\ 
  SSTc2dJ161222.7-371328  &  7177.5451 & N$^{c}$ & 3321 &    79   & 0.99  &  0.23  &  $\leq$8.0  &  ...    &      8.4  &   1.8 &  $\leq$0.2 & $\leq$0.2 & $\leq$0.2 & $\leq$0.2   & $\leq$0.2   &   ...   &   ...   \\ 
  SSTc2dJ161256.0-375643  &  7205.7677 & N$^{d}$ & 3544 &    66   & 0.54  &  0.25  &	    19.0 &   4.0   &   -112.5  &   1.8 &  $\leq$0.2 & $\leq$0.2 & $\leq$0.2 & $\leq$0.2   & $\leq$0.2   &   ...   &   ...   \\ 
\hline			        							     
\end{tabular}										      
\normalsize
* Mass and age derived from \citet{Siess2000} evolutionary models.\\
$^{a}$  Subluminous objects.\\
$^{b}$  Flat IR source \citep{Merin2008}. Subluminous with the ``optical'' luminosity. Mass and age derived adopting the bolometric luminosity of 0.18\,$L_{\sun}$ \citep{Evans2009}.\\
$^{c}$  Originally selected as Class\,II.\\
$^{d}$  Originally selected as Class\,III.
\end{table*}

\normalsize

\begin{landscape}
\setlength{\tabcolsep}{3pt}

\begin{table}[htb]
\caption[]{H$\alpha$, H$\beta$, \ion{Ca}{ii}, and \ion{Na}{i}~D$_{1,2}$ fluxes.}
\scriptsize
\begin{tabular}{lrrrrrrrrrrrrrrrrrr}
\hline
\hline
\noalign{\smallskip}
 Name	                  & $10\%W_{\rm H\alpha}$ &  err & $F_{\rm H\alpha}$ & err & $F_{\rm H\beta}$ & err& $F_{\rm CaII8498}$ & err & $F_{\rm CaII8542}$ & err & $F_{\rm CaII8662}$ & err & $F_{\rm CaII-K}$ & err & $F_{\rm NaI-D_2}$ & err & $F_{\rm NaI-D_1}$ & err \\  
                          & \multicolumn{2}{c}{(\kms)}   & \multicolumn{2}{c}{(erg\,cm$^{-2}$s$^{-1}$)}& \multicolumn{2}{c}{(erg\,cm$^{-2}$s$^{-1}$)} & \multicolumn{2}{c}{(erg\,cm$^{-2}$s$^{-1}$)}  & \multicolumn{2}{c}{(erg\,cm$^{-2}$s$^{-1}$)} & \multicolumn{2}{c}{(erg\,cm$^{-2}$s$^{-1}$)}& \multicolumn{2}{c}{(erg\,cm$^{-2}$s$^{-1}$)} & \multicolumn{2}{c}{(erg\,cm$^{-2}$s$^{-1}$)}& \multicolumn{2}{c}{(erg\,cm$^{-2}$s$^{-1}$)}\\   
\hline
\noalign{\smallskip}
  Sz66  		  &  475.4 &  22.8  & 1.58$\cdot 10^{7}$ &  1.94$\cdot 10^{6}$ &  3.08$\cdot 10^{6}$  &   3.93$\cdot 10^{5}$   &  8.26$\cdot 10^{5}$ & 1.14$\cdot 10^{5}$ &  1.06$\cdot 10^{6}$ &  1.19$\cdot 10^{5}$ &  1.01$\cdot 10^{6}$ &  1.20$\cdot 10^{5}$ &  5.62$\cdot 10^{5}$  &  3.04$\cdot 10^{4}$  &  1.53$\cdot 10^{5}$  & 4.68$\cdot 10^{4}$ &  1.00$\cdot 10^{5}$  &   3.85$\cdot 10^{4}$  \\ 
  AKC2006\,19		  &  228.5 &  45.7  & 2.57$\cdot 10^{6}$ &  3.94$\cdot 10^{5}$ &  8.91$\cdot 10^{5}$  &   1.41$\cdot 10^{5}$   &  5.66$\cdot 10^{4}$ & 1.74$\cdot 10^{4}$ &  6.43$\cdot 10^{4}$ &  2.81$\cdot 10^{4}$ &  5.43$\cdot 10^{4}$ &  1.74$\cdot 10^{4}$ &  2.03$\cdot 10^{5}$  &  2.12$\cdot 10^{4}$  &  1.23$\cdot 10^{4}$  & 9.10$\cdot 10^{3}$ &  1.39$\cdot 10^{4}$  &   1.22$\cdot 10^{4}$  \\ 
  Sz69  		  &  411.4 &  13.7  & 4.69$\cdot 10^{7}$ &  1.43$\cdot 10^{7}$ &  5.41$\cdot 10^{6}$  &   2.03$\cdot 10^{6}$   &  4.01$\cdot 10^{6}$ & 6.74$\cdot 10^{5}$ &  4.62$\cdot 10^{6}$ &  7.76$\cdot 10^{5}$ &  4.00$\cdot 10^{6}$ &  6.73$\cdot 10^{5}$ &  2.22$\cdot 10^{6}$  &  3.03$\cdot 10^{5}$  &  4.97$\cdot 10^{5}$  & 2.86$\cdot 10^{5}$ &  3.74$\cdot 10^{5}$  &   2.18$\cdot 10^{5}$  \\
\hline  																																							    
\end{tabular}																																							    
\label{Tab:fluxes}																																						    
\end{table}																																							    

\end{landscape}

\begin{landscape}

\addtocounter{table}{-1}																																					    

\begin{table}[htb]																																						    
\caption[]{{\it Continued.}}																																					     
\scriptsize																																							     
																	     
\begin{tabular}{lrrrrrrrrrrrrrrrrrr}
\hline  																																							     
\hline  																																							     
\noalign{\smallskip}																																						     
 Name	                  & $10\%W_{\rm H\alpha}$ &  err & $F_{\rm H\alpha}$ & err & $F_{\rm H\beta}$ & err& $F_{\rm CaII8498}$ & err & $F_{\rm CaII8542}$ & err & $F_{\rm CaII8662}$ & err & $F_{\rm CaII-K}$ & err & $F_{\rm NaI-D_2}$ & err & $F_{\rm NaI-D_1}$ & err \\ 
                          & \multicolumn{2}{c}{(\kms)}   & \multicolumn{2}{c}{(erg\,cm$^{-2}$s$^{-1}$)}& \multicolumn{2}{c}{(erg\,cm$^{-2}$s$^{-1}$)} & \multicolumn{2}{c}{(erg\,cm$^{-2}$s$^{-1}$)}  & \multicolumn{2}{c}{(erg\,cm$^{-2}$s$^{-1}$)} & \multicolumn{2}{c}{(erg\,cm$^{-2}$s$^{-1}$)}& \multicolumn{2}{c}{(erg\,cm$^{-2}$s$^{-1}$)} & \multicolumn{2}{c}{(erg\,cm$^{-2}$s$^{-1}$)}& \multicolumn{2}{c}{(erg\,cm$^{-2}$s$^{-1}$)}\\ 
\hline  																																							    
\noalign{\smallskip}																																						    
  SSTc2dJ160703.9-391112  &  164.5 &   4.6  & 9.11$\cdot 10^{5}$ &  1.80$\cdot 10^{5}$ &  3.50$\cdot 10^{5}$  &   9.95$\cdot 10^{4}$   &  4.21$\cdot 10^{4}$ & 1.56$\cdot 10^{4}$ &  2.84$\cdot 10^{4}$ &  1.69$\cdot 10^{4}$ & $\leq 1.33\cdot 10^{4}$ &  1.34$\cdot 10^{4}$ &  2.40$\cdot 10^{5}$  &  8.03$\cdot 10^{4}$  &    ...           &    ...             &     ...	           &	...	           \\  
  SSTc2dJ160708.6-391408  &  301.7 &  13.7  & 2.66$\cdot 10^{7}$ &  1.35$\cdot 10^{7}$ &  7.38$\cdot 10^{6}$  &   4.56$\cdot 10^{6}$   &  9.26$\cdot 10^{5}$ & 3.10$\cdot 10^{5}$ &  1.23$\cdot 10^{6}$ &  4.14$\cdot 10^{5}$ &  8.91$\cdot 10^{5}$ &  3.00$\cdot 10^{5}$ &  9.89$\cdot 10^{5}$  &  2.61$\cdot 10^{5}$  &  1.32$\cdot 10^{6}$  & 7.75$\cdot 10^{5}$ &  7.06$\cdot 10^{5}$  &   6.18$\cdot 10^{5}$  \\
\hline		
\end{tabular}	
\end{table}	

\normalsize

\end{landscape}

\setlength{\tabcolsep}{4pt}

\begin{table*}[htb]
\caption[]{Pa$\gamma$, Pa$\beta$, and Br$\gamma$ fluxes.}
\scriptsize
\begin{tabular}{lrrrrrr}
\hline
\hline
\noalign{\smallskip}
 Name	                  & $F_{\rm Pa\gamma}$ & err& $F_{\rm Pa\beta}$ & err & $F_{\rm Br\gamma}$ & err  \\  
                          & \multicolumn{2}{c}{(erg\,cm$^{-2}$s$^{-1}$)} & \multicolumn{2}{c}{(erg\,cm$^{-2}$s$^{-1}$)}  & \multicolumn{2}{c}{(erg\,cm$^{-2}$s$^{-1}$)}  \\  
\hline
\noalign{\smallskip}
  Sz66  		   & 	  1.22$\cdot 10^{6}$ &   2.36$\cdot 10^{5}$ &	1.56$\cdot 10^{6}$ &  2.63$\cdot 10^{5}$ &   1.79$\cdot 10^{5}$ &   7.45$\cdot 10^{4}$  \\ 
  AKC2006\,19		   &  $\leq 4.86\cdot 10^{4}$&   1.01$\cdot 10^{5}$ &	3.15$\cdot 10^{5}$ &  1.48$\cdot 10^{5}$ &   5.21$\cdot 10^{0}$ &   4.15$\cdot 10^{4}$  \\ 
  Sz69  		   & 	  8.06$\cdot 10^{5}$ &   1.32$\cdot 10^{5}$ &	1.54$\cdot 10^{6}$ &  2.36$\cdot 10^{5}$ &   1.29$\cdot 10^{5}$ &   3.49$\cdot 10^{4}$  \\ 
  Sz71  		   & 	  7.26$\cdot 10^{5}$ &   2.49$\cdot 10^{5}$ &	1.25$\cdot 10^{6}$ &  1.62$\cdot 10^{5}$ &   1.76$\cdot 10^{5}$ &   8.44$\cdot 10^{4}$  \\ 
  Sz72  		   & 	  3.13$\cdot 10^{6}$ &   3.68$\cdot 10^{5}$ &	4.77$\cdot 10^{6}$ &  4.72$\cdot 10^{5}$ &   7.44$\cdot 10^{5}$ &   1.38$\cdot 10^{5}$  \\ 
  Sz73  		   & 	  3.72$\cdot 10^{6}$ &   5.39$\cdot 10^{5}$ &	5.55$\cdot 10^{6}$ &  5.02$\cdot 10^{5}$ &   5.97$\cdot 10^{5}$ &   1.93$\cdot 10^{5}$  \\
\hline  																																							    
\end{tabular}																																							    
\label{Tab:NIR}																																						    
\end{table*}																																							    

\addtocounter{table}{-1}																																					    

\begin{table*}[htb]																																						    
\caption[]{{\it Continued.}}																																					     
\scriptsize																																							     
\begin{tabular}{lrrrrrr}
\hline  																																							     
\hline  																																							     
\noalign{\smallskip}																																						     
 Name	                   & $F_{\rm Pa\gamma}$ & err& $F_{\rm Pa\beta}$ & err & $F_{\rm Br\gamma}$ & err  \\  
                           & \multicolumn{2}{c}{(erg\,cm$^{-2}$s$^{-1}$)} & \multicolumn{2}{c}{(erg\,cm$^{-2}$s$^{-1}$)} & \multicolumn{2}{c}{(erg\,cm$^{-2}$s$^{-1}$)}   \\  
\hline 	     
\noalign{\smallskip} 																					  GQ~Lup             	   &       3.16$\cdot 10^{6}$ &   4.08$\cdot 10^{5}$ &   6.07$\cdot 10^{6}$    &  9.49$\cdot 10^{5}$ &   4.46$\cdot 10^{5}$    &   3.23$\cdot 10^{5}$    \\
  Sz76               	   &   $\leq 1.43\cdot 10^{5}$&   1.84$\cdot 10^{5}$ & $\leq 7.99\cdot 10^{4}$ &  2.45$\cdot 10^{5}$ &           ...~~~        &           ...~~~        \\
  Sz77               	   &       3.43$\cdot 10^{5}$ &   3.27$\cdot 10^{5}$ & $\leq 1.34\cdot 10^{5}$ &  4.10$\cdot 10^{5}$ & $\leq 2.10\cdot 10^{4}$ &   8.95$\cdot 10^{4}$    \\
  RXJ1556.1-3655     	   &       3.69$\cdot 10^{6}$ &   7.18$\cdot 10^{5}$ &   4.42$\cdot 10^{6}$    &  6.32$\cdot 10^{5}$ &   8.74$\cdot 10^{5}$    &   3.86$\cdot 10^{5}$    \\
  IM~Lup             	   &   $\leq 8.04\cdot 10^{4}$&   2.93$\cdot 10^{5}$ & $\leq 2.01\cdot 10^{5}$ &  6.05$\cdot 10^{5}$ & $\leq 1.09\cdot 10^{5}$ &   1.28$\cdot 10^{5}$    \\
  EX~Lup             	   &       1.57$\cdot 10^{6}$ &   2.54$\cdot 10^{5}$ &   1.37$\cdot 10^{6}$    &  2.16$\cdot 10^{5}$ &   1.86$\cdot 10^{5}$    &   1.00$\cdot 10^{5}$    \\
\hline					
\end{tabular}		
\end{table*}	

\normalsize

\color{black}

\newpage
\topmargin -1 cm
\twocolumn
\begin{appendix}

\section{The resolution of X-Shooter}
\label{sec:resolution}
\begin{figure}
\begin{center}
\parbox{18cm}{
\parbox{6cm}{
\includegraphics[width=7.5cm]{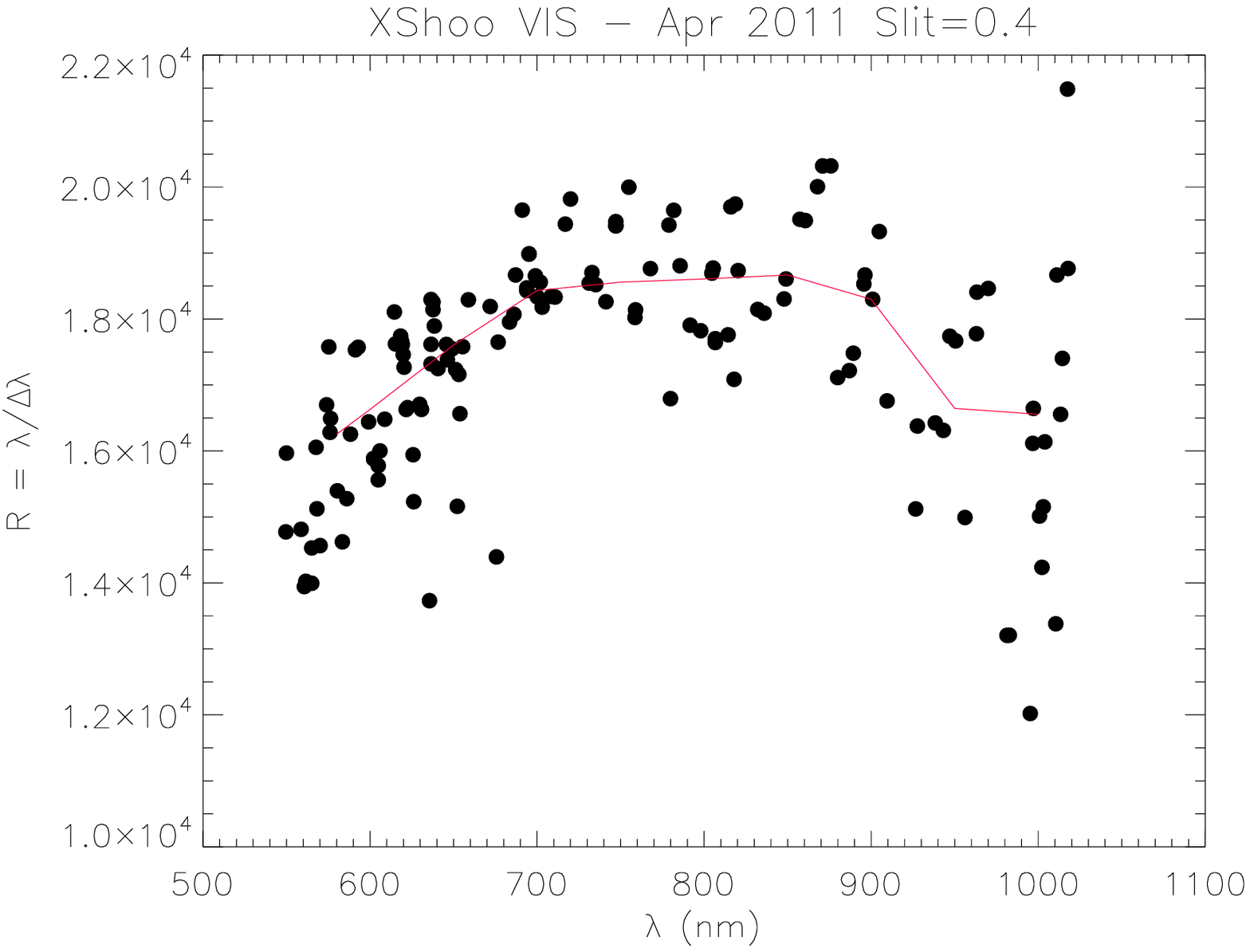}	
\includegraphics[width=7.5cm]{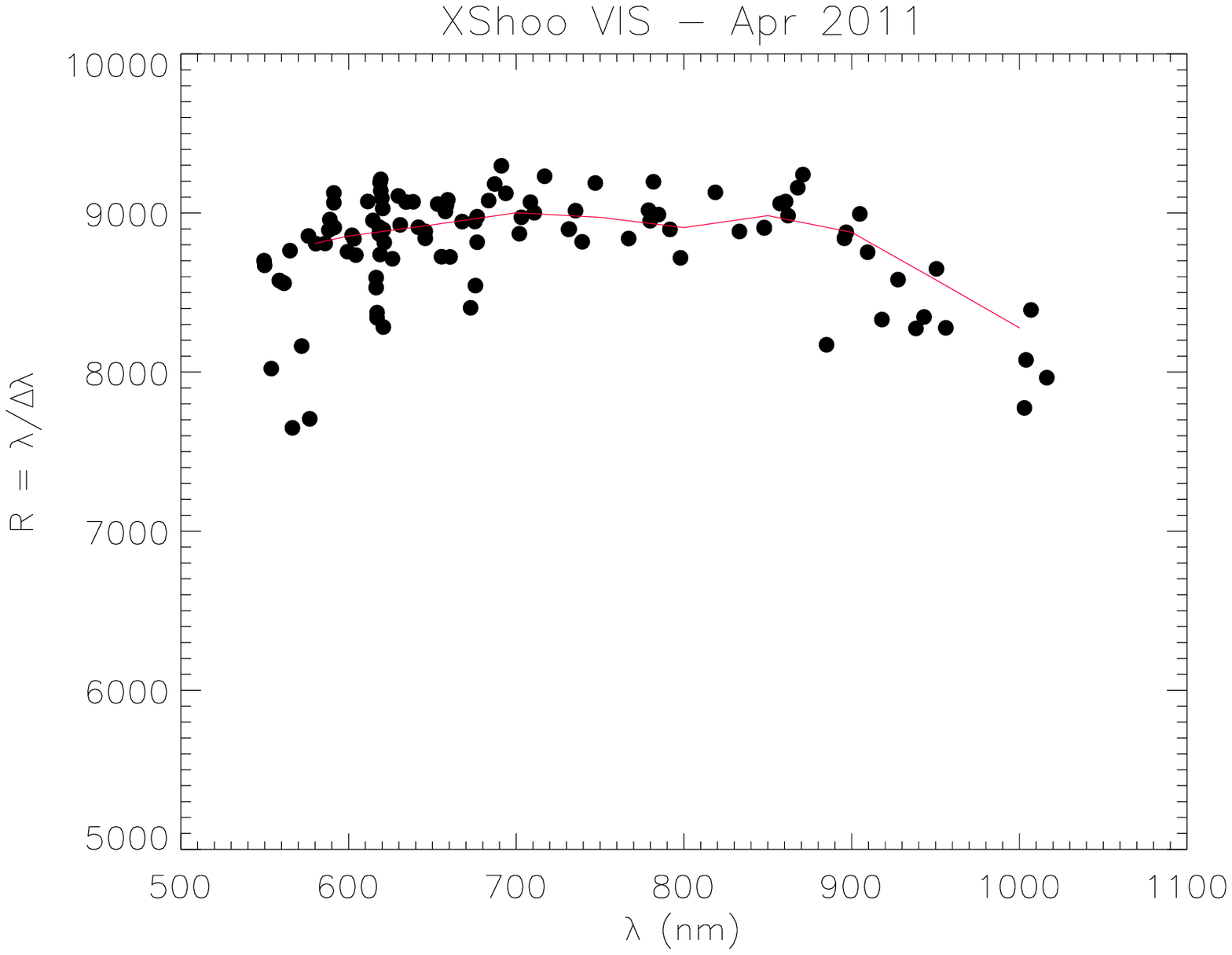}	
\includegraphics[width=7.5cm]{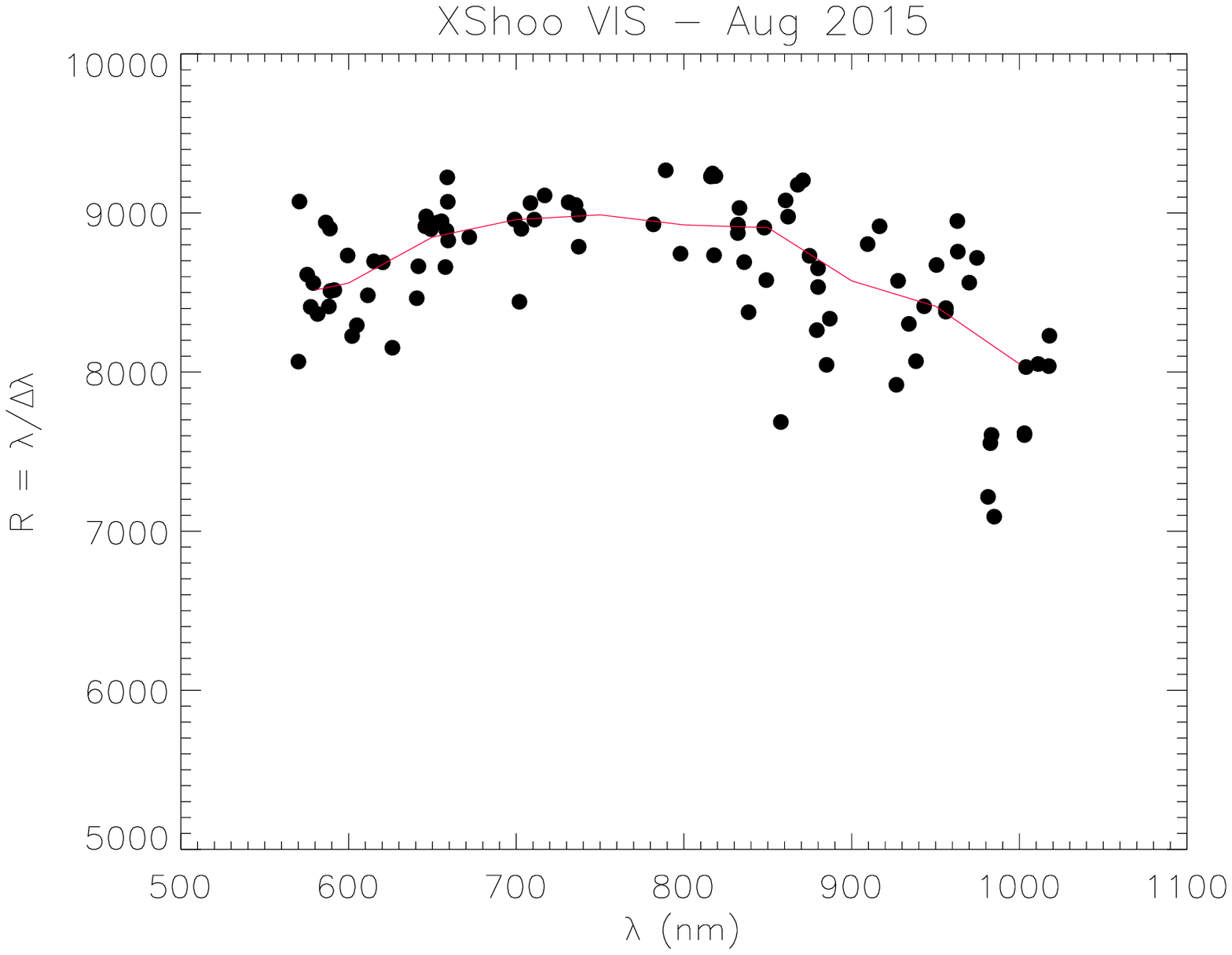}	
}
}
\caption{Resolution of X-Shooter VIS spectra with the 0$\farcs$4 slit ({\it upper panel}) and the 0$\farcs$9 slit ({\it middle and lower panels}) 
as evaluated from the FWHM of Th-Ar emission lines (dots). In each box, the continuous line is the running median with a bin size of 50 nm.
}
\label{Fig:resolution}
\end{center}
\end{figure}

To measure the resolution of X-Shooter in the VIS arm, we reduced some calibration-lamp spectra following the same steps as 
for the science frames. We extracted the spectra from the pre-reduced images summing up the same columns of 
the detector where the stellar spectra fall in typical seeing conditions.
We chose several tens of unblended emission lines that were fitted with Gaussians to obtain the line center $\lambda_{i}$ and 
the full width at half maximum (FWHM), $\Delta\lambda_{i}$.
The resolution $R_{i}=\lambda_{i}/\Delta\lambda_{i}$ is plotted as a function of wavelength in Fig.~\ref{Fig:resolution}
for VIS spectra acquired with the 0$\farcs$9 and 0$\farcs$4 slits.

We note that the resolution  with the 0$\farcs$9 slit, $R\simeq 8800$, is rather constant with $\lambda$ and decreases 
only slightly towards the edges of the spectrum. Moreover, it is basically unchanged between 2011 and 2015. 
For our analysis we adopted the value $R= 8400$, which is that of the region at $\lambda\simeq$\,9700\,\AA\  used
for the \vsini\  determination.

The resolution with the  0$\farcs$4 slit displays instead a higher variation with $\lambda$ ranging from $R\simeq 18\,500$
at the spectrum center to $R\simeq 16\,000$ near the edges.  We adopted the value $R=17\,000$ in our analysis.

\section{Application of ROTFIT to the X-Shooter spectra}
\label{sec:ROTFIT}

The template BT-Settl model spectra were first degraded in resolution to match that of the target spectrum and then resampled on its spectral 
points. Template and target spectra were then normalized to the local continuum or 
pseudo-continuum, paying attention to the spectral regions containing molecular bands, where the continuum level 
at wavelengths shorter than the band head was used as reference.

The code calculates the $\chi^2$ of the difference between observed and template
spectra, where the rotational broadening was considered by convolving template spectra with a rotational profile\footnote{ The rotational 
broadening kernel \citep[see, e.g.,][]{Gray2005} assumes a linear limb-darkened star, with the limb-darkening coefficients from 
\citet{Claretta}.} of increasing $v\sin i$ from 0 to 150\,\kms\  in steps of 1\,\kms. 

\begin{figure}
\begin{center}
\parbox{18cm}{
\parbox{6cm}{
\includegraphics[width=9cm]{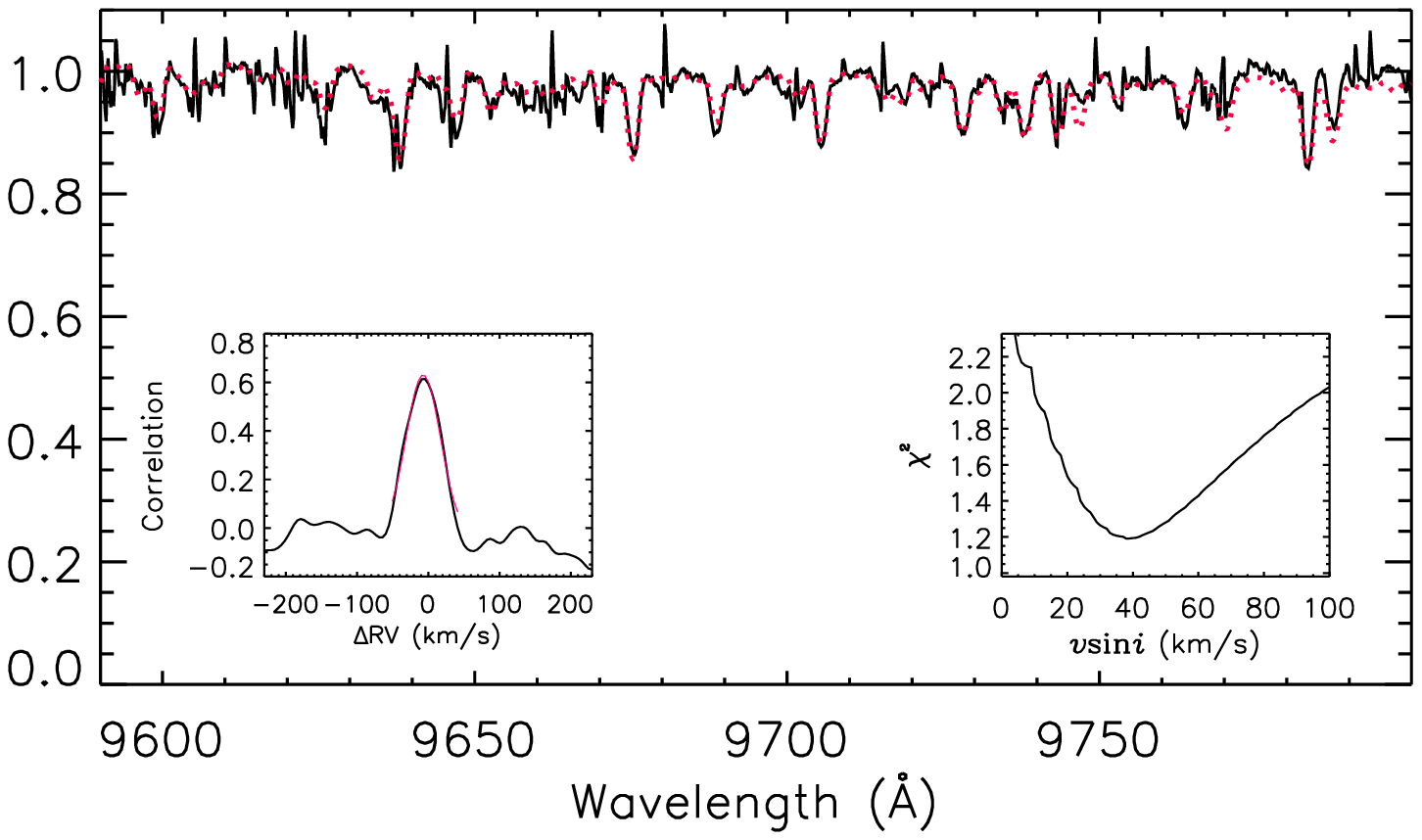}	
\includegraphics[width=9cm]{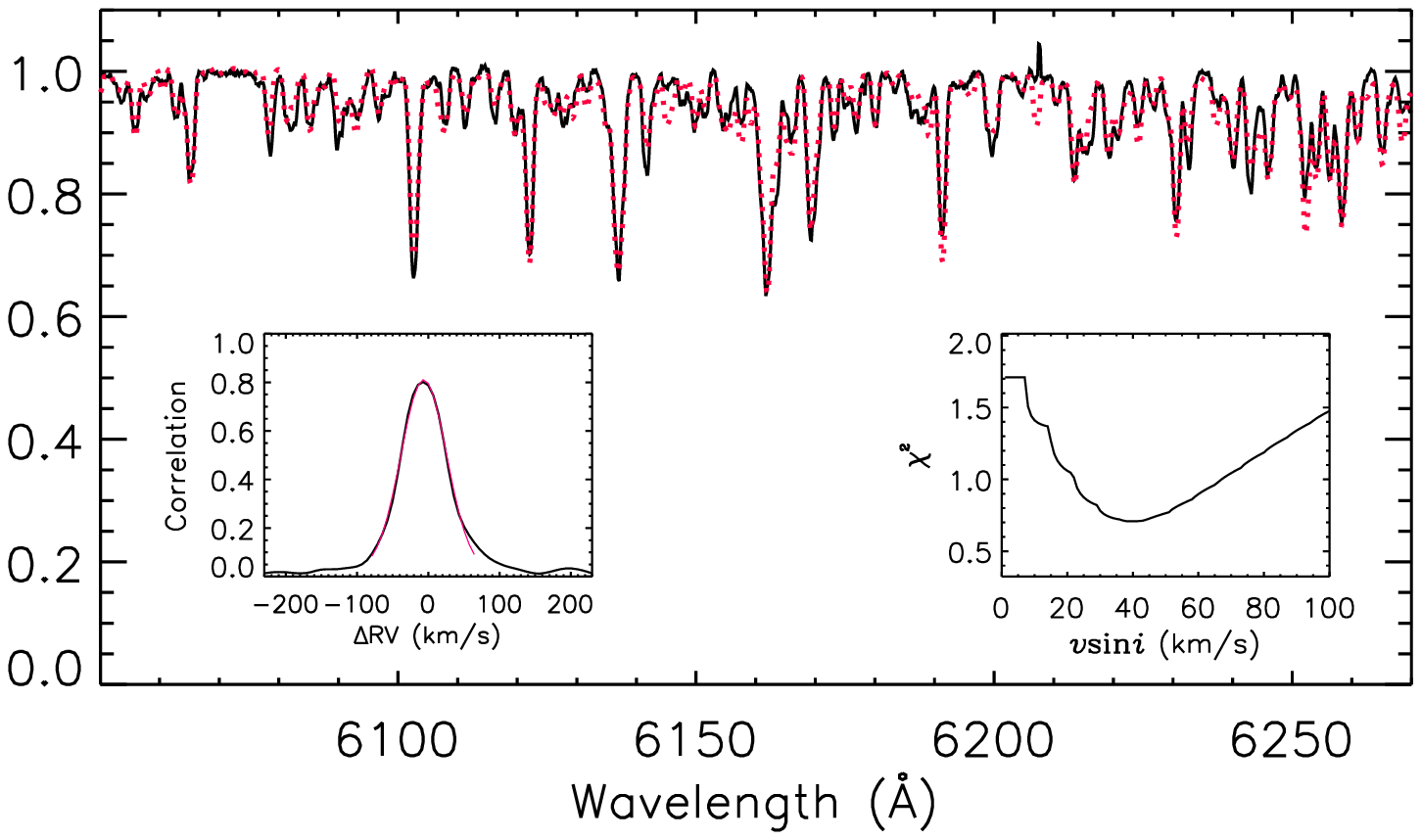}	
}
}
\caption{Continuum-normalized VIS X-Shooter spectrum of Sz\,68 in the two regions selected for
the measurement of \vsini\  (black full lines) along with the best fitting template (red dotted lines). 
The insets in each panel show the cross-correlation function (left boxes) with the Gaussian fit 
overplotted in red and the $\chi^2$ as a function of \vsini\  (right boxes).}
\label{fig:ROTFIT_vsini}
\end{center}
\end{figure}

\setlength{\tabcolsep}{4pt}

\begin{table}[htb]
\caption{Spectral regions analyzed with ROTFIT.}
\begin{tabular}{cllllll}
\hline
\hline
\noalign{\smallskip}
Range      &  Main features  & \teff & \logg & $r$ & \vsini &  RV  \\   
\noalign{\smallskip}
\hline
\noalign{\smallskip}
 9590--9800  & \ion{Ti}{i}, \ion{Cr}{i}      & Y$^{\rm a}$ & N  	 & Y	       & Y	     & Y  \\  
 8160--8222  & \ion{Na}{i}                   & Y	   & Y  	 & Y	       & N	     & Y  \\  
 7600--7720  & \ion{K}{i}                    & Y	   & Y  	 & Y	       & N	     & Y  \\  
 7020--7120  & TiO                           & Y	   & N           & Y	       & N	     & Y  \\  
 6050--6270  & \ion{Ca}{i}, \ion{Fe}{i}, TiO & Y	   & Y           & Y	       & Y$^{\rm b}$ & Y  \\  
 5320--5500  & \ion{Fe}{i}                   & Y$^{\rm b}$ & Y$^{\rm b}$ & Y$^{\rm b}$ & N           & N  \\  
 5120--5220  & \ion{Mg}{i}\,b                & Y$^{\rm b}$ & Y$^{\rm b}$ & Y$^{\rm b}$ & N           & N  \\  
 4400--4580  & \ion{Fe}{i}                   & Y$^{\rm b}$ & Y$^{\rm b}$ & Y$^{\rm b}$ & N           & N  \\  
\hline
\end{tabular}
\label{Tab:SpecRegions}
{\scriptsize
\begin{list}{}{}									
\item[$^{\mathrm{a}}$]~Low sensitivity to this parameter. 
\item[$^{\mathrm{b}}$]~For the best exposed spectra of stars hotter than about 3500\,K. 
\end{list}
}
\end{table}

We selected a few spectral regions that are particularly suitable for the determination of one or more stellar parameters 
(Table~\ref{Tab:SpecRegions}). One of the main requirements is that these regions should be free, 
as much as possible, of strong emission lines produced by accretion or chromospheric activity. 
We mainly used the VIS spectra, but some portions of the UVB ones were used as well 
for the objects with spectra  well exposed; these are mostly  K-type stars and all earlier than M4.
In each spectral interval, the sensitivity to a given parameter normally changes with the characteristics of the target
and mainly depends on the SpT (i.e. \teff). Therefore, we have chosen spectral regions that are usable in the 
largest possible range of SpT. 

\subsection{\vsini}
 The best intervals to perform an accurate determination of $v\sin i$ with ROTFIT are those 
devoid of very broad lines and strong molecular bands, but encompassing several strong absorption lines.
A suitable region for the GK-type stars ranges from 6050\,\AA\  to 6270\,\AA, where some \ion{Ca}{i} lines and a 
number of unblended lines, mostly of iron-group elements are present. These lines are also good diagnostics of temperature and gravity. 
However, a TiO molecular band starts to be visible for \teff$\leq$\,4000\,K and strongly affects the spectra of stars 
cooler than about 3500\,K, for which reason we did not use that part of the spectrum for the determination of \vsini\  for cool objects, 
but it is still useful to measure \teff\ in these cases.
Another wavelength range suitable to measure \vsini\ with X-Shooter VIS spectra is between 9600 and 9800 \AA\footnote{The 
final value of $v\sin i$, which is reported in Table~\ref{Tab:param}, is that one derived from the redder of the two 
spectral regions or the weighted average of the values derived in the two regions at 9700\,\AA\  and 6200\,\AA\  for 
the stars hotter than about 3500\,K.}. It includes several sharp absorption lines of \ion{Ti}{i} and \ion{Cr}{i} that are 
strong enough down to \teff\,$\simeq$\,2900\,K. 
Moreover, the intensity of these lines changes mildly with the SpT, thus they can also be used as temperature 
indicators, although their sensitivity to \teff\  is not as high as that of other lines or molecular bands.
 
The results of the ROTFIT analysis for a rather rapid rotator (Sz\,68) are illustrated in Fig.~\ref{fig:ROTFIT_vsini}.
 The upper and lower panels show the observed spectrum around 9700\,\AA\  and 6150\,\AA, respectively. The best fitting template, 
i.e. the synthetic BT-Settl spectrum rotationally broadened, resampled and wavelength-shifted on the target spectrum that produces 
the minimum $\chi^2$, is overplotted with a red dashed line in both panels. The CCF and the $\chi^2$ as a function of \vsini\ 
are also displayed in the insets.

\subsection{\teff\ and \logg}
For the determination of the atmospheric parameters, the \vsini\  was kept fixed to the value derived through the analysis of 
the 9700-\AA\ and 6150-\AA\ spectral regions.
To determine the gravity we selected two spectral regions containing the \ion{Na}{i} doublet at $\lambda\approx 8190$\,\AA\  
and the \ion{K}{i} doublet at $\lambda\approx 7660-7700$\,\AA, whose wings are very 
sensitive to the electronic pressure and therefore to \logg.
In addition, we analyzed the spectral segment from 7020 to 7150\,\AA\, which  contains three TiO molecular bands 
strongly sensitive to \teff\  for cool stars and substellar objects.

\begin{figure*}
\begin{center}
\vspace{-.5cm}
\includegraphics[width=8.5cm,height=4.4cm]{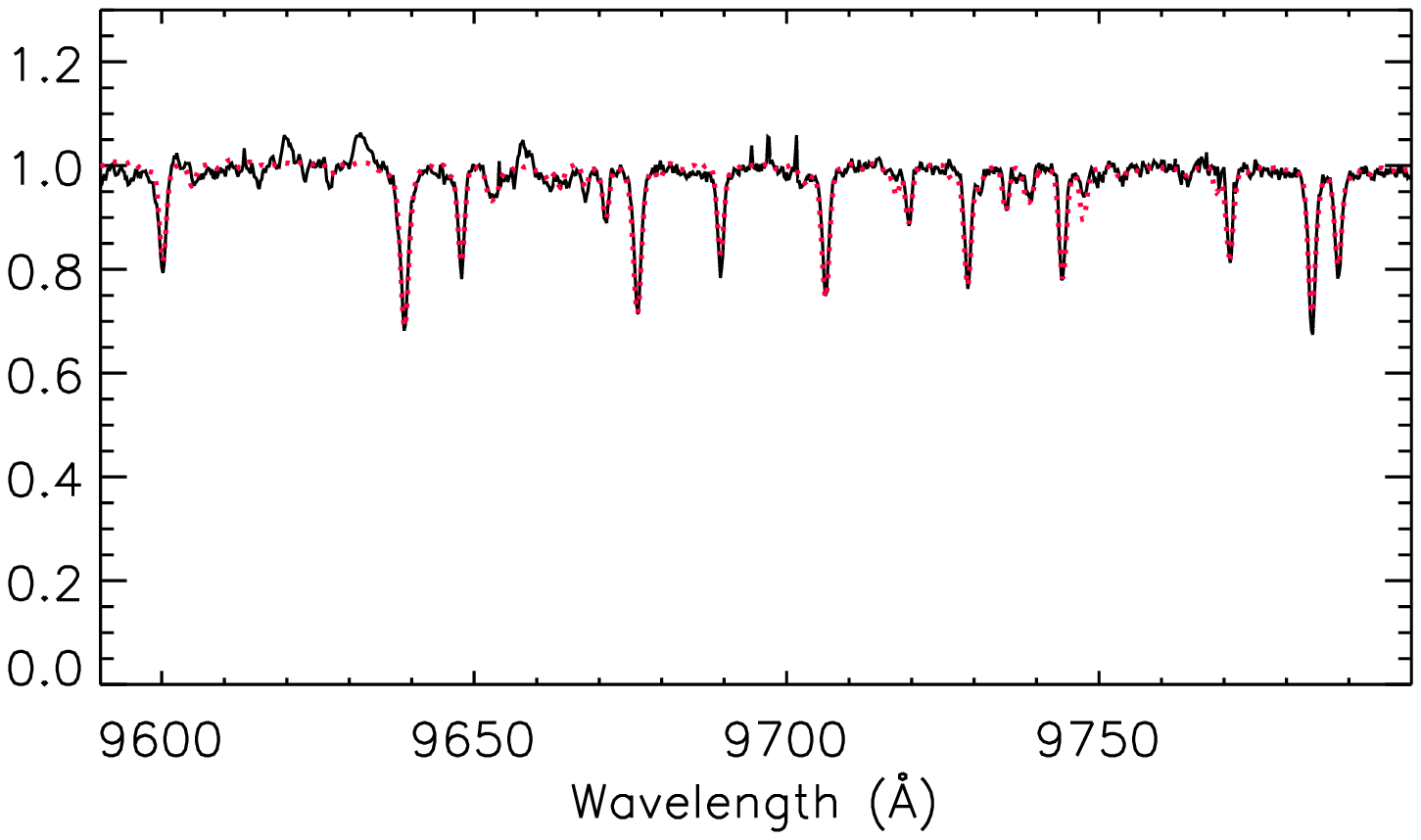}	
\vspace{-.5cm}
\includegraphics[width=7.2cm,height=4.4cm]{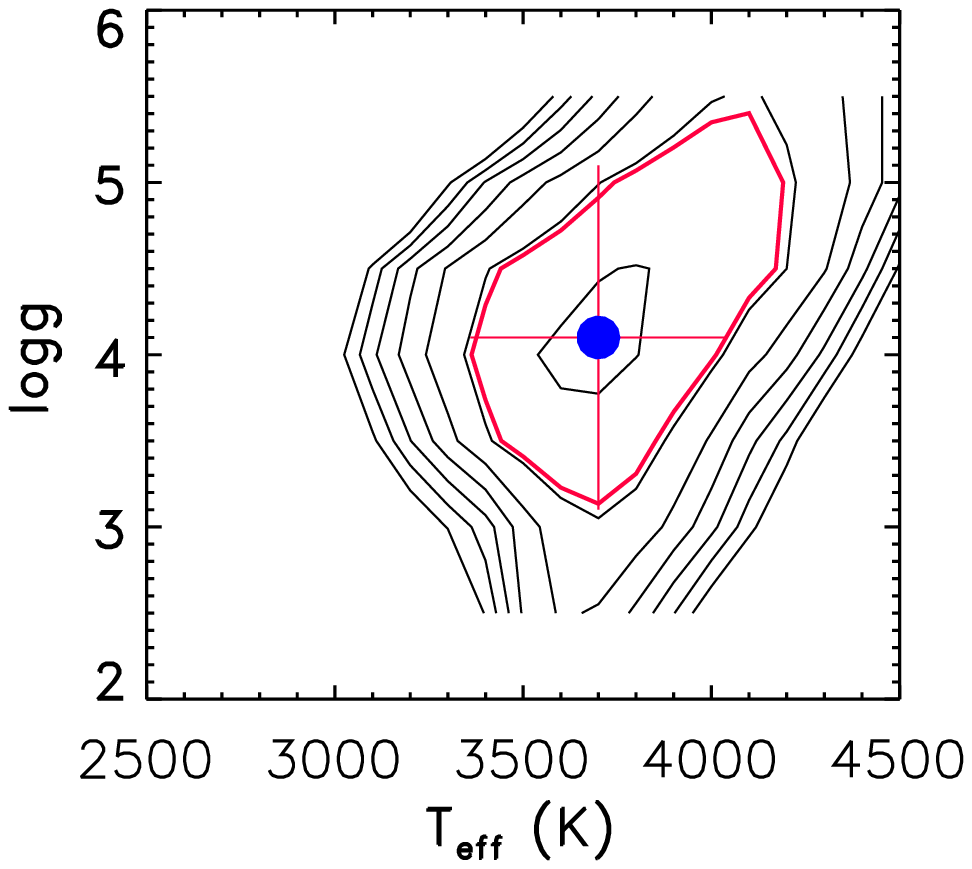}	
\vspace{-.2cm}
\includegraphics[width=8.5cm,height=4.4cm]{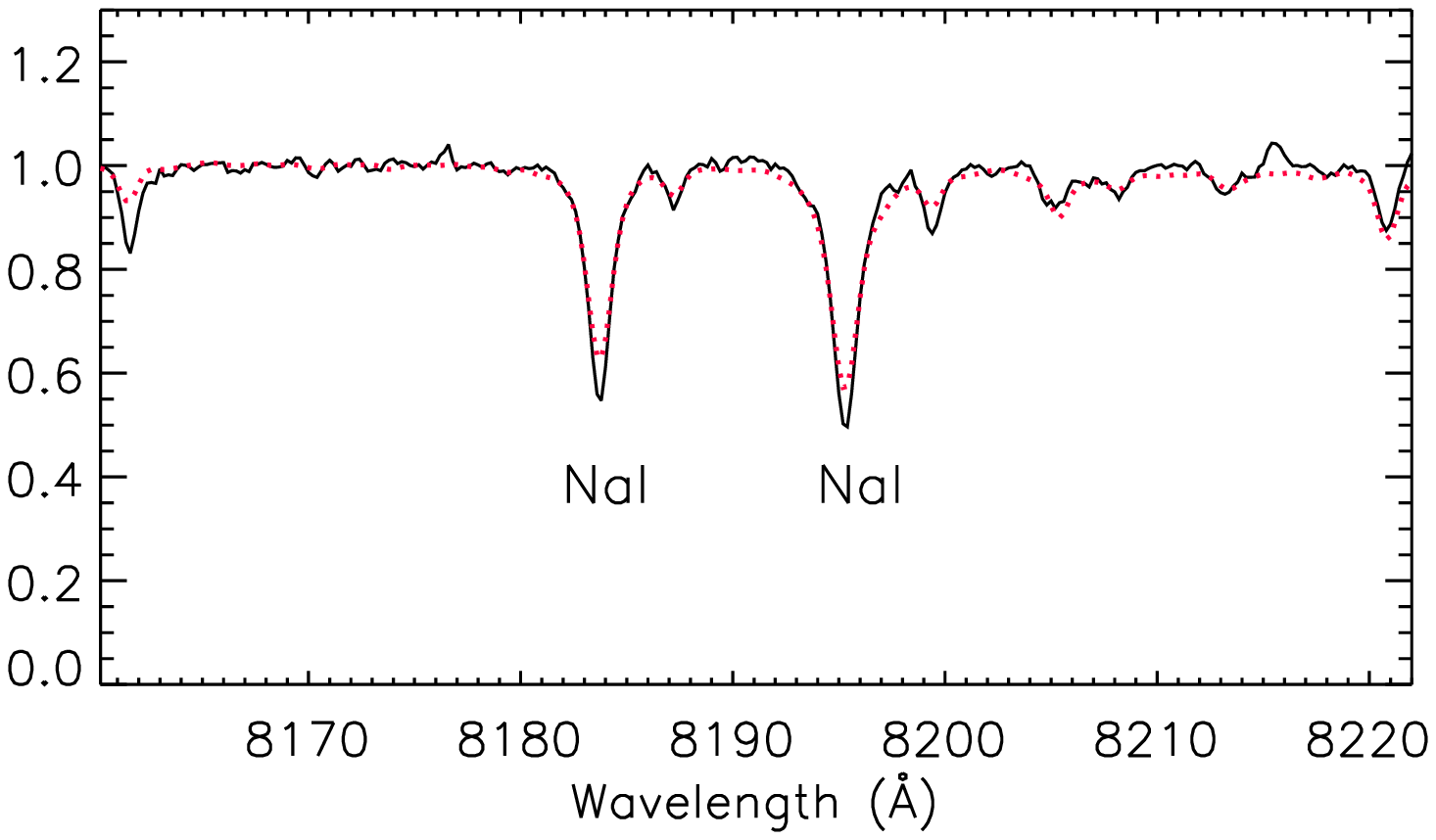}	
\vspace{-.2cm}
\includegraphics[width=7.2cm,height=4.4cm]{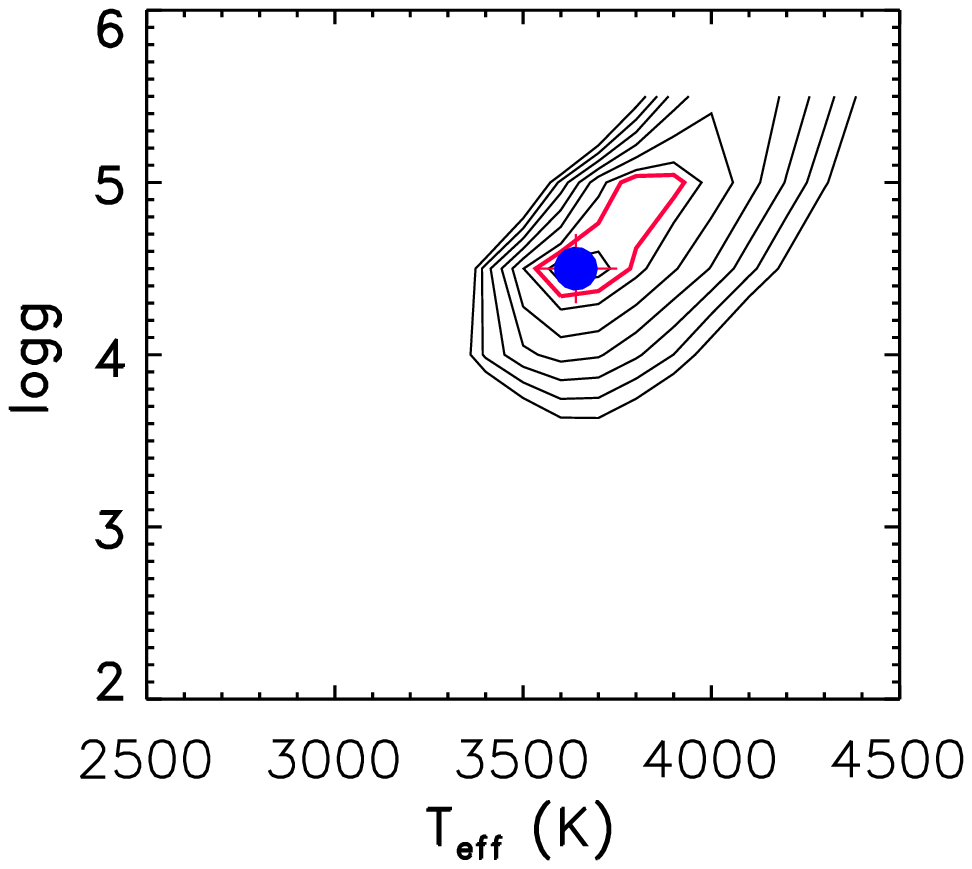}	
\vspace{-.2cm}
\includegraphics[width=8.5cm,height=4.4cm]{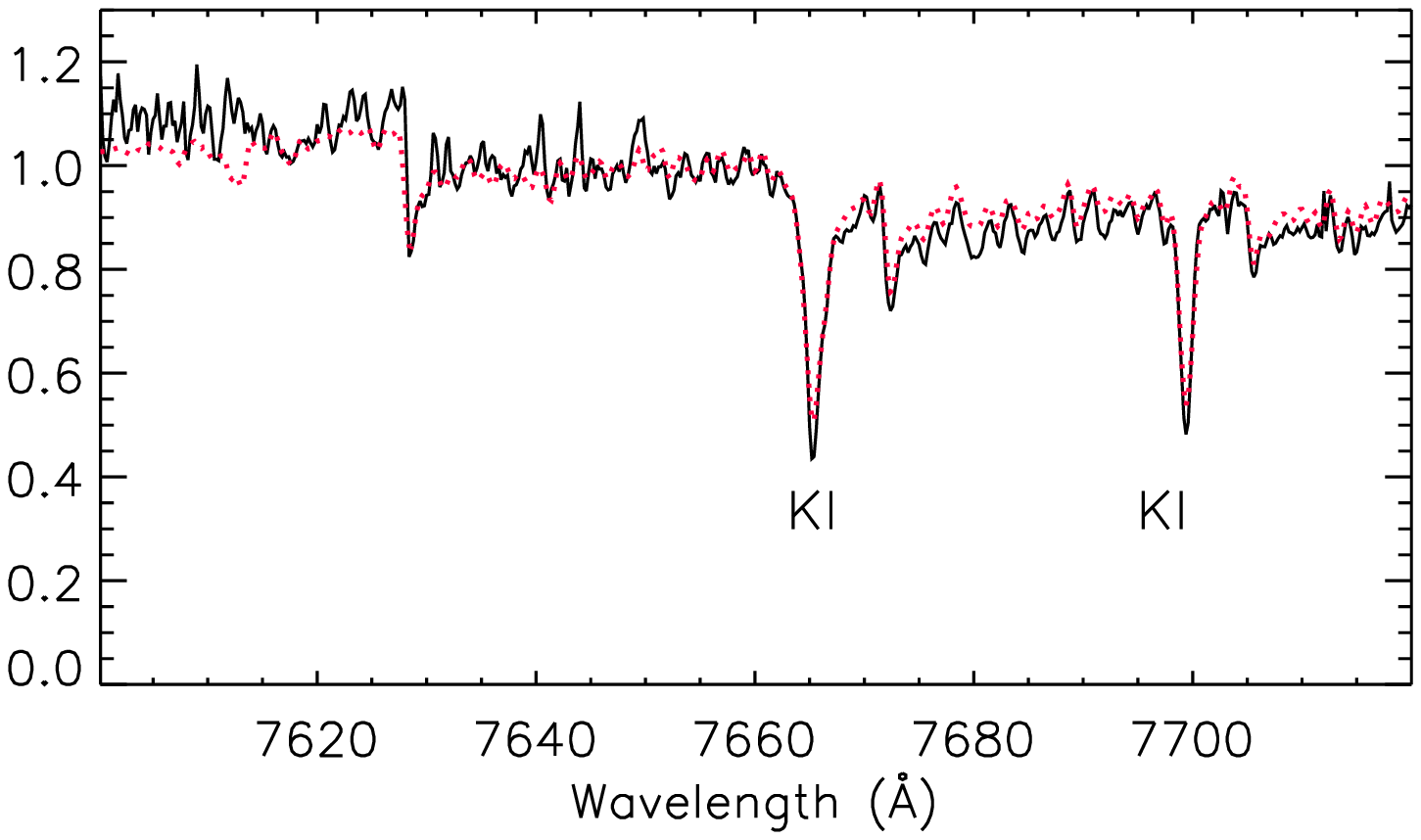}	
\vspace{-.2cm}
\includegraphics[width=7.2cm,height=4.4cm]{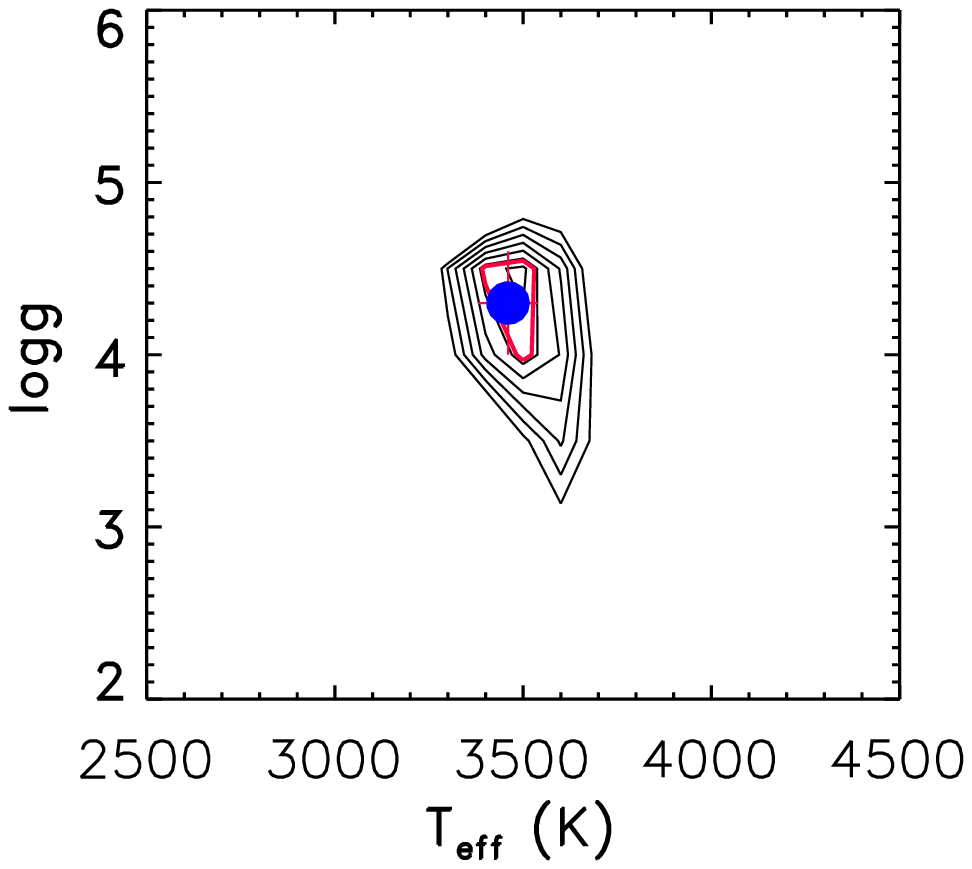}	
\vspace{-.2cm}
\includegraphics[width=8.5cm,height=4.4cm]{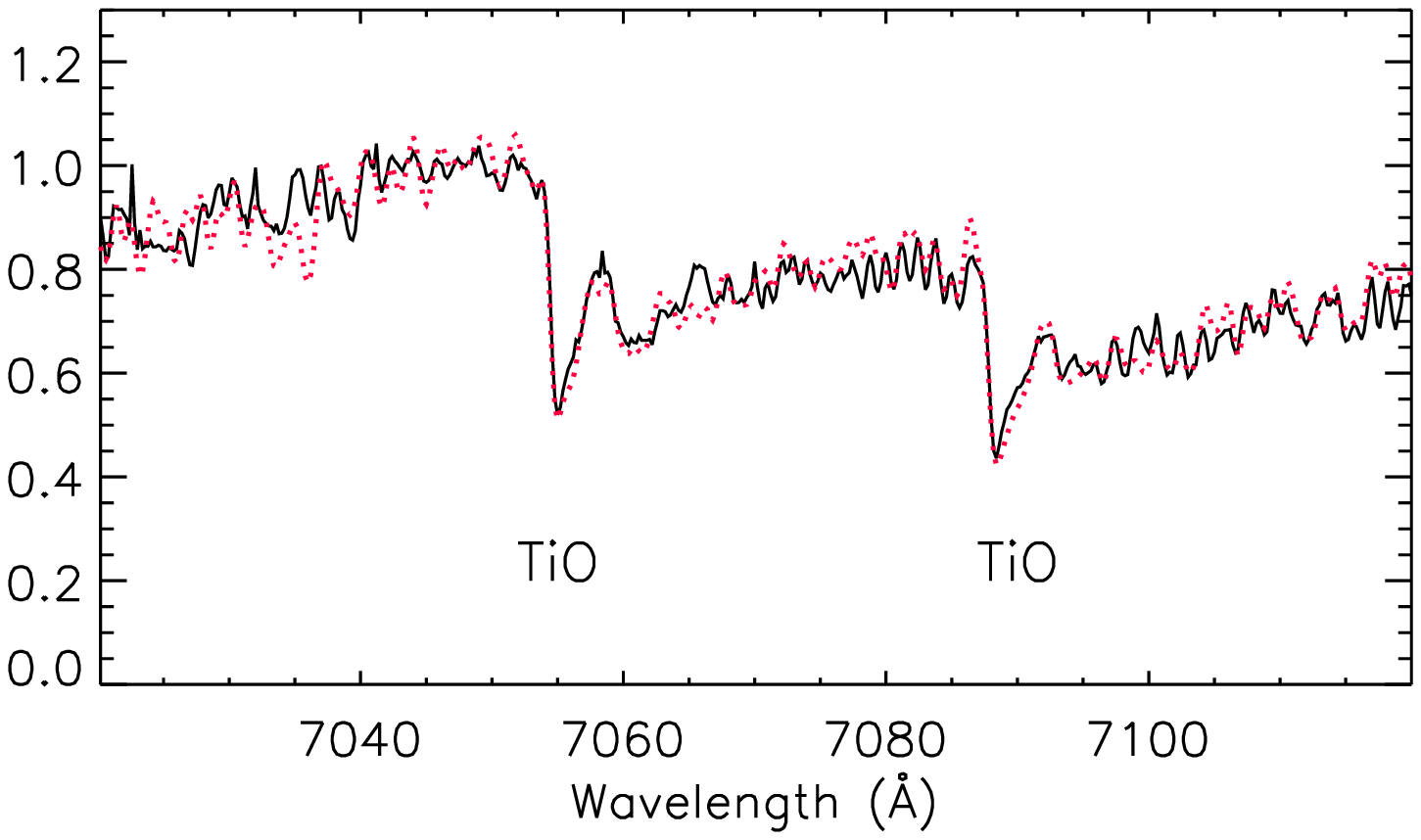}	
\vspace{-.2cm}
\includegraphics[width=7.2cm,height=4.4cm]{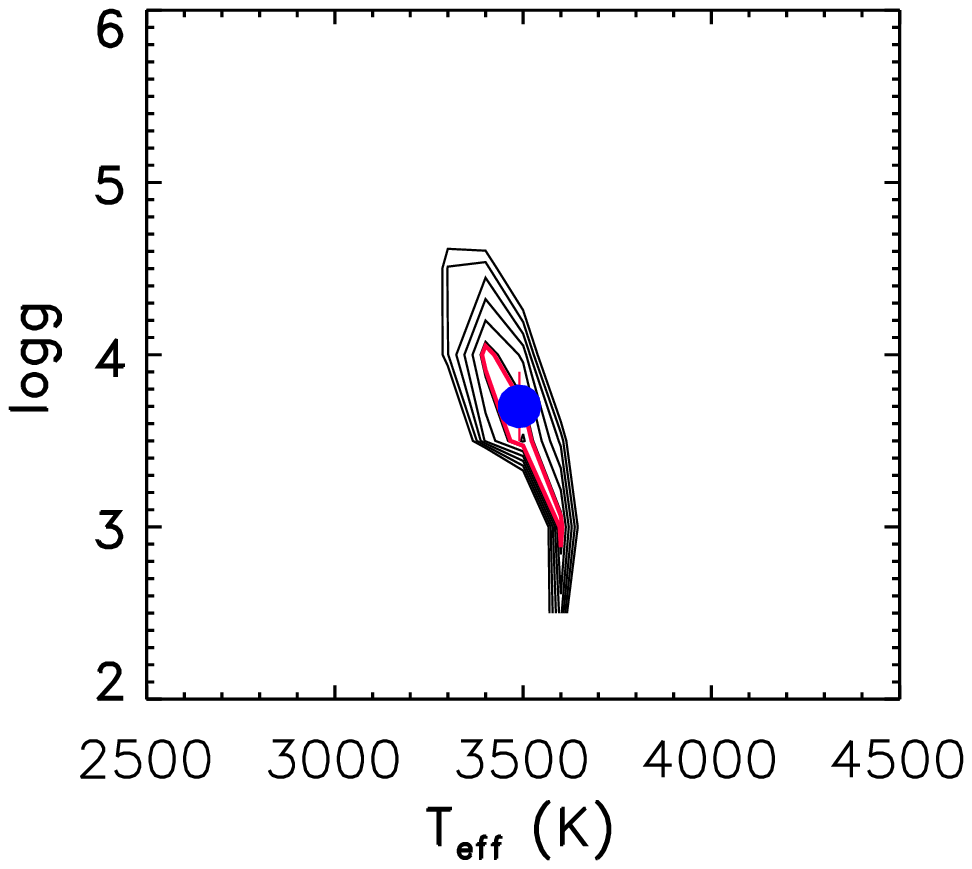}	
\vspace{-.2cm}
\includegraphics[width=8.5cm,height=4.4cm]{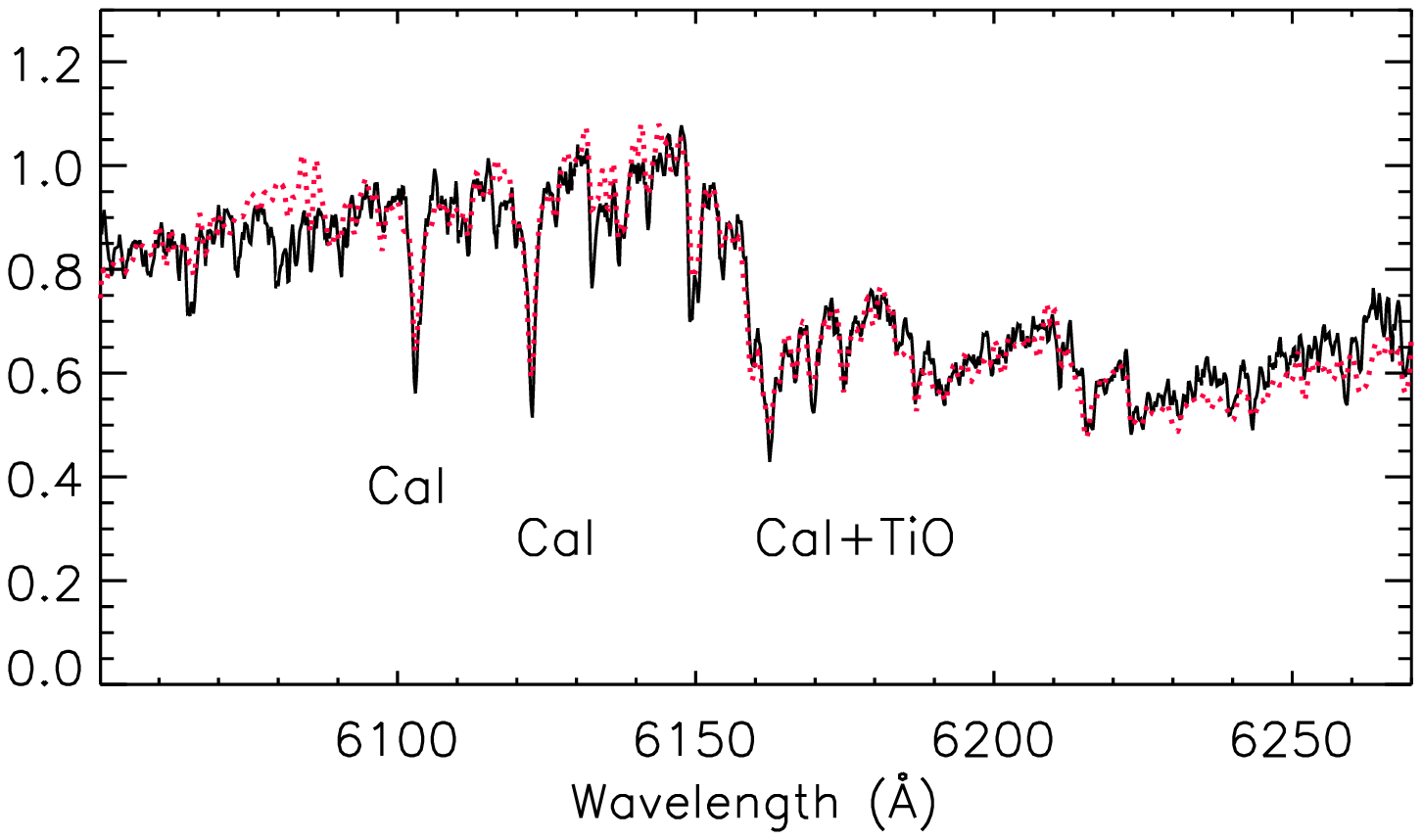}	
\includegraphics[width=7.2cm,height=4.4cm]{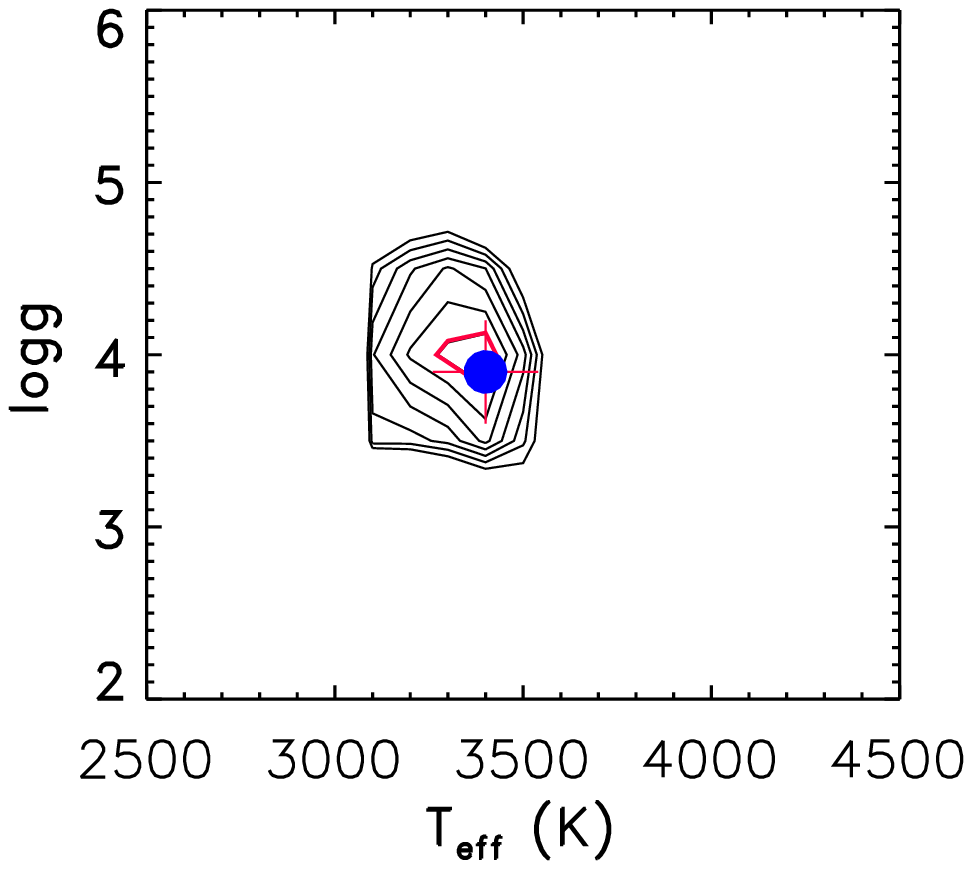}	
\caption{{\it Left panels)} Continuum-normalized VIS X-Shooter spectrum of Sz\,117 in the five regions analyzed 
with ROTFIT (black full lines) with  overplotted the best fitting template (red dotted lines). {\it Right panels)} $\chi^2$ contour maps
in the \teff--\logg\  plane. In each panel, the blue dot marks the best values of \teff\  and \logg\  while the 1$\sigma$ 
confidence level is denoted by the thick red contour. The errorbars on \teff\  and \logg\  are also marked.  }
\label{fig:ROTFIT}
\end{center}
\end{figure*}

Figure~\ref{fig:ROTFIT} displays an example of the results of the ROTFIT analysis for an M star (Sz\,117).
The left panels show the observed spectrum and the best fitting template,  similarly to Fig.~\ref{fig:ROTFIT_vsini}.
The contour maps of the $\chi^2$ for each spectral region are shown in the right panels, where the 1$\sigma$ confidence 
level is marked with a thick red line. 
For a better determination of the \teff\  and \logg\  values we have performed 
a quadratic interpolation onto  a grid finer than the one of the models, with 10-K and 0.1-dex steps for \teff\  and \logg, respectively.

\begin{figure}
\begin{center}
\includegraphics[width=9cm]{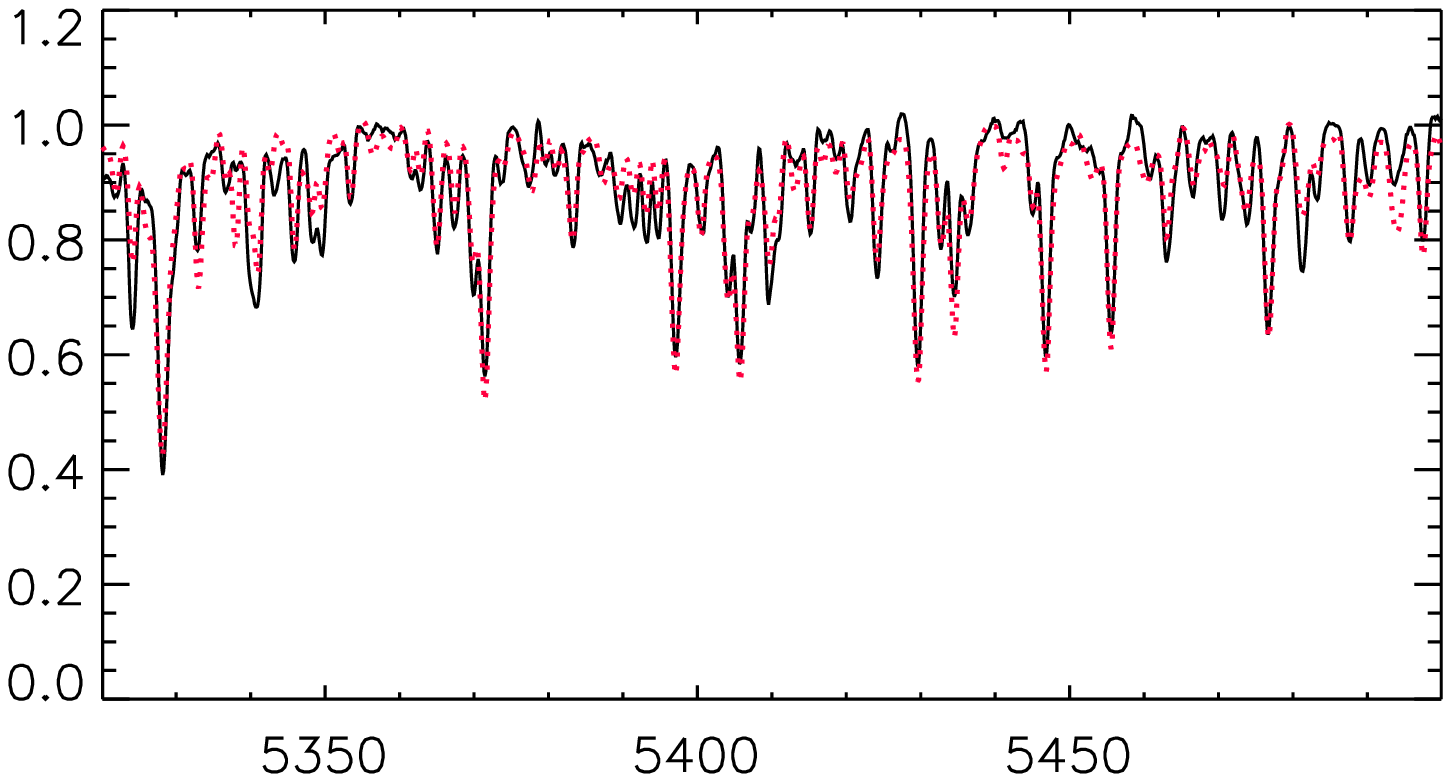}	  
\includegraphics[width=9cm]{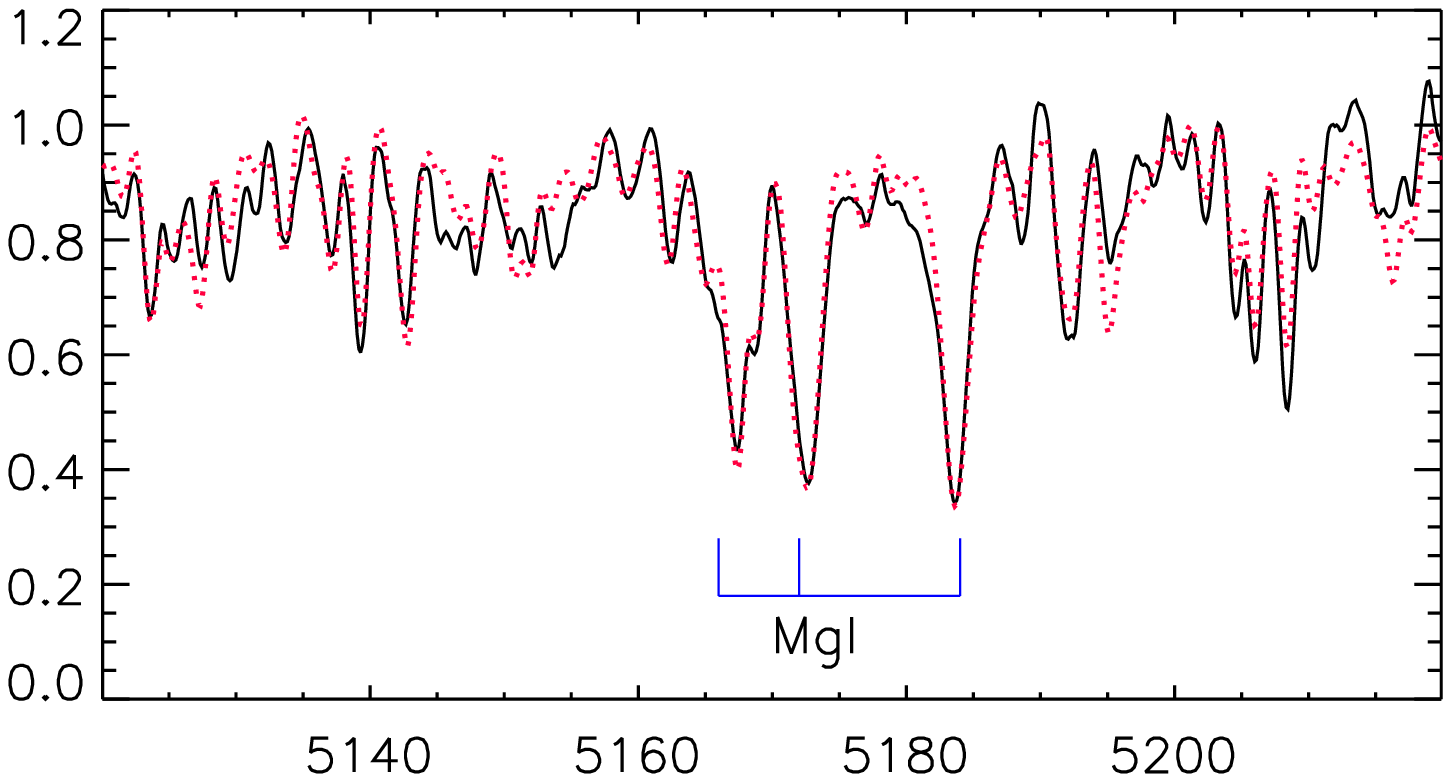}	  
\includegraphics[width=9cm]{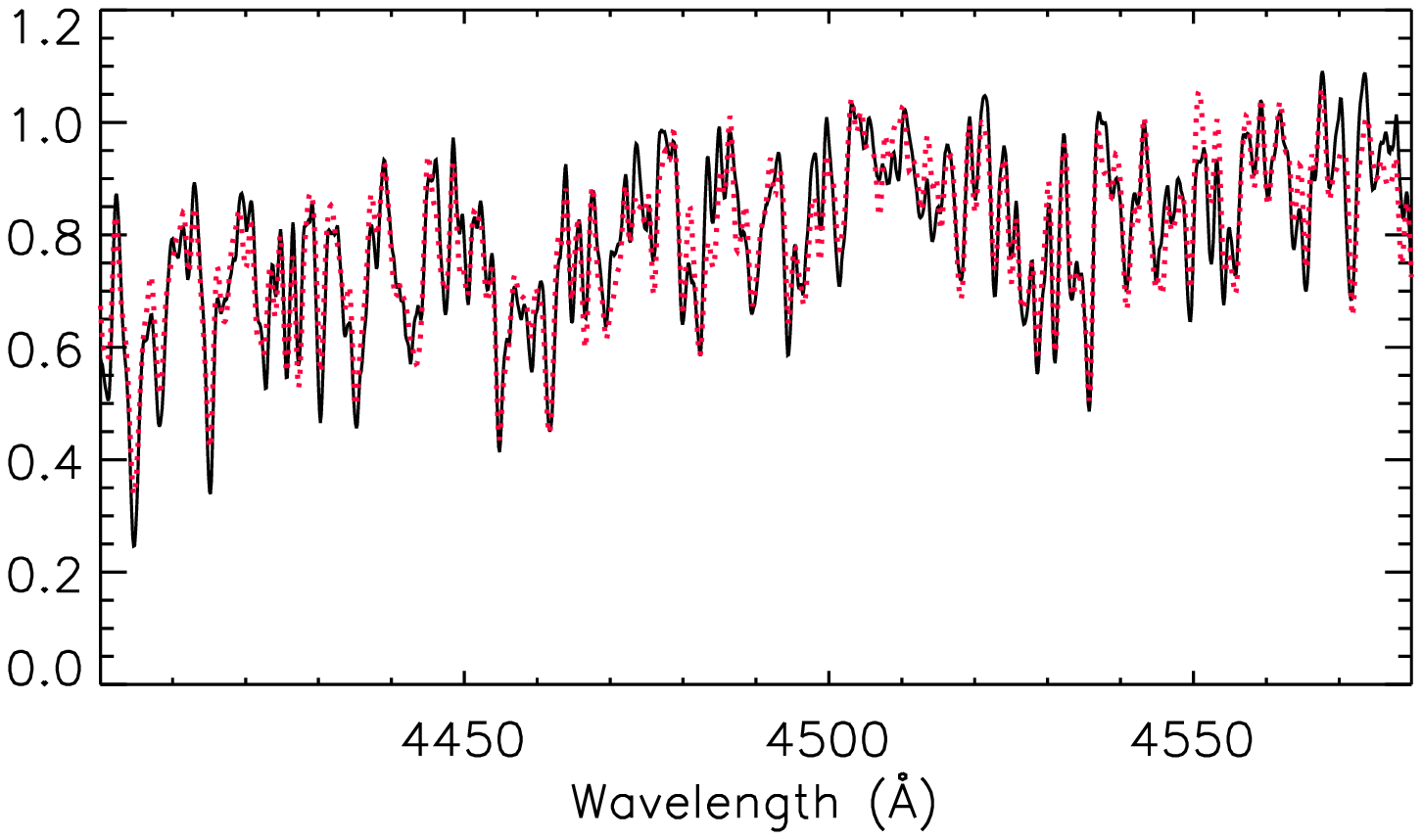}	  
\caption{UVB X-Shooter spectrum of Sz\,68  (black lines) in the three regions that
have been analyzed with ROTFIT. In each box, the best fitting template is
displayed superimposed with dotted red lines. }
\label{fig:ROTFIT_UVB}
\end{center}
\end{figure}

As mentioned above, for the hotter and best exposed sources we have also analyzed three spectral segments in the UVB 
spectra that cover the 5300--5500\,\AA, 5100--5250\,\AA, and 4400--4600\,\AA\  ranges. The spectral features they contain 
carry information about \teff\  and \logg\  and allow to evaluate the veiling at these shorter wavelengths. 
As an example, we show the fit of the UVB spectrum of Sz\,68 (K2) in Fig~\ref{fig:ROTFIT_UVB}. 
 However, given the lower resolution of the UVB spectra, we did not use the latter to measure \vsini\ and RV.

The adopted values of $\log g$ and \teff\ are the weighted averages of the results for each spectral segment according to
Table~\ref{Tab:SpecRegions}. These values are reported in Table~\ref{Tab:param} along with their errors and the veiling at five wavelengths.

For SSTc2dJ160034.4-422540 we failed to derive reliable atmospheric parameters because of the very low S/N of its spectrum.
The VIS spectrum of this object does not display emission lines, which makes it unlikely to be a YSO.
The reddest part of the VIS spectrum instead presents a number of molecular absorption bands which are indicative of a very low \teff.	
We have thus performed a customized analysis of the whole spectral range from  8500 to 11\,000\,\AA, which provided us with 
an estimate for \teff\  and a value of RV. The spectral features present in this region do not provide solid constraints for
\logg, although this object has likely a low gravity. More details and a plot of its VIS spectrum can be found
in Appendix\,\ref{sec:peculiar}.

\subsection{Veiling}

The spectra of Class~II sources, which typically undergo mass accretion from the circumstellar disk, can be affected by a
different amount of veiling,  which normally decreases at longer wavelengths, but it can still be noticeable in the 
IYJ bands \citep[e.g.,][]{Fischer2011}.
 For this reason, we also left the veiling free to vary. 
This was accomplished by adding to the templates, for each iteration, a wavelength-independent veiling in each spectral segment.
This assumption is justified by the limited spectral ranges that we analyzed, which are, at most, 200\,\AA\  wide. 
Therefore, the ``veiled'' template is given by:
\begin{equation}
\label{eq:veiling}
\biggl(\frac{F_\lambda}{F_C}\biggr)_{r}=\frac{\frac{F_\lambda}{F_C} + r}{1 + r}\,,
\end{equation}
where $F_\lambda$ and $F_C$ represent the line and continuum fluxes, respectively, and the veiling $r$ was let free to vary from 0.0 to 5.0 in 0.1 steps.

\begin{figure}
\begin{center}
\includegraphics[width=8.5cm,height=15cm]{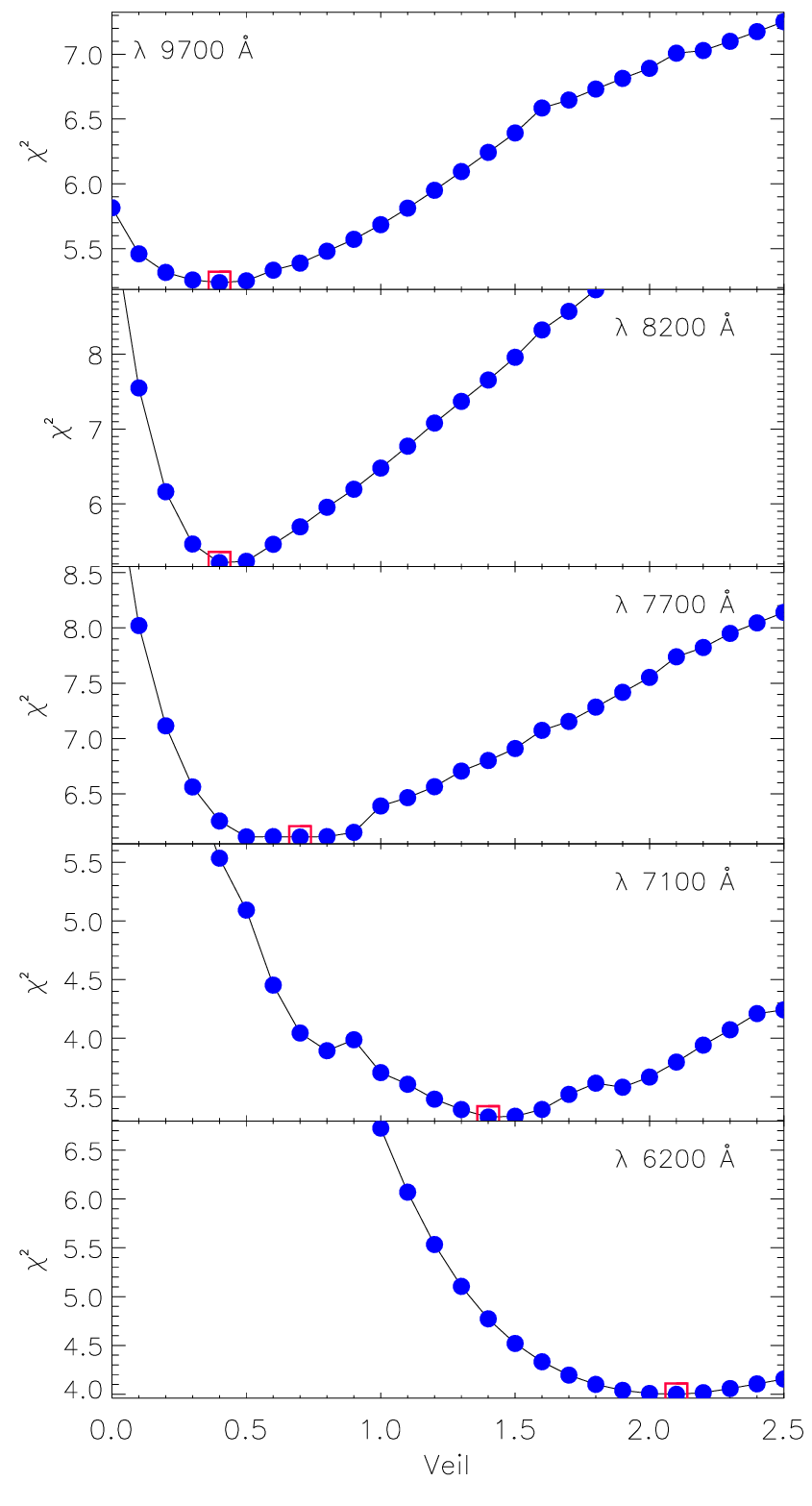}	
\caption{$\chi^2$ values as a function of veiling for Sz\,88A in different spectral regions. We note the strong increase 
of the veiling at the shorter wavelengths. }
\label{fig:veil}
\end{center}
\end{figure}

In cases of highly veiled spectra we often noticed the enhancement of $r$ with decreasing wavelength (see Fig.~\ref{fig:veil} for an example). 
This behavior is normally observed in strongly accreting objects \citep[e.g.,][]{Hartigan1991,Fischer2011} and is
interpreted as optically thick emission from the heated photosphere below the accretion shock \citep[][]{Calvet1998}. 
 The decrease of the veiling at longer wavelengths is explained by the slow decline of the continuum accretion flux, which can be be
approximated with a blackbody of $T\sim$\,10,000\,K, and the steep rise of the photospheric flux from visible to near-IR wavelengths 
for late-type stars.

\section{Monte Carlo simulations for \vsini}
\label{sec:MonteCarlo}

\begin{figure}[th]
\includegraphics[width=8cm]{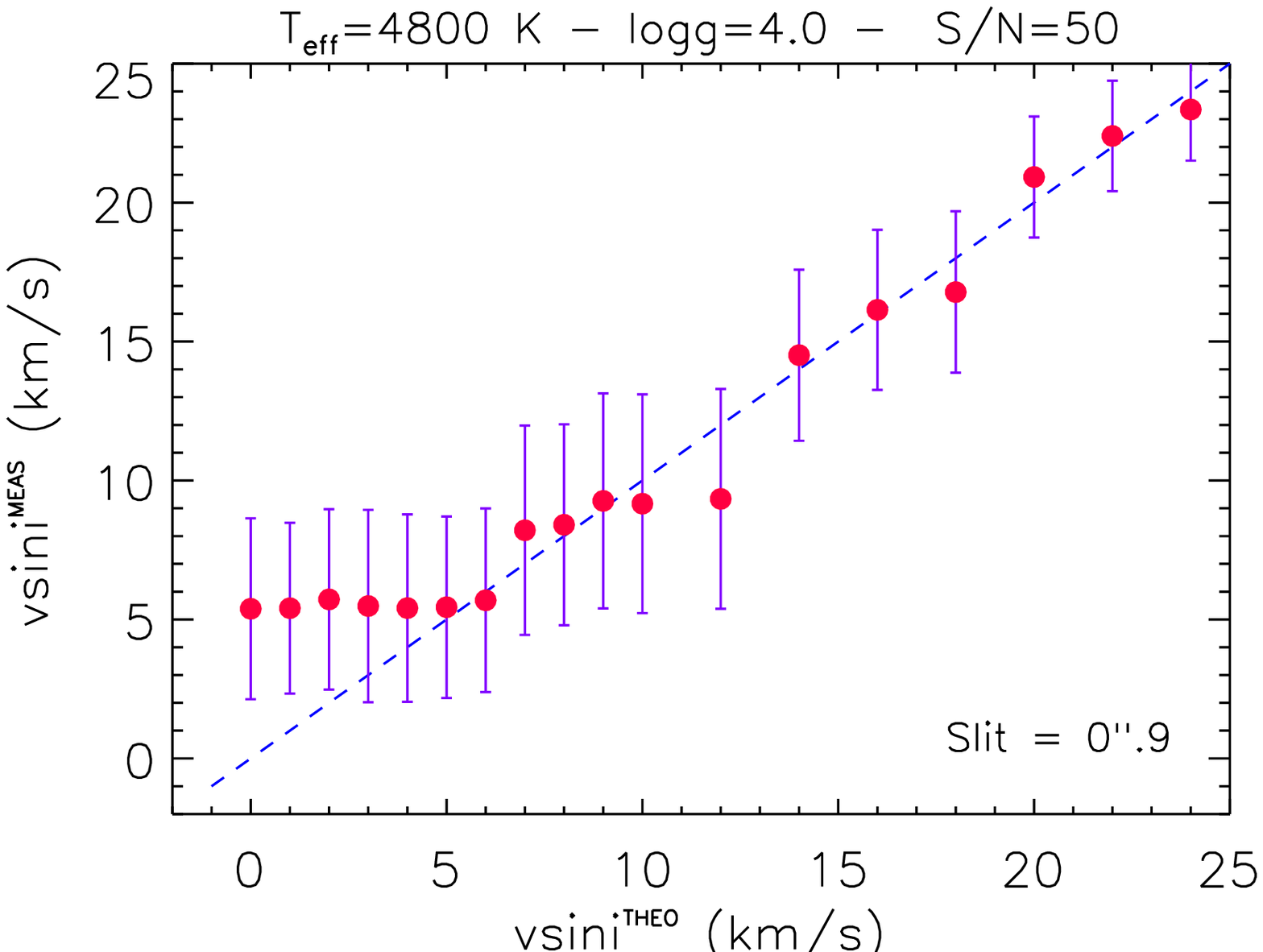}	
\vspace{.4cm}
\includegraphics[width=8cm]{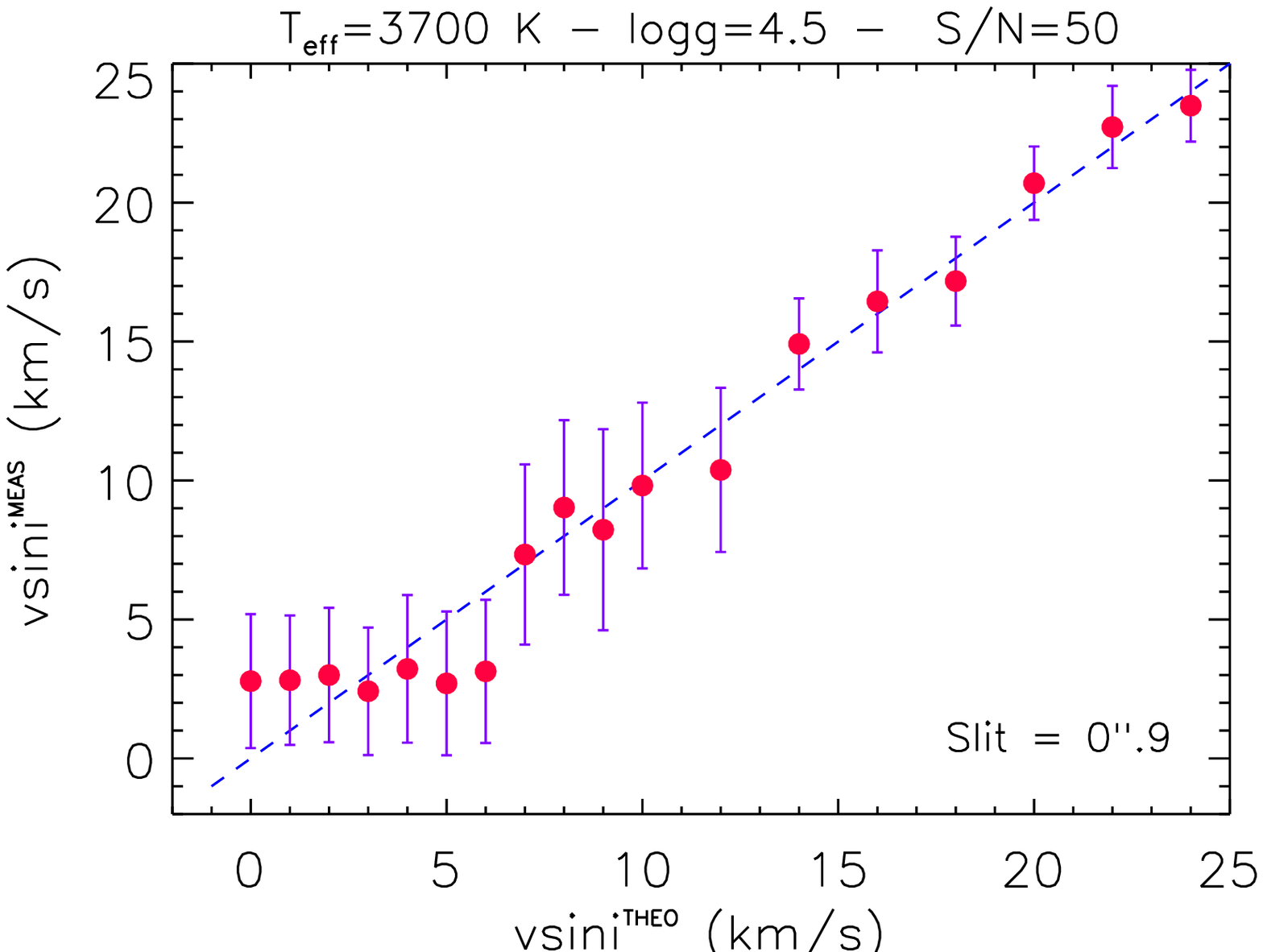}	
\includegraphics[width=8cm]{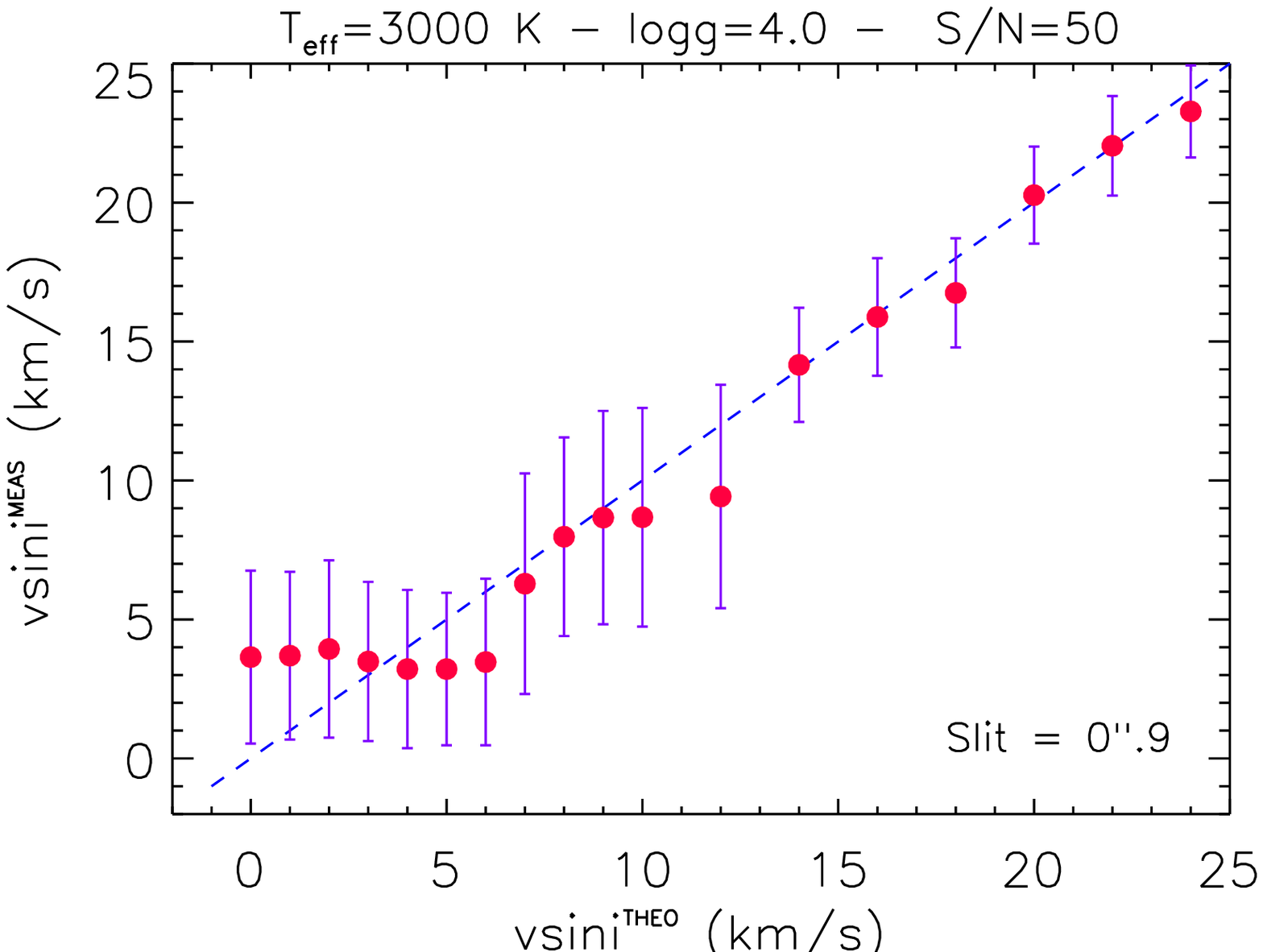}	
\caption{Results of the Monte Carlo simulations on $v\sin i$ made with synthetic spectra at three \teff\  for S/N=50 and $R=8400$ (slit width 0$\farcs$9). 
The average $v\sin i$ values measured with our procedure (dots) are plotted against the ``theoretical'' $v\sin i$ to which the spectra have been 
broadened. The errorbars represent the standard deviations of the 200 simulations at each \vsini.
The 1:1 relation is plotted with a dashed line.}
\label{Fig:MonteCarlo}
\end{figure}

We run Monte Carlo simulation to evaluate the resolution in \vsini\  that can be reached with these data and our analysis code. 
To this aim, we built synthetic spectra at the same resolution and sampling as the X-Shooter ones.
We broadened these spectra from 0 to 24\,\kms\  in steps 
of 1 or 2\,\kms\  and made 200 simulations per each \vsini,  adding a random noise corresponding to signal-to-noise ratios S/N=20, 50, and 100. 
These simulations were made with BT-Settl spectra corresponding to \teff\,=\,2700, 3000, 3700, and 4800\,K.
We applied the code ROTFIT to each simulated spectrum, analogously to what we did with the target spectra, to determine the \vsini\  from
the spectral region centered at 9700\,\AA.
As apparent in Fig.~\ref{Fig:MonteCarlo}, where only some examples are shown for the slit width 0$\farcs$9 ($R\simeq8400$), we do not resolve 
\vsini\ values lower than 8 \kms. Thereafter, the measured  \vsini\  starts to grow with the simulated \vsini\  and smoothly follows the one-to-one line.
Therefore, we adopted 8\,\kms\  as upper limit in this configuration. 	
The \vsini\  errors measured with ROTFIT on the target spectra were in most cases quite comparable with those inferred by the aforementioned
Monte Carlo simulations, while for a few spectra they were larger.
In these cases we have adopted the error of the simulation closest in \vsini, \teff\  and S/N  to the target spectrum.   

\begin{figure}[th]
\includegraphics[width=8cm]{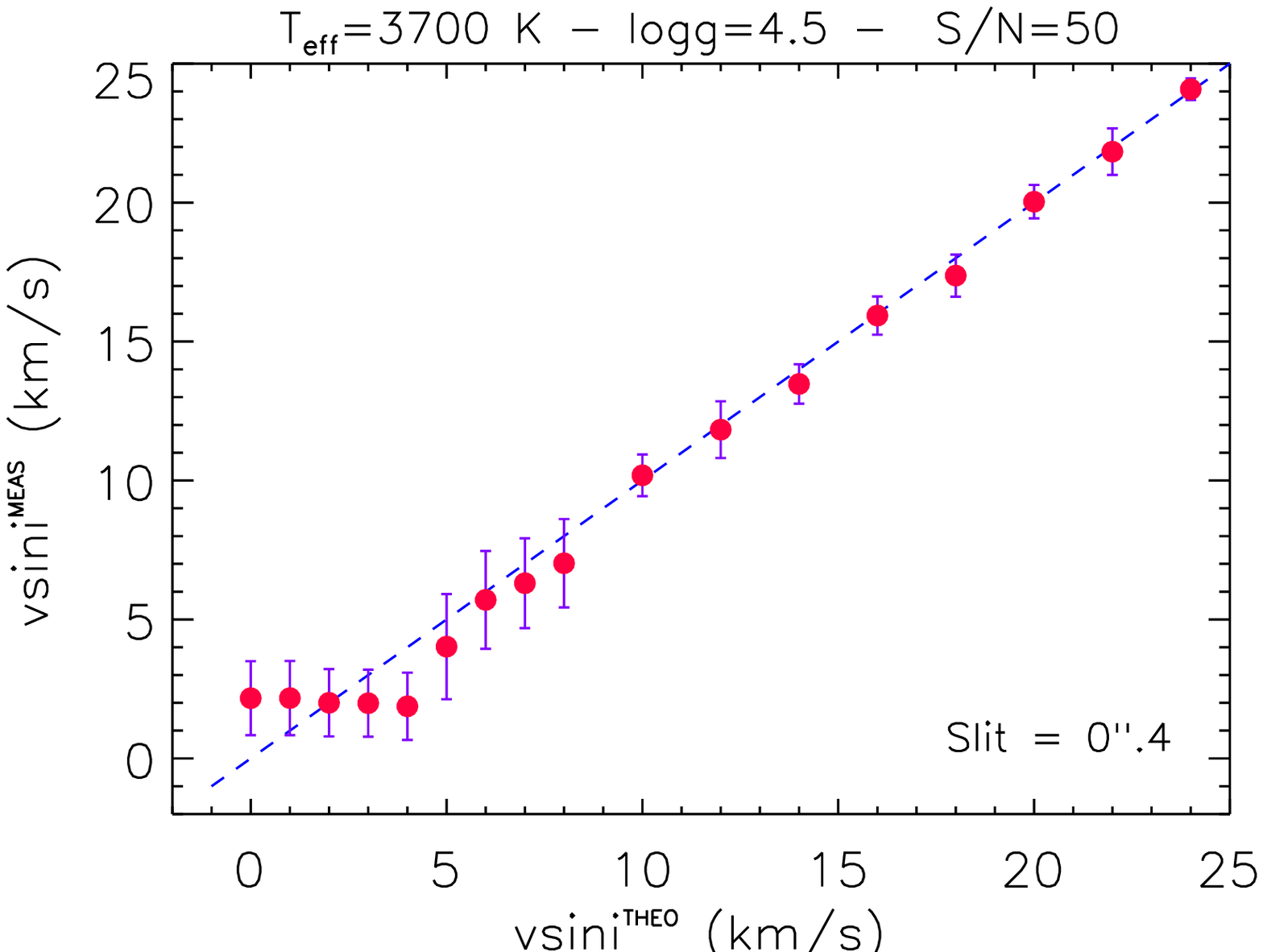}	
\caption{Results of the Monte Carlo simulations on $v\sin i$ for S/N=50 and $R=17\,000$ (slit width 0$\farcs$4). The meaning of the symbols is the same as
in Fig.~\ref{Fig:MonteCarlo}.}
\label{Fig:MonteCarlo_high}
\end{figure}

We also made some tests for the high-resolution X-Shooter spectra (slit width 0$\farcs$4, $R\simeq$\,17\,000). In this case we obtained lower
\vsini\ errors and we estimated an upper limit of 6\,\kms\ (Fig.~\ref{Fig:MonteCarlo_high}).
  
\section{Notes on individual objects}
\label{sec:peculiar}

\subsection{SSTc2d\,J160034.4-422540}
\begin{figure}
\begin{center}
\includegraphics[width=9cm]{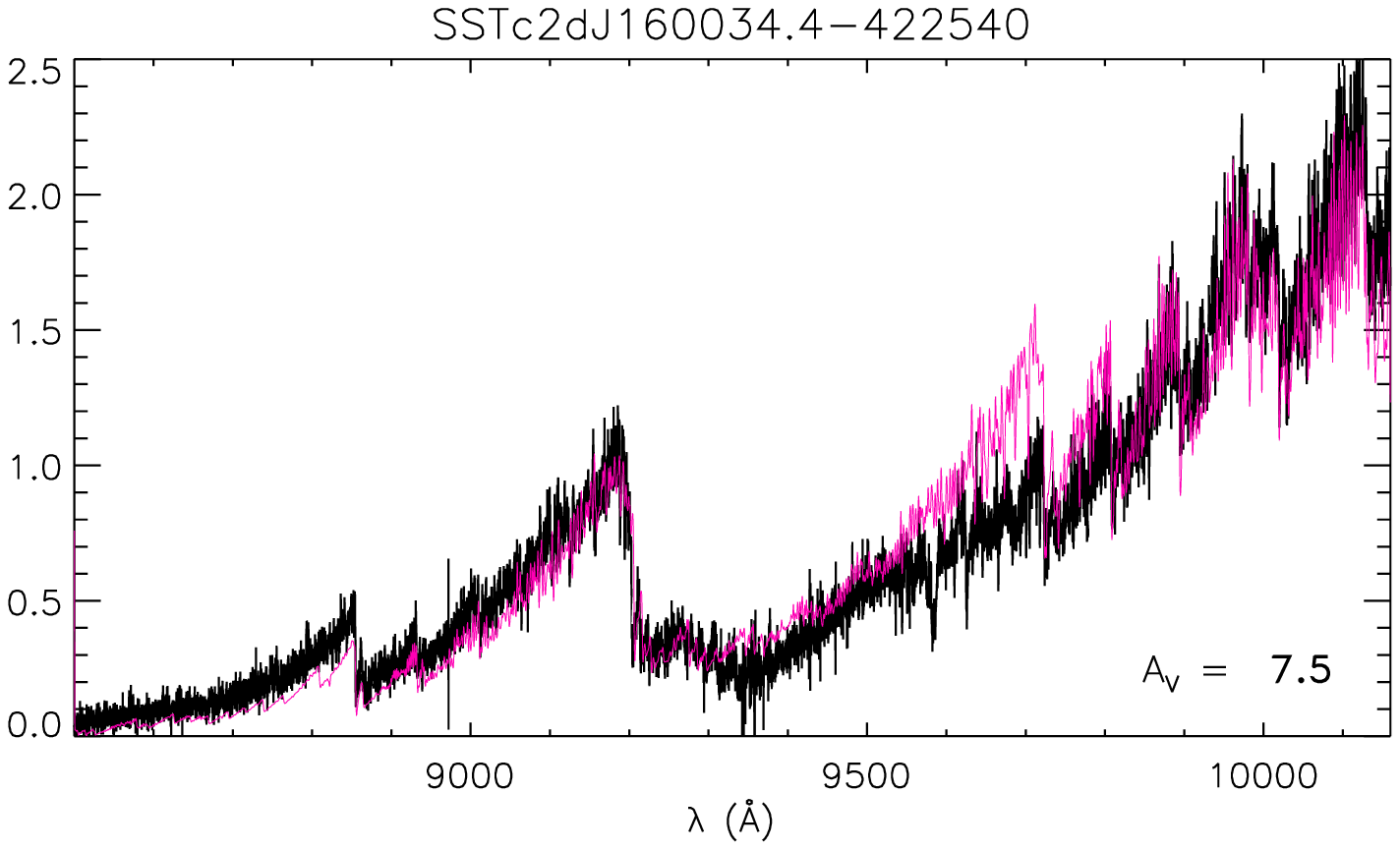}	
\caption{VIS X-Shooter spectrum of SSTc2dJ160034.4-422540  scaled to the level of the TiO molecular-band head at 9170\,\AA\ (black full line) along with the 
best fitting template with a \teff=2000\,K and \logg=0 (red line). An extinction $A_V=7.5$\,mag minimizes the $\chi^2$.}
\label{fig:SSTc2dJ160034.4-422540}
\end{center}
\end{figure}

SSTc2d\,J160034.4-422540 is a faint and very red (and/or reddened) source.  
The VIS spectrum does not show emission lines and has a very low signal in the spectral regions analyzed with ROTFIT, 
which makes the results of this code unreliable for this object. However, the reddest part of the VIS spectrum has  
a S/N sufficient to display the shape of the continuum and the molecular bands. 
We thus analyzed the flux-calibrated spectrum ranging from  8500 to 11\,000\,\AA, with a method similar to the one adopted by \citet{Alcala2006}.
In particular, we scaled to unity both the target and template spectra at 9170\,\AA, which is the wavelength of the head of a strong TiO molecular band. 
 We artificially reddened the template spectra with the extinction law $A_{\lambda}/A_{V}$ of \citet{Cardelli1989}, adopting the average interstellar 
value for  total-to-selective extinction, $R_V=3.1$. The extinction in the $V$~band, $A_{V}$, was let free to vary. 
We kept the parameters of the model spectrum and the $A_{V}$ value that minimize the $\chi^2$ of the fit (see Fig.~\ref{fig:SSTc2dJ160034.4-422540}). 
As the shape of the TiO bands is not very sensitive to the gravity, \logg\ is not well constrained by this  fit, but, however, the atmospheric
parameters found with this method (\teff=2000\,K, \logg=0) suggest a very cool giant star.

\subsection{SSTc2d\,J160708.6-391408}
\label{flat_source}
This source is one of the  brightest YSOs with a flat SED in Lupus \citep{Merin2008, Evans2009}.
Its spectrum includes many permitted 
and forbidden emission lines, but the five VIS spectral segments listed in Table~\ref{Tab:SpecRegions} are
free from emission lines and have a S/N sufficient to be analyzed with ROTFIT. 

\begin{figure}
\begin{center}
\includegraphics[width=9cm]{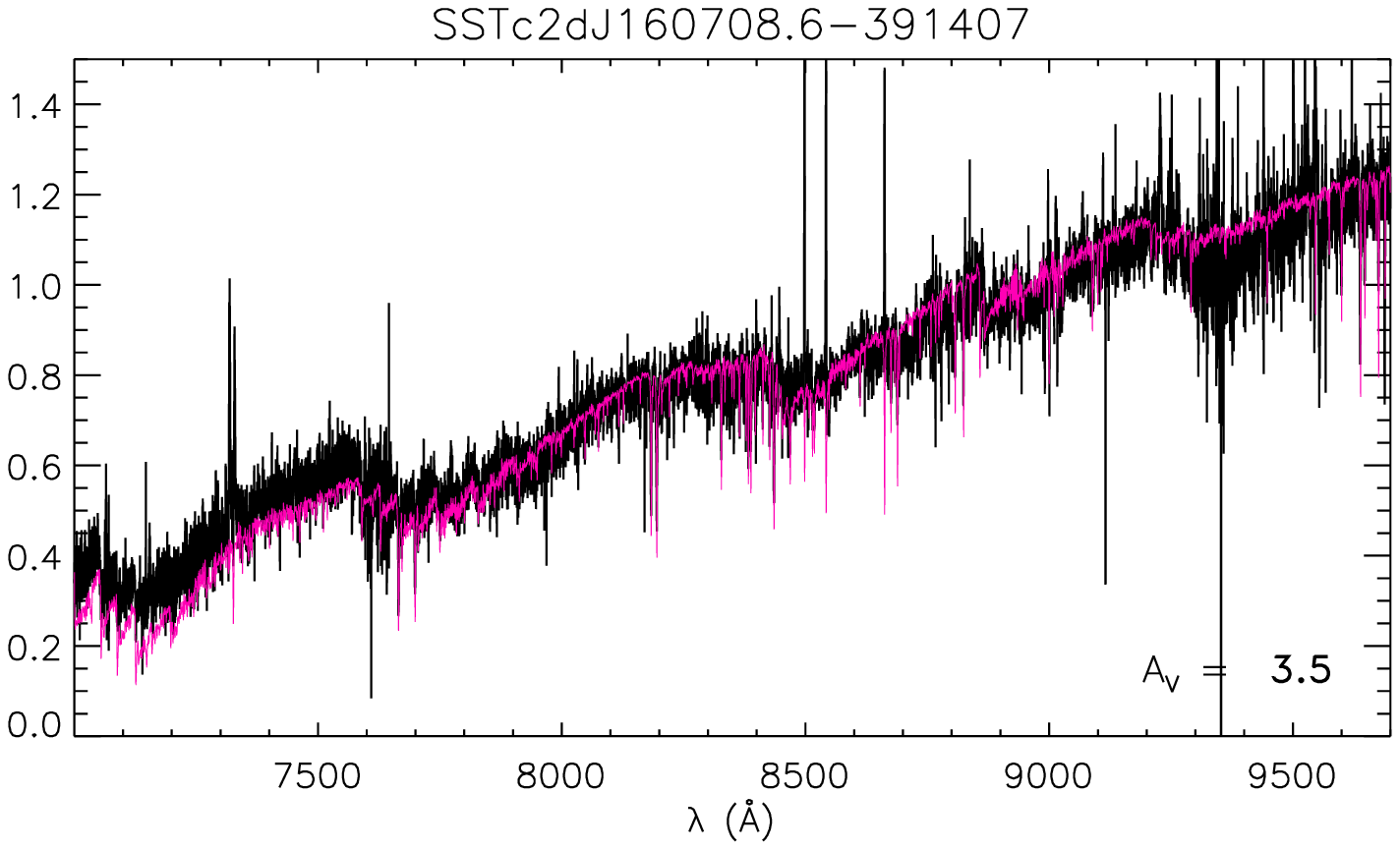}	
\caption{VIS X-Shooter spectrum of SSTc2dJ160708.6-391408  scaled to unity at 9170\,\AA\ (black full line). 
The the best fitting template (\teff=3450\,K) reddened to $A_V=3.5$\,mag is overplotted with a red line.}
\label{fig:SSTc2dJ160708.6-391408}
\end{center}
\end{figure}

We found \teff=\,3474$\pm$206\,K, which corresponds to an M2 spectral type according to the \citet{PecautMamajek2013}
calibrations. \citet{Alcala2017} found a later spectral type (M5) but their results can also be consistent
with an M3 type, the uncertainty resulting from the strong accretion and veiling of the spectrum.
Our \teff\ determination is more consistent with the M1.75 spectral classification of \citet{muzic14}.
We also estimated \teff\ and extinction with the same method used for SSTc2dJ160034.4-422540.  
The results are shown in Fig.~\ref{fig:SSTc2dJ160708.6-391408} where the observed spectrum is best fitted
with a template with \teff=3450\,K and an extinction $A_V=3.5$\,mag.

We dereddened the flux calibrated spectrum of SSTc2dJ160708.6-391408 (obtained merging the UVB, VIS, and NIR spectra) according 
to the \citet{Cardelli1989} law. Integrating the flux over all the X-Shooter wavelength range and multiplying
for $2\pi d^2$ ($d=200$\,pc) we obtain a luminosity \Lstar$=0.032$\Lsun. 
However, it is difficult to distinguish the photospheric flux from that coming 
from the circumstellar matter in the SEDs of flat sources; therefore, the above measure must be considered as 
a rough estimate of the stellar luminosity. 
Nevertheless, this value makes the object rather subluminous in the HR diagram
(green filled square in Fig.~\ref{fig:HR}).

Since flat sources are believed to be YSOs still surrounded by an infalling envelope \citep{calvet94}, 
SSTc2d\,J160708.6-391408 may be in this evolutionary stage in which the energy output from the central star 
can heat the envelope. Therefore, the radiation emitted by the central object would be partly reprocessed 
by the envelope, i.e. it will emerge redistributed to longer wavelengths. In these conditions the bolometric luminosity of 
0.18\,\Lsun, as derived by \citet{Evans2009}, would be a better estimate of the total energy emitted by the
central object. Adopting this value, SSTc2d\,J160708.6-391408 rises to the position marked by the arrowhead 
in Fig.~\ref{fig:HR}, in good agreement with the other Lupus members.
Moreover, as emphasized by \citet{Alcala2017}, the value of 0.18\,\Lsun would bring the ratio of accretion to 
bolometric luminosity in agreement with the YSOs of similar mass.

\subsection{SSTc2dJ160708.6-394723 and SSTc2dJ161045.4-385455}
\label{lithium_rich_gia}
The two non-members with Lithium, namely SSTc2dJ160708.6-394723 and SSTc2dJ161045.4-385455, are likely lithium-rich giants. 
The first star has  \teff\,$\simeq$\,4650\,K and \logg\,$\simeq$\,2.6\,dex that locate it in the RGB region of the \logg\--\teff\ diagram, close
to the red clump locus (Fig.~\ref{Fig:LoggTeff}). For the second object we measured instead a larger gravity \logg\,$\simeq$\,3.4$\pm 0.9$\,dex, 
which is more consistent with a subgiant or RG stage, taking into account the large error.  

The origin of lithium-rich giants is still debated and different hypotheses have 
been put forward to explain the high lithium content in their atmospheres. The main scenarios that have been proposed are the regeneration 
of lithium in later evolutionary stages by the Cameron-Fowler mechanism \citep{CameronFowler1971}, the engulfment of Li-rich substellar 
companions \citep[e.g.,][and references therein]{Alcala2011b,Casey2016} or inhibition of mixing because of fossil magnetic fields
of their Ap-type progenitors \citep[e.g.,][]{Smiljanic2016}. 

\end{appendix}

\end{document}